\documentclass[12pt]{article}
\usepackage{graphicx}
\usepackage{amsmath}
\usepackage{amssymb}
\usepackage{latexsym}
\usepackage{hyperref}
\usepackage[sort]{cite}

{\catcode`\|=\active 
  \gdef\Braket#1{\left<\mathcode`\|"8000\let|\bravert {#1}\right>}}
\def\bravert{\egroup\,\vrule\,\bgroup}
\newcommand{\be}{\begin{eqnarray}}
\newcommand{\ee}{\end{eqnarray}}
\newcommand{\bea}{\begin{eqnarray}}
\newcommand{\eea}{\end{eqnarray}}
\newcommand{\ben}{\begin{equation}}

\newcommand{\del}{\partial}

\newcommand{\nn}{\nonumber}

\numberwithin{equation}{section}

\newsavebox{\ns}
\newsavebox{\dbrane}
\newsavebox{\dbshort}

\usepackage{epsfig}
\usepackage{amssymb}

\def\appendix{{\newpage\section*{Appendix}}\let\appendix\section%
        {\setcounter{section}{0}
        \gdef\thesection{\Alph{section}}}\section}

\newcommand\ba{\begin{eqnarray}}
\newcommand\ea{\end{eqnarray}}

\def\Dslash{\,\,{\raise.15ex\hbox{/}\mkern-12mu D}}
\def\Dbarslash{\,\,{\raise.15ex\hbox{/}\mkern-12mu {\bar D}}}
\def\delslash{\,\,{\raise.15ex\hbox{/}\mkern-9mu \partial}}
\def\delbarslash{\,\,{\raise.15ex\hbox{/}\mkern-9mu {\bar\partial}}}
\def\pslash{\,\,{\raise.15ex\hbox{/}\mkern-9mu p}}
\def\calDslash{\,\,{\raise.15ex\hbox{/}\mkern-12mu {\cal D}}}

\newcommand{\hh}{{1\over 2}}

\renewcommand{\ll}{_}
\newcommand{\uu}{^}
\newcommand{\pp}{\partial}

\renewcommand{\exp}[1]{{\rm exp}\left( #1 \right)}

\renewcommand{\d}{\delta}
\newcommand{\m}{\mu}

\renewcommand{\dag}{{}^\dagger{}}

\renewcommand{\m}{\mu}
\newcommand{\n}{\nu}
\newcommand{\s}{\sigma}
\renewcommand{\t}{\tau}
\newcommand{\G}{\Gamma}
\newcommand{\g}{\gamma}

\renewcommand{\o}{\omega}

\newcommand{\sqd}{^2}

\renewcommand{\hh}{{1\over 2}}

\newcommand{\eee}[1]{\ba{#1}\ea}
\renewcommand{\th}{\theta}
\renewcommand{\t}{\tau}

\renewcommand{\b}{\beta}

\newcommand{\st}{{}^*}
\newcommand{\D}{\Delta}

\newcommand{\pr}{^\prime {}}

\newcommand{\apr}{{\alpha^\prime} {}}

\newcommand{\IZ}{\relax\ifmmode\mathchoice
{\hbox{\cmss Z\kern-.4em Z}}{\hbox{\cmss Z\kern-.4em Z}}
{\lower.9pt\hbox{\cmsss Z\kern-.4em Z}} {\lower1.2pt\hbox{\cmsss
Z\kern-.4em Z}}\else{\cmss Z\kern-.4em Z}\fi} \font\cmss=cmss10
\font\cmsss=cmss10 at 7pt
\newcommand{\inbar}{\,\vrule height1.5ex width.4pt depth0pt}
\newcommand{\IC}{{\relax\hbox{$\inbar\kern-.3em{\rm C}$}}}
\newcommand{\IQ}{{\relax\hbox{$\inbar\kern-.3em{\rm Q}$}}}
\newcommand{\IP}{\relax{\rm I\kern-.18em P}}
\newcommand{\Ione}{{\relax\hbox{$\inbar\kern-.39em{\rm 1}$}}}

\newcommand{\ed}{\dot{e}}

\renewcommand{\l}{\lambda}

\newcommand{\cc}{{\cal C}}

\renewcommand{\cc}{{c_1}}

\newcommand{\phb}{{\bar{\phi}}}

\newcommand{\psb}{{\bar{\psi}}}

\renewcommand{\o}{\omega}

\newcommand{\ct}{\tilde{c}}

\renewcommand{\cc}{c}

\renewcommand{\pr}{{}^\prime{}}

\newcommand{\pst}{\tilde{\psi}}

\newcommand{\cl}{{\cal L}}

\newcommand{\IR}{\relax{\rm I\kern-.18em R}}
\def\blfootnote{\xdef\@thefnmark{}\@footnotetext}

\renewcommand{\cc}[1]{\cite{#1}}

\newcommand{\bm}{\begin{matrix}}

\newcommand{\co}{{\cal O}}

\newcommand{\rr}[1]{(\ref{{#1}})}
\newcommand{\bbb}{\ba}
\renewcommand{\eee}{\ea}
\newcommand{\een}[1]{\label{#1}\ea}

\newcommand{\xxx}{(\xx)}

\newcommand{\cn}{{\cal N}}

\newcommand{\heading}[1]{\ \\ \noindent {\bf #1} \nopagebreak \\  \nopagebreak }

\def\lrdd{\left(\,  }
\def\rrdd{\, \right)}
\def\lsqq{\left[ \,}
\def\rsqq{\, \right]}

\newcommand{\kket}[1]{\left | {#1} \right \rangle }

\def\bi{\begin{itemize}}
\def\ei{\end{itemize}}

\def\ed{\end{document}}

\def\cc{{\cal C}}

\renewcommand{\rr}[1]{(\ref{#1})}

\def\ct{{\cal T}}

\def\cc{\,}

\newcommand{\nnno}[1]{  {{#1}\atop {#1}   }  }

\def\nno{\nnno{\circ}}

\renewcommand\xxx{\nn\\ &&\ \nn\\  }
\newcommand\XXX{ \ee \be }

\def\tlc{{T^{\rm LC}}}

\def\glc{G^{\rm LC}}

\def\ggb{\bar{G}}

\def\qqt{\tilde{Q}}
\def\qbt{\tilde{\bar{Q}}}
\def\qtb{\bar{\tilde{Q}}}
\def\qqb{\bar{Q}}
\newcommand{\fourtrans}[5]{Q\cc{#1} = {#2} \ , &\qquad& 
\qqt\cc{#1} = {#3} \ , \xxx \qqb\cc{#1} = {#4} \ , &\qquad&
\bar{\qqt}\cc{#1} = {#5}}
\def\chb{\bar{\chi}}

\newcommand{\vplus}{{v^+}}
\newcommand{\vminus}{{v^-}}
\newcommand{\stopdoc}{\cheatsheet \end{document}}
\newcommand{\zzz}{\nn\\ &&}
\def\gb{\bar{G}}

\def\bbq{{{\bf Q}_{\psi}}}
\def\calt{\tilde{c}}

\def\calt{\tilde{c}}
\def\gamb{{{\bar{\gamma}}}}

\def\cfive{c_5}                                                       
\def\cfivebar{\bar c_5}                                                      
\def\bfive{b_5} 
\def\bfivebar{\bar b_5} 
\def\cfour{c_4}
\def\cfourbar{\bar c_4} 
\def\bfour{b_4}
\def\bfourbar{\bar b_4} 
\def\cthree{c_3} 
\def\cthreebar{\bar c_3} 
\def\bthree{b_3} 
\def\bthreebar{\bar b_3} 
\def\ctwo{c_2} 
\def\ctwobar{\bar c_2} 
\def\btwo{b_2} 
\def\btwobar{\bar b_2} 
\def\cone{c_1} 
\def\conebar{\bar c_1} 
\def\bone{b_1} 
\def\bonebar{\bar b_1} 
\def\dypfive{{y_5^+}'} 
\def\dymfive{{y_5^-}'}
\def\dupfive{{u_5^+}'}
\def\dumfive{{u_5^-}'}

\def\dupfour{{u_4^+}'}
\def\dumfour{{u_4^-}'}

\def\dupthree{{u_3^+}'}
\def\dumthree{{u_3^-}'}

\def\duptwo{{u_2^+}'}
\def\dumtwo{{u_2^-}'}

\def\dupone{{u_1^+}'}
\def\dumone{{u_1^-}'}
\def\dvpone{{v_1^+}'}
\def\dvmone{{v_1^-}'}
\def\centperp{c^\perp}
\def\cbos{c_{\rm bose}}
\def\balt{{\hat b}}
\def\calt{{\hat c}}
\def\clcbos{{c^{\rm LC}_{\rm bose}}}
\def\tlcbos{{T^{\rm LC}_{\rm bose}}}
\def\tperp{{T^\perp}}
\def\jperp{{J^\perp}}

\def\standard{standard }
\def\TachGrad{ B  }
\def\jcurrent{ {\cal J} }
\def\normx{e^{-i\frac{\TachGrad }{\sqrt{2}} u^+}}
\def\normxb{e^{i\frac{\TachGrad }{\sqrt{2}} u^+}}
\def\rcharge{  {\bf r}  }
\def\tildercharge{ \tilde{\bf r}  }
\def\OMEGA{   }
\def\OMEGABAR{   }
\def\tripleprime{ ''' }
\begin{document}

\begin{titlepage}
\begin{flushright}
arXiv:0709.2166 [hep-th]
\end{flushright}
\vspace{15 mm}
\begin{center}
  {\Large \bf  Supercritical ${\cal N} = 2$ string theory  
}  
\end{center}
\vspace{6 mm}
\begin{center}
{ Simeon Hellerman and Ian Swanson }\\
\vspace{6mm}
{\it School of Natural Sciences, Institute for Advanced Study\\
Princeton, NJ 08540, USA }
\end{center}
\vspace{6 mm}
\begin{center}
{\large Abstract}
\end{center}
\noindent
The ${\cal N}=2$ string is examined in dimensions
above the critical dimension ($D=4$) in a linear dilaton background.  
We demonstrate that string states in this background propagate in a single
physical time dimension, as opposed to two such dimensions present when the dilaton
gradient vanishes in $D=4$.  
We also find exact solutions describing dynamical dimensional reduction and
transitions from ${\cal N}=2$ string theory to bosonic string theory via closed-string 
tachyon condensation.
\vspace{1cm}
\begin{flushleft}
September 16, 2007
\end{flushleft}
\end{titlepage}
\tableofcontents
\newpage

\section{Introduction}
In a series of recent papers \cite{previous,previous2,previous3,previous4}, a number of
exactly solvable string backgrounds were presented, describing novel dynamical 
transitions between various different string theories.  These transitions 
are initiated by closed-string tachyon condensation in some parent theory.
The tachyon condensate forms a bubble of new vacuum that expands
outward from a nucleation point at the speed of light.  The transition to new vacua
is driven by the evolution of target-space time, terminating 
in the late-time limit in a distinct final theory.  Because of the exact solvability 
of these models, these dynamical transitions can be studied in detail.
These solutions provide precise connections between theories that were previously
thought to be entirely disparate.  In particular, we are now able to dynamically
connect theories in different numbers of spacetime dimensions, with varying degrees 
of worldsheet and spacetime supersymmetry.

The mechanisms studied in \cite{previous,previous2,previous3,previous4} can 
be classified into two major categories.  In one class of transitions, the parent theory
is formulated in $D> D_{\rm crit}$ supercritical spacetime dimensions.  
Following condensation of a closed-string tachyon, an expanding bubble of lower-energy
vacuum forms in which string states with excitations along a certain number 
of spatial dimensions are driven out from the bubble region.  
Through this mechanism, some number $n$ of 
spatial dimensions of the parent theory are removed completely from 
the dynamics in the late-time limit, so that the final theory lives in 
a total of $D-n$ dimensions.  From the point of
view of the worldsheet theory, $n$ coordinate embedding fields feel a worldsheet
potential in the region of tachyon condensate that becomes infinitely sharply
peaked at the origin.  These fields can be integrated out of the theory, with the
consequence that the effective couplings in the final theory are renormalized
by loop diagrams.  By simple diagrammatic arguments, however, it can be shown 
that quantum corrections to the connected correlators are zero beyond one-loop
order, and quantum renormalizations can be computed exactly.  The result is that 
the dilaton gradient and string-frame metric receive quantum corrections that ultimately
compensate for the change in central charge due to the loss of $n$ spatial dimensions.

In the second class of transitions, which we call c-duality, 
the parent theory is type 0 superstring 
theory, where the tachyon (which remains in the spectrum for the
type 0 GSO projection) couples to the
worldsheet theory as a $(1,1)$ superpotential.  In this case, tachyon condensation nucleates 
a phase bubble wherein worldsheet supersymmetry is broken spontaneously, signaled 
by the presence of a particular $F$ term in the supersymmetry algebra.  In the late-time
limit, worldsheet supersymmetry is broken completely, and the final phase of the theory
is purely bosonic string theory (with an additional current algebra).  In fact,
the final state of this transition realizes a particular mechanism discovered 
by Berkovits and Vafa in 1993 \cite{bv}, in which the bosonic string can be
embedded in the solution space of the $\cn = 1$ superstring.

Taken together, these mechanisms serve to connect a much broader space of string
theories to the well-known supersymmetric moduli space (or duality web) of critical
superstring theory.  The pictures that emerge are semi-infinite lattices of connected
theories, living in any number of target-space dimensions and possessing varying amounts
of supersymmetry.  In this paper we expand this landscape even 
further to include the supercritical $\cn = 2$ string as a parent theory.  
One interesting aspect of this theory is that, unlike the
critical case, the theory exhibits only a single physical timelike direction.
As will be shown below, the R-symmetry component
of the super-Virasoro conditions in the linear dilaton background 
act to eliminate one of the two timelike directions that are present in the sigma
model prior to the application of the constraints.
In particular, we demonstrate that supercritical $\cn=2$
string theory exhibits both dimension-quenching and c-duality transitions.  
As a corollary, we show explicitly how the bosonic string is embedded in the solution space 
of $\cn = 2$ string vacua.

The $\cn = 2$ string is an interesting object in its own right.  Historically, this 
theory was discovered by Ramond and Schwarz \cite{N2schwarz} in classifying the set
of gauge algebras that admit a Virasoro subalgebra.  It comprises a consistent,
self-dual string theory \cite{OV1,OV2}, and is critical in four real dimensions
(albeit with $2+2$ signature).  While a large number of interesting properties of the 
$\cn = 2$ string have been described, the theory as a whole has remained somewhat 
more removed from direct phenomenological considerations.
We hope that by connecting the $\cn = 2$ string with a large collection of 
distinct string theories, new light may be shed on the broader 
role of this theory in quantum gravity.

In Section \ref{basics} we establish conventions and notation by briefly reviewing the parent 
$\cn = 2$ string theory in supercritical dimensions.  
In Section \ref{tachyon_condensation} we study generic aspects of tachyon condensation 
in the parent theory.  Section \ref{dimred} describes dimension-quenching transitions
driven by tachyon condensation.   In Section \ref{transition}, we describe the 
c-duality transition from $\cn = 2$ string theory to purely bosonic string theory.
Section \ref{cov} provides in detail the explicit variable redefinition that embeds the bosonic
string in the space of vacua of the $\cn =2$ string.  
Section \ref{BRST} describes the BRST quantization of this embedding at tree level in the
string coupling.
The final section contains a summary of our results and a brief discussion of future research 
directions.  A number of useful equations and definitions are recorded in the Appendices.

\section{$\cn =2$ string theory in a linear dilaton background}
\label{basics}
The worldsheet action of $\cn = 2$ string theory admits
the $\cn = 2$ superconformal algebra as a gauge symmetry.  On the cylinder,
the right-moving $\cn = 2$ algebra is characterized by the following 
commutation relations:
\bbb
&&
\lsqq L\ll m , L\ll n 
\rsqq = (n - m) L\ll{m+n} + \frac c {12} \d\ll{m,-n}
(m\uu 3 - m) \ ,
\xxx
&&
\{G\ll r, G\ll s \} = \{\gb\ll r, \gb\ll s \} = 0  \ ,
\xxx
&&
\{ G\ll r, \gb \ll s \} 
	= 2 L\ll{r+s} +  (r - s) J \ll{r+s}
	+ {c\over{12}} (4 r\sqd - 1) \d\ll{r,-s}  \ ,
\xxx
&&
\lsqq
L\ll m , J\ll n 
\rsqq
= - n J\ll{m+n}  \ ,
\qquad \qquad
\lsqq  J \ll m, J\ll n  \rsqq = {c\over{3}} m \d\ll{m,-n}  \ ,
\xxx
&&
\lsqq  L\ll m, G\ll r  \rsqq = \bigl(\hh m - r\bigr) G\ll{n+r}  \ ,
\qquad \qquad
\lsqq  L\ll m, \gb\ll r   \rsqq = \bigl(\hh m - r\bigr) \gb\ll{m+r}  \ ,
\xxx
&&
\lsqq  J \ll m, G\ll r   \rsqq  =  G\ll{m+r}  \ , 
\qquad \qquad 
\lsqq   J \ll m , \gb\ll r  \rsqq = - \gb\ll{m+r}  \ .
\eee
$G\ll m$ and $\bar G\ll m$ are modes of the complex $\cn = 2$ supercurrent,
while $J\ll m$ and $L\ll m$ are modes of the R-current and Virasoro generators, respectively.
The simplest backgrounds of the ${\cal N} = 2$ string have
flat string-frame metric and constant dilaton gradient.
Relative to the linear dilaton backgrounds of bosonic, type 0, type II or heterotic string
theory, the linear dilaton background of $\cn = 2$ string theory has reduced
spatial symmetry, due to the transformation properties of the
spacetime embedding coordinates under worldsheet supersymmetry.
It is natural to employ complex bosonic embedding coordinates $\phi\uu\m$, 
with $\m$ running from $0$ to $D\ll c - 1 \equiv \hh D - 1$ ($D\ll c$ denotes the
number of complex dimensions, or half the number real dimensions $D$).  
Along with their left- and right-moving fermionic superpartners $\pst\uu\m, \psi\uu\m$, the
bosons transform under worldsheet supersymmetry according to the algebra presented 
in Appendix~\ref{SUSY}.

\heading{Signatures}
The $\phi\uu\m$ coordinates comprise the lowest components of $(2,2)$ chiral multiplets. 
As such, they 
have an intrinsic complex structure defined by their supersymmetry transformations. 
Worldsheet supersymmetry restricts the sigma model metric to be K\"ahler with respect
to this complex structure.  Among other things, this means that the real and
imaginary parts of a single complex dimension must have the same signature.
Since we are formulating our theory in Lorentzian spacetime, we demand 
that the direction 
\be
X\uu 0 \equiv {\rm Re}\, \phi\uu 0
\ee 
be timelike.
This would seem to indicate that the presence of a second timelike dimension
\be
Y\uu 0 \equiv {\rm Im}\, \phi\uu 0
\label{udefinition}
\ee
is inevitable (along with the exotic phenomena that typically 
accompany theories with multiple timelike directions).
In a background with constant dilaton, this is indeed the case: the critical
$\cn = 2$ string with vanishing dilaton gradient exhibits 2+2 signature.
We will show, however, that the presence of a timelike or lightlike dilaton
gradient renders the second time direction pure gauge.  In such a background,
the second timelike direction does not constitute
an independent degree of freedom in the physical Hilbert space.

\heading{Symmetries}
We define the dilaton dependence on the bosonic directions by
\be
\Phi = \hh \lrdd W\ll\m \phi\uu\m + W\st\ll\m \phb\uu\m \rrdd \ .
\ee
The complex dilaton gradient $W\ll\m = 2 \del \ll{\phi\uu\m} \Phi$
contributes to the central charge as 
\be
c^{\rm dilaton} = 3 \apr W\uu\m W\st\ll\m \ .
\ee  
To cancel the worldsheet Weyl anomaly, the dilaton 
contribution to the central charge must satisfy 
$c^{\rm dilaton} = 6 - {{3 D}\over 2}= 6 - 3 D\ll c$, 
so we obtain the following condition:
\bbb
W\uu\m W\st\ll\m = {2\over\apr} \lrdd 1 - {D\over 4} \rrdd
= {2\over{\apr}} \lrdd 1 - {{D\ll c}\over 2} \rrdd  \ .
\eee
For $D > 4$, this means $W\ll\m$ must be timelike.  
We can use a $U(D\ll c - 1,1)$ transformation
to set $W\ll i = {\rm Im}\, W\ll 0 = 0$ for $ i = 1,\cdots, D\ll c - 1$.  
Furthermore, without loss of generality, we are free to choose
$W\ll 0 < 0$.  This fixes the following conditions:
\bbb
W\ll 0 = W\ll 0 \st = - s  \ ,
\label{sdefinition0}
\eee
with
\bbb
 s = \sqrt{{D -4}\over{2 \apr}} = \sqrt{{D\ll c - 2}\over{\apr}} \ ,
\label{sdefinition}
\eee
so that the dilaton itself satisfies $\Phi = - s ~{\rm Re}\, \phi\uu 0 = - s X\uu 0  $.

Ignoring the dilaton, the spatial symmetry comprises a semidirect product of
the translation group with the rotation group $U(D\ll c)$.
This group is broken by the dilaton to a 
semidirect product of residual translations, with the residual 
$U(D\ll c - 1)$-dimensional rotational group acting on the spatial
directions.  The $(2 D - 1)$-dimensional group of
residual translations includes $D-2$ real translations in the spatial directions,
as well as a translation in the $ Y\uu 0 =  {\rm Im}\,  \phi\uu 0$ direction.  
The latter will turn out to be unphysical, so the residual symmetry group will be 
generated by $2D-2$ real translations, and $U(D\ll c - 1)$ rotations of the spatial directions.

\subsection{States on the cylinder:  NS states in \standard picture}
Let us now discuss the physical states and operators 
of the ${\cal N} = 2$ string in this background.  We will
begin by discussing the physical state conditions at the level of old covariant
quantization (OCQ), omitting for the moment
any reference to Fadeev-Popov ghosts or BRST quantization.
One important set of physical states of the ${\cal N} = 2$ string is
the set of Neveu-Schwarz (NS) states in \standard picture.  
These correspond to superconformal primaries
of weight $0$.  The requirement is that the nonnegative modes of
$G,\ \bar G,\ J$ and $L$ must annihilate physical states.

To establish the connection between $W\ll\m$ and the string coupling, we
identify the $SL(2,\IC)$-invariant state.  This state
is particularly easy to determine in a supersymmetric theory, as it is annihilated by both
$G\ll{-\hh}$ and $\gb\ll{-\hh}$.
To this end, and to establish basic conventions, we define the usual Fourier mode 
expansions for fields on the cylinder:
\be
\phi_m & = & -\frac{1}{\pi\sqrt{2\apr}} \int d\s e^{-i m \s^+} \del_+ \phi(\s) \ ,
\xxx
\bar \phi_m & = & -\frac{1}{\pi\sqrt{2\apr}} \int d\s e^{-i m \s^+} \del_+ \bar\phi(\s) \ ,
\ee
(note that by $\bar \phi_m$, we mean the $m^{\rm th}$ mode of the conjugated field $\phi$, 
and not the conjugate of the mode $\phi_m$ itself).
Lightcone coordinates are defined with the convention $\s^\pm \equiv \frac{1}{\sqrt{2}}(-\s^0 \pm \s^1)$.
The center-of-mass values for the $\phi$ fields are defined to be
\be
\phi_{\rm CM} = \frac{1}{2\pi} \int d\s \phi \ ,
\qquad
\phb_{\rm CM} = \frac{1}{2\pi} \int d\s \phb \ ,
\label{COM}
\ee
with conjugate momenta given by
\be
p_\mu  =  P_{\phi^\mu_{\rm CM}} = 
	\frac{1}{\apr}\eta_{\mu\nu}\, {\dot{\bar{\phi}}^\nu_{\rm CM}} \ ,
\qquad
\bar p_\mu  =  p_{\bar \phi^\mu_{\rm CM}} = 	
	\frac{1}{\apr}\eta_{\mu\nu}\, {{\dot{ {\phi}}}_{\rm CM}}^\nu \ .
\ee
The canonical commutation relations are
\be
\left[ \dot \phi(\s) , \bar \phi(\t) \right] = -2\pi i \apr \delta(\s - \t) \ ,
\ee
which means
\be
\left[ \del_+ \phi(\s) , \del_+ \bar \phi (\t) \right] = -\pi i \apr \delta'(\s - \t) \ .
\ee
With these conventions, we recover the standard commutation relation for the modes
$\phi_m$ and $\bar \phi_m$:
\be
\left[ \phi_m , \bar \phi_n \right] = m\, \delta_{m,-n} \ .
\label{comm1}
\ee
The Fourier modes of $\psi$ and $\psb$ on the cylinder can be similarly defined,
with the corresponding anticommutation relation
\be
\{ \psi\ll r, \psb\ll s\} = \d\ll{r,-s} \ .
\label{comm2}
\ee
We can therefore write a consistent mode expansion of the
generators of the $\cn = 2 $ algebra:
\bbb
L\ll m &=& \sum\ll n  \phi\ll n \phb\ll{m-n} 
+ \hh \sum\ll r (m-2r) 
 \psi\ll r \psb\ll {m - r}  
\nn\\
&&
+ {{i \sqrt {2\apr}}\over 4}~m~\lrdd W\ll\m \phi\uu\m\ll m
+ W\ll\m^* \phb\uu\m\ll m \rrdd +  {\apr\over 8}  W\uu *\ll\m W\uu\m  ~\d\ll{m,0} \ ,
\xxx
G\ll r &=& \sqrt{2}~\sum\ll n \phb\ll n \psi\ll{r - n} +
i r \sqrt{\apr} W\ll\m \psi\uu\m\ll r   \ ,
\xxx
\gb\ll r &=& \sqrt{2}~\sum\ll n \phi\ll n \psb\ll{r - n} + i r \sqrt{\apr}
W\st\ll\m\psb\uu\m\ll r  \ ,
\xxx
J\ll m &=&   \psi\ll r \psb\ll{m-r}  + 
{{i\sqrt {2\apr}}\over 2}~\lrdd W\ll\m \phi\uu\m\ll m
- W\ll\m^* \phb\uu\m\ll m \rrdd \ .
\eee
At this point we note that all operators are assumed to be normal-ordered.

Using the basic commutators in Eqns.~\rr{comm1} and \rr{comm2},
it is straightforward to verify 
that this set of generators satisfies the $\cn = 2$ algebra with central charge
$c = c\uu{\rm free} + c\uu{\rm dilaton},$ where 
\bbb
c\uu{\rm dilaton} = 3 \apr W\uu *\ll\m W\uu\m \ .
\label{cdilaton0}
\eee
(Of course, we will eventually restrict to the form of $W\ll \m$ in 
Eqns.~(\ref{sdefinition0},\ref{sdefinition}) above.)
The zero-modes $\phi\uu\m\ll 0$ and $\phb\uu\m\ll 0$ 
are related to the translators for the overall
center-of-mass values in Eqns.~\rr{COM} by
\bbb
\phi_0^\m = 
\sqrt{{{\apr}\over 2}}
 \eta\uu{\m\n} p\ll{\phb\uu\n} \ ,
\qquad
\phb^\m \ll 0 = 
\sqrt{{{\apr}\over 2}} 
\eta\uu{\m\n} p\ll{\phi\uu\n} \ ,
\eee
where the momenta $p\ll{\phi\uu\n}$ and $p\ll{\phb\uu\n}$ are defined
according to
\bbb
p\ll{\phi\uu\m} \equiv - i {{\pp}\over{\pp \phi\uu\m\ll{\rm CM}}}\ ,
\qquad
p\ll{\phb\uu\m} \equiv - i {{\pp}\over{\pp \phb\uu\m\ll{\rm CM}}}\ .
\eee
The $G\ll{-\hh}$ and $\gb\ll{-\hh}$ conditions yield the two
independent constraints
\bbb
p\ll{\phi\uu\m} =  {{i W\ll\m}\over 2} = - i \pp\ll{\phi\uu\m} \ ,
\qquad
p\ll{\phb\uu\m} =  {{i W\st\ll\m}\over 2}= - i  \pp\ll{\phb\uu\m} \ ,
\eee
so the identity state has a wavefunction of the functional form
$\exp{ - \hh W\ll\m \phi\uu\m - \hh W\st\ll\m \phb\uu\m}$.

\heading{R-symmetry constraint}
The presence of the dilaton background changes the
nature of the R-symmetry constraints.
As an example, let us consider the oscillator
ground state.  In a background with vanishing dilaton
gradient, the $J_0$ condition is vacuous when acting on the  
ground state.  In contrast, 
the $J\ll 0 $ condition in a background with timelike linear dilaton
imposes the following condition on the ground state:
\be
 p_{Y\uu 0} = 0
\ee
(where $Y\uu 0$ is defined in Eqn.~\rr{udefinition}).
The R-symmetry constraint therefore eliminates one dimension 
as a degree of freedom in the string wavefunction.
For a general state, we obtain the following condition:
\be
 p_{Y\uu 0} 
  = {2\over{\apr s}}  (N_\psi - N_{\psi^\dagger}) 
  = {4\over{\apr s}} \bbq \ ,
\ee 
where $ N_\psi $, for example, counts the number of excited $\psi$
oscillators, and $\bbq \equiv (N\ll \psi - N\ll\psb)$.

We conclude that $Y\uu 0$ does not constitute an independent
degree of freedom in the Hilbert space of the $\cn = 2$ string in a 
linear dilaton background.  Although the multiplet structure of the
$\cn = 2$ supersymmetry forces us from the outset to adopt two timelike directions, 
the gauge constraints of the $\cn =2$ algebra in this 
background leave only a single time coordinate.

\heading{On-shell conditions for normalizable states}
Let us now consider the on-shell condition for string states.
In the old covariant quantization of the ${\cal N} = 2$ string,
the $L\ll 0$ physical state condition is just $L\ll 0 = 0$, so the
on-shell condition is
\bbb
  \eta\uu{\m\n} p\ll{\phi\uu\m} p\ll{\phb\uu\n} + 
{2\over{\apr}} E\uu{\rm osc} +
{1 \over 4} W\ll\m\st W\uu\m = 0 \ ,
\eee
where $E^{\rm osc}$ is the total oscillator energy of the state.
We will assume throughout this section that our states are normalizable in the spatial
directions $\phi\uu 1,\cdots , \phi\uu{D\ll c - 1}$.
The wavefunction of a physical state is therefore of the form
\bbb
\Psi \propto \exp{ i \o X\uu 0 + i \o\pr Y\uu 0 + i K\ll a  \phi\uu a
+ i K\st\ll a \phb\uu a} ,
\eee
where $\o$ and $\o'$ are not assumed to be real.
For such a wavefunction, the on-shell condition reads:
\bbb
- {1\over 4} \lrdd \o\sqd + \o\uu{\prime 2} \rrdd + K\uu {a*} K\uu a + {2\over{\apr}}
E\uu{\rm osc} + {1\over 4} W\st\ll\m W\uu\m = 0 \ .
\eee
If there are $N\ll\psi$ excited $\psi$ oscillators
and $N\ll\psb$ excited $\psb$ oscillators, we obtain
\bbb
\o\pr = p\ll{Y\uu 0} = {2\over{\apr s}} \lrdd N\ll\psi - N\ll\psb \rrdd 
		= {4\over{\apr s}} \bbq  \ ,
\eee
where we have substituted the form for $W\uu\m$ determined above in 
Eqn.~\rr{sdefinition0}.  This implies
\bbb
\o\sqd = 4 K\uu{a*} K\uu a  + {{8 E\uu{\rm osc}}\over{\apr}} - {{16 \bbq \sqd}\over {s\sqd\apr\sqd}}
- s\sqd \ ,
\eee
where the magnitude of $W\uu\m$ is given explicitly in Eqn.~\rr{sdefinition}.

If $\o\sqd \geq 0$, the state is oscillatory and does not represent an
instability.   Let us consider the possibility that
$\o\sqd  < 0$, where the state could potentially represent an
instability.  The physically relevant criterion for stability \cite{previous,evaofer} is
whether or not the wavefunction of the state can grow more quickly than $\exp{-\Phi} = 
\exp{ s X\uu 0}$.
That is, if $\o\sqd = - \G\sqd$, with $\G$ real and positive,
then the state represents an instability if and only if
$\G > s$.  We will now see that, as in the examples studied in \cite{previous,evaofer}, 
a physical state that is normalizable in the spatial directions never 
represents a true instability.

Since we are considering NS states in the \standard picture, each 
$\psi$ or $\psb$ carries an energy of at least $E\uu{\rm osc} = \hh$, 
and there are $D\ll c - 2$ species of $\psi$ fermions transverse to the 
lightcone.\footnote{Light-cone fermions $\psi\uu\pm$ and their
complex conjugates are removed by the fermionic constraints and
null-state equivalences generated by $G,\bar{G}$.}
One attains the lowest energy per unit $\bbq$-charge
by filling fermi surfaces of each species to equal height.
We conclude that the inequality
\be
E\uu{\rm osc} \geq {{2 \bbq\sqd}\over {D\ll c - 2 }}
\ee
must hold.  This, in turn, implies
\bbb
\o\sqd \geq 4 K\uu{a*} K\uu a
+ {{16 \bbq\sqd}\over{\apr(D\ll c - 2)}}  - {{16 \bbq \sqd}\over {s\sqd\apr\sqd}}
- s\sqd  \ .
\eee
The two terms containing $\bbq\sqd$ cancel in the above equation, 
since $s\sqd = {{D\ll c - 2}\over{\apr}}$.  This leaves the condition
\bbb
\o\sqd \geq 4 K\uu{a*} K\uu a - s\sqd  \ .
\eee
We have assumed the state is normalizable in the spatial directions,
so the quantity $K\uu{a*} K\uu a$ is real and positive.  We therefore have that
\bbb
\o\sqd \geq - s\sqd \ ,
\eee
so an exponentially growing state can grow no faster than $\exp{ s X\uu 0}$, precisely
saturating the stability bound.

\subsection{Tachyons and worldsheet superpotentials}
We have seen that the second timelike coordinate $Y\uu 0$ is purely a gauge artifact in
supercritical $\cn = 2$ string theory.  Excitations of $Y\uu 0$ never give rise to 
negative-norm states, a second independent timelike degree 
of freedom in the wavefunction, or physical instabilities.  
This stands in contrast to the $\cn = 2$ string with 
vanishing dilaton gradient, in which there is a true second timelike direction that produces 
a spectrum with energy unbounded from below
(despite the absence of negative-norm states).

We will now discuss a second important way in which the $\cn = 2$ string 
in $D>4$ differs from its
critical counterpart.  In the supercritical $\cn = 2$ string 
there are additional physical states
at non-standard picture that have no counterpart in the critical theory.
These states are non-normalizable, and are easy to understand in the language of operators:
they correspond to superpotential deformations of the Lagrangian that preserve the full
$\cn = 2$ superconformal symmetry.  

\heading{Operators}
Any Lagrangian perturbation preserving the full $\cn = 2$
algebra corresponds to a BRST-invariant operator (when integrated over the worldsheet).
All NS states in \standard picture correspond to Lagrangian perturbations that
come from $\cn = 2$ superconformal primaries $\co$ of weight $(0,0)$ 
and chiral R-charges $(0,0)$, integrated over all four Grassmann coordinates:
\bbb
\cl = \int d\th\ll + d\th\ll +\dag d\th\ll - d\th\ll - \dag {\cal O} \ .
\eee
In other words, NS states in the standard picture represent full superspace perturbations.

In the presence of a linear dilaton background, it is easy to see that there are also
half-superspace perturbations preserving the full $\cn = 2$ superconformal algebra.
Consider a matter operator $\co$ of weight $(h,\tilde h) = (\hh,\hh)$ and chiral R-charges 
$(\rcharge ,\tildercharge ) = (-1,-1)$  that
is annihilated by $\gb\ll{-\hh}$ and $\tilde{\gb}\ll{-\hh}$, in addition
to being primary under the full $\cn = (2,2)$ superconformal algebra.
We would then expect the perturbation
\bbb
\Delta \cl\ll{\rm chiral} = \int d\th\ll + d\th\ll - \co 
\eee
to correspond to an allowed physical state, 
though not necessarily one that is normalizable.
It is useful to understand what such a state looks like in 
integrated and fixed picture, verifying its
BRST invariance explicitly.  This will be the essential
goal in the remainder of this section.

The BRST current can be conveniently defined 
by collecting expressions for the superconformal
generators of the ${\cn} = 2$ ghost sector.  The ghost
sector of the ${\cn} = 2$ string in conformal gauge
consists of the usual reparametrization $bc$ ghosts, a
complex $\beta \g$ system, and a ghost system $\balt\calt$
of weights $(1,0)$, corresponding to the
Fadeev-Popov ghosts of the R-symmetry.  The superconformal
generators for these sectors are:
\be
T_{\rm ghost} & = & 
	2 i  b\, c'  
	+ i  b'\, c 
	+i  \balt \, \calt' 
	+\frac{3}{2}  \left(
		\bar{\beta }\, \gamma '
		+ \beta \, \bar{\gamma }'
		\right) 
	+\frac{1}{2}  \left(
		\bar{\beta }'\, \gamma 
		+ \beta' \, \bar{\gamma }
		\right)  \ ,
\xxx
J_{\rm ghost} & = & 
	i   \bar{\beta }\, \gamma  
	-i  \beta \, \bar{\gamma } 
	-2 i c'\, \balt
	-2 i c\, \balt'	\ ,
\xxx
G_{\rm ghost} & = &
	\sqrt{2}\left(
	2  \balt\, \gamma '	 
	-i  b\, \gamma 
	+i  c\, \beta '
	+ \balt'\, \gamma
		\right) 
	+\frac{1}{\sqrt{2}}\left(
	{3 i} c'\, \beta
	- \calt\, \beta
	\right) \ ,
\xxx
\bar G_{\rm ghost} & = & 
	\sqrt{2}\left(
	i  b\, \bar\gamma 
	+ 2  \balt\, \bar \gamma '
	-i  c\, \bar \beta '
   	+ \balt'\, \bar\gamma 
		\right)
	-\frac{1}{\sqrt{2}} 
	\left( 
	{3 i} c'\, \bar\beta 
	+ \calt\, \bar\beta
	\right) \ .
\ee
Here and below, the prime notation on fields indicates the action of the
lightcone derivative $\del_+$.
With these generators in hand, the BRST current can be written as
\be
j_{\rm BRST} = c \, {\bf T} 
	+ \frac{1}{2} \calt \, {\bf J}
	+ \frac{i}{\sqrt{2}} \bar\gamma \, {\bf G}
	-\frac{i}{\sqrt{2}} \gamma \,  \bar{\bf G} \ ,
\ee
where the bold-faced quantities ${\bf T},~{\bf J},~{\bf G}$ 
are defined to be the usual generators in the physical sector 
plus one half the corresponding quantity in the ghost sector:  
\be
{\bf T} &=& T + \frac{1}{2} T_{\rm ghost} \ ,
\xxx
{\bf J} &=& J + \frac{1}{2} J_{\rm ghost} \ ,
\xxx
{\bf G} & = & G + \frac{1}{2} G_{\rm ghost} \ .
\label{moddefs}
\ee
The left-moving current $\tilde{j}\ll {\rm BRST}$ can be defined similarly.

\heading{Vertex operators for short multiplets }
The familiar physical states of the $\cn = 2$ string are generic, non-BPS (or non-short)
chiral primaries.  These operators are not annihilated by $G_{-\hh}$ or $\bar G_{-\hh}$.
(For brevity, we will generally suppress all left-moving fields and refer only to the right-moving
ghost and matter structure; exceptions will be noted explicitly.)
For non-BPS $\cn = 2$ superconformal primary
matter operators of weight $0$ and R-charge $0$, the correct ghost dressing is 
$\d (\g) \d(\bar{\g}) c $.  (I.e., the dressed operators are BRST invariant.)
Tachyonic excitations do not correspond to generic physical states,
but rather to BPS primaries, denoted here by $\co$.
We now wish to find the appropriate ghost dressing for $\co$.

Since the operators of interest lie in the NS sector, it will be convenient to
avoid bosonizing the superghosts and work directly with the $\beta \g $ and 
$\bar\b \bar\g$ systems. 
In terms of a generic function $f$ of $\bar\gamma$,
we have the following commutation relation involving the BRST charge $Q_{\rm BRST}$:
\be
\left[ Q_{\rm BRST}, c f(\gamb ) \right] & = & 
  \frac{i}{2 }
 c \calt \gamb f'(\gamb)
+i \gamb \g f(\gamb)
 + c c' \left( f(\gamb) + \frac{1}{2} \gamb f'(\gamb) \right)  \ .
\ee
Combining the above with a general matter 
chiral primary $\co$ of weight $h$ and R-charge $\rcharge$, we obtain
\be
\left[ Q_{\rm BRST}, c f(\gamb ) \co  \right] & = & 
 c c' \left( \frac{1}{2} \gamb f'(\gamb) + (1-h) f(\gamb ) \right) \co 
 + i \gamb \g f(\gamb) \co
\xxx
&&
-\frac{i}{\sqrt{2}} c \gamb f(\gamb) \Delta {\cal L} 
 + \frac{i}{2 } c \calt (\gamb f'(\gamb ) -\rcharge f(\gamb) ) \co \ ,
\ee
where
\be
\Delta {\cal L} \equiv G_{-\hh} \co  \ .
\ee
Setting the weight $h = \hh$ and R-charge $\rcharge = -1$, we 
see that the ghost dressing with
$f(\bar\g)  = \delta (\bar\g) $ renders the operator $\co$ 
BRST-invariant.\footnote{Here we have used the identities
$x\, \d(x) = 0$ and $x\, \d\pr(x) = - \d(x)$. }
Likewise, anti-BPS primaries annihilated by $G\ll{-\hh}$ with weight $h = \hh$ and 
R-charge $\rcharge = 1$ must be dressed with $c\,\delta(\g)$.


\heading{Existence of short physical operators}
In the supercritical $\cn = 2$ string, short operators exist that
satisfy the physical state conditions.  The simplest is $\co$, which we take to 
be of the form
$w(\phi)$.   An arbitrary holomorphic function $w$ will be annihilated by 
$\bar{G}\ll{-\hh}$, and will be automatically primary with respect to the
$\cn = 2$ algebra.   The sole conditions remaining to be satisfied are that 
$h = \hh$ and $\rcharge = -1 $.   
A holomorphic function of scalars has no singularities with
itself, so its anomalous and canonical dimensions both vanish.  The only contribution to
the weight is therefore due to the linear dilaton:
\bbb
L\ll 0 \, w(\phi) =  \frac{\apr}{4} W^{*\mu} \del_\mu w(\phi ) 
	= \frac{\apr s}{4}\del_{\phi^0} w(\phi ) \ .
\eee
The R-charge is also determined by the dilaton term:
\bbb
J_0\,  w(\phi)  =   \rcharge_w w(\phi)  = -w(\phi)  = -\frac{\apr s}{2} \del_{\phi^0} w (\phi)\ .
\eee
%
%
These conditions are saturated by the form
\bbb
w(\phi) = \exp{ \TachGrad \phi\uu 0 / \sqrt{2} }~{\bf W}(\phi\uu a) \ ,
\eee
where $a = 1,\cdots,D\ll c - 1$, and 
\be
\TachGrad =  \frac{2 \sqrt{2} }{s\apr} \ .
\ee
Note that $\TachGrad$ is positive:  the holomorphic tachyon can only increase 
exponentially in the direction of weak string coupling.
This condition is automatically enforced, and differs from the cases studied 
in \cite{previous,previous2,previous3,previous4}, where the condition was chosen 
to render the background solvable.

\heading{States of short multiplets}
States corresponding to the BPS primaries we have specified above 
are in the Fock vacuum of
the $\beta,~ \gamb$ system, but in the identity sector
of the 
conjugate system.  In other words, the states  
are annihilated by all positive modes of the $\bar{\beta}\ll{r},~\g\ll {r}$ oscillators,
except for $\g\ll{ \hh}$. 
They are, however, annihilated by the raising operator $\bar{\beta}\ll{-\hh}$.  
The properties of the conjugate states 
are the same, with the roles of $\g,~\bar{\beta}$
interchanged with the roles of $\gamb,~\beta $.  There is a shift in the $\phi\uu 0$ momentum
of the state relative to that of the corresponding operator, by virtue of the Liouville
term.  Explicitly, the wavefunction takes the functional form 
\bbb
\kket{\exp{-  W\ll\m^* \phb\uu\m / 2 } f(\phi) }
= \kket{\exp{ s \phb\uu 0 / 2} f(\phi) } 
\eee
for a short operator, and
\bbb
\kket{\exp{-  W\ll\m \phi\uu\m / 2} f(\phb) }
= \kket{\exp{  s  \phi\uu 0  / 2 } f(\phb) }
\eee
for an antishort operator, 
where we have taken an arbitrary holomorphic function $f(\phi)$.
The on-shell condition then requires that $f(\phi)$ be of
the form
\bbb
f(\phi) = {\rm exp}\left[ \left( \frac{s}{2} +  {2\over{\apr s}} \right) \phi\uu 0 
			\right]\, {\bf W}(\phi\uu a) \ .
\eee

\heading{GSO projections and holomorphic tachyons}
The subject of consistent GSO projections for the ${\cal N} = 2$ string
is a rich and interesting one.  Even in the critical dimension $D = 2D\ll c = 4$
there are a large number of options from which one can choose. 
For a more detailed exposition on this subject, the reader is referred to
\cite{N2GSO}, for example.  
In dimensions $D>D\ll c$, the set of allowed GSO projections is
undoubtedly even more intricate.  Since we are interested 
in studying tachyon condensation, we will adopt the diagonal GSO
projection $(-1)\uu{F_W} = 1$, which is the simplest
GSO projection that leaves all tachyons present in the space of physical states.  
This projection acts with a $(-1)$ on 
all left- and right-moving fermions simultaneously.
For modular invariance to hold, one needs to include in 
the spectrum a single Ramond-Ramond sector, in which
all worldsheet fermions are periodic.  With this inclusion,
the diagonal GSO projection is always modular invariant 
for any 2D theory of free fermions.

\section{Condensation of holomorphic tachyons}
\label{tachyon_condensation}
We aim to study the physics of holomorphic tachyon condensation in
supercritical $\cn = 2$ string theory.
As noted, we will adopt the diagonal GSO projection, leaving the
holomorphic tachyons as allowed deformations.
The off-shell supersymmetry transformations of the $D\ll c$ chiral multiplets are given in 
Appendix~\ref{SUSY}.   
From these rules, we can construct a supersymmetric kinetic action.  Modulo
total derivatives, we have
\be
{\cal L}\ll{\rm kin} &=& Q\tilde{Q}\bar{Q}\tilde{\bar{Q}}
\lrdd - {1\over{2\pi\apr}} \phi\uu\m\phb\ll\m \rrdd
\xxx
   &=& {1\over{2\pi}} F\uu\m \bar{F}\ll\m - {i\over{\pi}} 
\pst\pp\ll + \tilde{\psb} - {i\over\pi}
\psi\pp\ll - \psb 
\xxx
&&
	+ {1\over{\pi\apr}}
(\pp\ll + \phi\uu\m)(\pp\ll - \phb\ll\m) 
+ {1\over{\pi\apr}}
(\pp\ll + \phb\uu\m)(\pp\ll - \phi\ll\m)\ .
\label{kinaction}
\ee

The OPE of the fundamental $\phi$ fields reads:
\be
\phi^\m (\sigma) \bar \phi^\n (\tau) \sim -\frac{\apr}{2}
	\log 
	\left| (\sigma^+ - \tau^+) (\sigma^- - \tau^-) \right| \eta^{\m\n} \ ,
\ee
where, as usual, $\sim$ indicates equivalence up to nonsingular terms.
The $\psi$ fields admit the OPE
\be
\psi(\s)^\m \psb(\t)^\n \sim \psb(\s)^\m \psi(\t)^\n \sim \frac{i}{\s^+ - \t^+ }\eta^{\m\n}\ .
\ee
Using the above equations, it is straightforward to verify that the local 
superconformal currents in this theory satisfy the OPEs recorded in 
Appendix \ref{appope}.

To make contact with the notation employed in \cite{previous,previous2,previous3,previous4},
we introduce the usual dilaton gradient $V_\mu$, which is given in terms of $W_\mu$ by
\be
V_\mu = \frac{W_\mu}{\sqrt{2}} \ .
\ee
This theory admits the following stress tensor and complex supercurrent:
\bbb
T &=& - {2\over\apr}    \pp\ll + \phi\uu\m \pp\ll + \phb\ll\m   
	+ {i\over 2}   \left(
		\psi\uu\m\pp\ll + \psb\ll\m   
		+ \psb\uu\m \pp\ll + \psi\ll\m 
		\right)   
	+ {1\over{\sqrt{2}}}  \left(
		V\ll\m \pp\ll + \sqd \phi\uu\m 
		+V\ll\m\st \pp\ll + \sqd \phb\uu\m 
		\right)   \ ,
\xxx
G &=& 
{{2\OMEGA}\over{\sqrt{\apr}}} \psi\uu\m \pp\ll + \phb\ll\m
- \OMEGA\sqrt{2\apr} V\ll\m \pp\ll + \psi\uu\m \ .
\eee
We have chosen a normalization of $V\ll\m$ in the stress tensor such 
that the linear dilaton contribution to the central charge takes the form 
determined above (see, e.g., Eqn.~\rr{cdilaton0}):
\be
c\uu{\rm dilaton} = 6\apr V\ll\m\st V\uu\m = 3\apr W\ll\m\st W\uu\m \ .
\ee  
There is also an R-current defined by
\bbb
J &=&  \psi\uu\m\psb\ll\m  - i \sqrt{2} V\ll\m\pp\ll + \phi\uu\m
+ i \sqrt{2} V\st\ll\m \pp\ll + \phb\uu\m \ .
\eee

At this point, we restrict $V\ll \m$ (and hence $W\ll\m$) to lie entirely in the real timelike direction:
\bbb
&& V\ll + = V\ll - = V\ll + \st = V\ll - \st = - {q\over{\sqrt{2}}}\ ,
\een{cdilaton}
with $q$ taken to be real and greater than zero.
Here, $q$ is related to the quantity $s$ defined above (Eqn.~\rr{sdefinition}) 
by\footnote{In addition to $V\ll \m$, we have introduced $q$ to make contact with 
the notation in \cite{previous,previous2,previous3,previous4}.}
\be
 q = \frac{s}{\sqrt{2}}\ .
\ee
This sets the dilaton decreasing toward the future. 
For later convenience, we also note the component values $V_0 = V_0^* = -q$.

The critical central charge for the ${\cal N} = 2$ string is
$c_{\rm total} = 6$.  In $D\ll c $ complex dimensions there is
a contribution of $\Delta c = 2D\ll c$ from the scalars and
$\Delta c = D\ll c $ from their fermionic superpartners.  
With the assignment in Eqn.~\rr{cdilaton}, the 
total dilaton contribution to the 
central charge is $c^{\rm dilaton} = -6\apr q\sqd$. 
The magnitude of the dilaton gradient must satisfy $2\apr V^2 = -(D_c - 2)$, 
so we set 
\bbb
q = \sqrt{{{D\ll c - 2}\over{2\apr}}}\ .
\eee

In what follows, we wish to consider interaction terms in the 
kinetic Lagrangian in Eqn.~\rr{kinaction}, obtained by perturbing 
with a superpotential (plus its Hermitian conjugate).  
For a general superpotential $w$, we have
\bbb
{\cal L}\ll{\rm int} 
	&=& Q\tilde{Q}\lrdd - {i\over{2\pi}} w \rrdd
+ {\rm h.c.}
\xxx
	&=& {{i\apr}\over{2\pi}}
(\del_\m\del_\n w) \psi\uu\m\pst\uu\n - {{\sqrt{\apr}}\over{2\pi}}
\del_\m w  F\uu\m + {\rm h.c.}
\eee
By analogy with \cite{previous,previous2,previous3,previous4}, we 
will choose particular forms of the superpotential to 
obtain transitions from the $\cn=2$ parent theory in supercritical
dimensions to theories with reduced spatial dimensions, reduced supersymmetry,
or a combination of the two.  We start by describing {\it dimension-quenching} 
transitions, where condensation of a holomorphic tachyon removes spatial 
degrees of freedom from the theory.

\section{Dimension-reducing transitions}
\label{dimred}
In this section we will describe dynamical dimensional reduction in
supercritical $\cn = 2$ string theory, driven by closed-string tachyon condensation.
In particular, we will describe the condensation of 
a holomorphic tachyon with exponential dependence on a
lightlike direction and quadratic dependence on some number of transverse dimensions.
We therefore focus on tachyon perturbations of
the general form
\bbb
\ct  = \exp{\TachGrad\phi\uu +} \, {\bf q}\ll{ab} \, \phi\uu a \phi\uu b \ ,
\eee
where ${\bf q}\ll{ab}$ is a quadratic form in the holomorphic
coordinates $\phi\uu a,~ a \in \{2,\cdots, D\ll c - 1\}$.  

For instance,
to reduce the number of spacetime dimensions by one complex dimension,
we can choose ${\bf q}\ll{ab}\phi\uu a\phi\uu b \equiv \hh \m\uu{\prime\prime} ~\phi\ll 2\sqd$.  
The resulting worldsheet superpotential is
\bbb
w =  \ct =  {\m\over{2\apr}}~\exp{\TachGrad\phi\uu +}\phi\ll 2\sqd \ ,
\eee
which in components gives the following worldsheet F-term coupling:
\bbb
{\cal L}\ll{\rm int}  = - {{\sqrt{\apr}}\over{2\pi}} F\uu\m \del\ll{\phi\uu\m}w
- {{\sqrt{\apr}}\over{2\pi}} \bar{F}\uu\m \del\ll{\phb\uu\m} w\st  \ .
\eee
Defining $M\equiv \m~\exp{\TachGrad \phi\uu +}$, the bosonic worldsheet 
potential takes the form
\bbb
V\ll{\rm ws} = {{ \m\sqd}\over{2\pi\apr}}~\exp{\TachGrad(\phi\uu + + \phb\uu + )} |\phi\ll 2|\sqd
= {{|M|\sqd}\over{2\pi\apr}}~ |\phi\ll 2|\sqd \ ,
\eee
and the Yukawa coupling is given by
\be
{\cal L}\ll{\rm Yuk} &=& {{i\m}\over{2\pi}}~\exp{\TachGrad\phi\uu +} \lrdd \psi\ll 2 \pst\ll 2
+ \TachGrad\phi\ll 2 \pst\uu + \pst\ll 2 + \TachGrad \phi\ll 2 \psi\ll 2 \pst\uu + 
+ {{\TachGrad \sqd}\over 2} \phi\ll 2\sqd \psi\uu + \pst\uu + \rrdd
~ + {\rm h.c.} 
\xxx
&=& {{i M}\over{2\pi}} \lrdd \psi\ll 2 \pst\ll 2
+ \TachGrad\phi\ll 2 \pst\uu + \pst\ll 2 + \TachGrad \phi\ll 2 \psi\ll 2 \pst\uu + 
+ {{\TachGrad \sqd}\over 2} \phi\ll 2\sqd \psi\uu + \pst\uu + \rrdd
~ + {\rm h.c.} 
\ee
The elements of the $\phi\ll 2$ multiplet can be integrated out, treating the
elements of the $\phi\uu +$ multiplet as fixed source terms.  The 
path integral over the $\phi\ll 2,~\phb\ll 2$ multiplets is purely
Gaussian, so it can be computed exactly.
The effect of
integrating out the $\phi\ll 2$ multiplet is to generate effective
terms involving fields in the $\phi\uu +$ multiplet (and their
conjugates).   Note that these effective terms have no dependence on the fields in
the $\phi\uu -,~\phb\uu -$ multiplets.  As a result, there are no
operator ordering ambiguities in terms that are generated by integrating out
the $\phi\ll 2,~\phb\ll 2$ multiplets.

The simplicity of the worldsheet theory can be understood at the level of
Feynman diagrams.  The $\phi\ll 2,~\phb\ll 2$ multiplets 
admit unoriented propagators.  The $\phi\uu\pm,~\phb\uu\pm$
multiplets have oriented propagators, however, directed from $+$ to $-$ fields.  
If we draw the $\phi\ll 2,~\phb\ll 2$ propagators with solid lines,
and the $\phi\uu\pm,~\phb\uu\pm$ propagators with dotted, oriented lines,
all connected Feynman diagrams have
the structure of a single solid line segment or loop, with outgoing
dotted lines attached at a number of vertices.   There can be no closed loops
involving dotted lines, because the incoming end of the propagator can never
join itself to a vertex.  As a result,
the 2D worldsheet theory is exactly solvable at the quantum level
(for further details, see Ref.~\cite{previous2}).  Notably, quantum corrections 
to the classical theory vanish beyond one-loop order.

The computation of effective interactions in the theory with $\phi\ll 2,~\phb\ll 2$
integrated out can be performed
exactly.  The result is that coefficients of effective scalar couplings of
dimension $\Delta$ and R-charge $\rcharge $ scale
as $\exp{(2 - \D)\TachGrad X\uu + +  {i\over 2}  \rcharge \TachGrad Y\uu +}$,
in the limit where $X\uu + \to \infty$.
Here we have defined 
\be
\phi\uu + \equiv X\uu + + i Y\uu + \ .
\ee  
In the limit of large $X\uu +$, only marginal and relevant couplings can
survive.  Since the path integral over $\phi\ll 2,~\phb\ll 2$
and their superpartners is Gaussian, there are no nonperturbative
quantum corrections, so all effective operators generated
by integrating out the massive degrees of freedom are necessarily D-terms.
It follows that no relevant couplings are generated.  Only marginal 
couplings with $\Delta = 2$ are the only terms generated that survive
in the limit of late light-cone time, $X\uu + \to \infty$.

Employing the techniques described in \cite{hellerman2,previous2},
we calculate the renormalization of the string-frame metric $G_{\m\n}$ and dilaton
gradient generated by integrating out $\phi\ll 2$ and its superpartners.  We obtain
the following results:
\bbb
\Delta G\ll{X\uu + X\uu +} &=&  \Delta G\ll{Y\uu + Y\uu +} =  {{\TachGrad \sqd\apr}\over 2} \ ,
\xxx
\Delta G\ll{X\uu + Y\uu +} &=& 0 \ ,
\xxx
\Delta \Phi &=& \hh \TachGrad X\uu + \, + \, {\rm const.}
\eee 
As in \cite{hellerman2,previous2},
the effect is to renormalize the dilaton contribution to the central charge by
an amount
\bbb
\Delta c\uu{\rm dilaton} =  3 \ ,
\eee
which precisely compensates the loss of central charge due to integrating out the 
chiral multiplet containing $\phi\ll 2,~ \psi\ll 2,~\pst\ll 2$.
Note that the renormalization of the worldsheet metric indeed respects $(2,2)$ supersymmetry.
In particular, the renormalized kinetic term for the bosons
can be written in a form that makes this
manifest:
\bbb
\Delta {\cal L}_{\rm kin} = {1\over{2 \pi\apr}}  \Delta G\ll{X\uu + X\uu +}
( \pp\ll + \phi\uu +  \pp\ll - \phb\uu +   +  \pp\ll - \phi\uu +  \pp\ll + \phb\uu + )\ .
\eee
This comes from the superspace integral
\be
\Delta {\cal L}_{\rm kin} =   
	Q \tilde Q \bar Q \tilde\bar Q  
	\left( -\frac{1}{4\pi \apr } \Delta G\ll{X\uu + X\uu +} \phi^+ \bar \phi^+  \right)    \ .
\ee

The process of dimension quenching described in \cite{previous2,previous4} 
is therefore realized in a precisely analogous manner.  
As described in \cite{previous2,previous4}, however, reducing to 
the minimal dimension raises certain
subtle issues that reach beyond the scope of the present paper.   
For example, there is a nonzero null linear dilaton in the 
late-time limit of our theory:  it would be useful to understand the 
precise relationship between this limit and the 
corresponding theory with no linear dilaton, which exhibits spacetime signature $(2,2)$.


\section{Transition from $\cn = 2$ to bosonic string theory}
\label{transition}
We now turn to transitions (called c-dualities in \cite{previous4}) 
that connect $\cn = 2$ string theory dynamically with bosonic string theory. 
Similar to the transitions starting from $\cn = 1$ parent theories \cite{previous3}, 
the basic ingredient is the condensation of a lightlike tachyon. 
In the system at hand, we let the holomorphic tachyon profile depend on a 
lightlike combination of $\phi\uu 0$ and $\phi\uu 1$.  
The result is that the worldsheet potential for the bosonic fields 
vanishes identically, as do loop and multivertex tree diagrams. 
As in the transitions described in 
\cite{previous,previous2,previous3,previous4}, time evolution in the target space
drives a renormalization group flow on the worldsheet, dressed with an exponential of the
lightlike tachyon profile.  This renders the perturbation as a whole strictly
scale invariant:  the deformed theory is exactly conformal, 
both perturbatively and nonperturbatively in $\apr$.

Specializing to the superpotential 
\be
w \equiv \tilde{\m} \cc\exp{\TachGrad\phi\uu +} 
\label{c_duality_superpotential}
\ee
gives the interaction Lagrangian
\bbb
{\cal L}\ll{\rm int} &=&
- {{i\m}\over{2\pi}} \psi\uu + \pst\uu +\cc\exp{\TachGrad\phi\uu +}
-  {i {\bar{\m}}\over{2\pi}} \psb\uu + \tilde{\psb}\uu +\exp
{\bar{\TachGrad}\phb\uu +}
\xxx
&&
+ {{\m}\over{2\pi\TachGrad \sqrt{\apr}}} F\uu + \cc\exp{\TachGrad \phi\uu +}
+ {{\bar{\m}}\over{2\pi\bar \TachGrad \sqrt{\apr}}} \bar{F}\uu + \exp{\bar{\TachGrad}\phb\uu +}\ ,
\een{LINT}
where we have defined
\bbb
\m \equiv \TachGrad \sqd\apr \cc\tilde{\m}\ , 
\qquad \qquad 
\bar{\m}
\equiv \bar{\TachGrad}\sqd\apr \cc\tilde{\bar{\m}}    \ .
\eee
By further setting
\bbb
M\equiv \m\cc\exp{\TachGrad\phi\uu +} \ ,
\qquad 
\bar{M}\equiv \bar{\m} \cc\exp{\bar{\TachGrad}\phb\uu +} \ ,
\eee
we can write Eqn.~\rr{LINT} as
\bbb
{\cal L}\ll{\rm int} &=&
- {{i M}\over{2\pi}} \psi\uu + \pst\uu +\cc
- i {{\bar{ M }}\over{2\pi}} \psb\uu + \tilde{\psb}\uu +
+ {{ M}\over{2\pi\TachGrad \sqrt{\apr}}} F\uu + 
+ {{\bar{M}}\over{2\pi\bar\TachGrad \sqrt{\apr}}} \bar{F}\uu + \ .
\eee
Solving the equations of motion for the auxiliary fields, we
find
\bbb
F\uu + = \bar{F}\uu + = 0\ ,
\qquad
F\uu - = {{\bar{M}}\over{\bar\TachGrad \sqrt{\apr}}} \ , 
\qquad 
\bar{F}\uu - = 
{{M}\over{\TachGrad \sqrt{\apr}}}\ .
\een{sbosfterm}
The worldsheet potential therefore vanishes, and 
we obtain the following Lagrangian, split into 
lightcone and transverse contributions:
\bbb
{\cal L} &=&  
{\cal L}\uu{\perp} +
{i\over{\pi}} \biggl[
\pst\uu +\pp\ll + \tilde{\psb}\uu -
+ \pst\uu -\pp\ll + \tilde{\psb}\uu +
 + \psi\uu + \pp\ll - \psb\uu - 
 + \psi\uu - \pp\ll - \psb\uu +
- {{ M}\over{2}} \psi\uu + \pst\uu +\cc
-  {{\bar{ M }}\over{2}} \psb\uu + \tilde{\psb}\uu +  
\biggr]
\xxx
&&
\kern-15pt
- {1\over{\pi\apr}}
\biggl[
(\pp\ll + \phi\uu +)(\pp\ll - \phb\uu -) 
+ (\pp\ll + \phi\uu -)(\pp\ll - \phb\uu + ) 
+ (\pp\ll + \phb\uu + )(\pp\ll - \phi\uu - )
+ (\pp\ll + \phb\uu -)(\pp\ll - \phi\uu +)
\biggr]\ .
\eee
The transverse Lagrangian  ${\cal L}\uu\perp$ takes the form
\bbb
{\cal L}\uu\perp \equiv 
{1\over{\pi\apr}}
(\pp\ll + \phi\uu a )(\pp\ll - \phb\uu a ) 
+ {1\over{\pi\apr}}
(\pp\ll + \phb\uu a)(\pp\ll - \phi\uu a )\ ,
\eee
with $a$ running from $2$ to $D_c -1$.  

The marginality condition for the tachyon is obtained by demanding that 
$\int d\theta_+ d\theta_- \exp{ \TachGrad \phi^+ }$ is weight $(1,1)$.
This translates to the requirement that
$\TachGrad q = {2/{\apr}}$, and we will henceforth assume this
relationship. (In particular, note that $\TachGrad $, along with $q$, is real and 
positive.)   At this stage, the equations of motion appear as
\bbb
\begin{array}{cc}
\pp\ll - \psi\uu - = \hh\bar{M}\tilde{\psb}\uu + \ ,
& \qquad 
\pp\ll + \pst\uu - = - \hh \bar{M} \psb\uu + \ ,
\\ & \\
\pp\ll - \psb\uu - =  \hh M \pst\uu + \ ,
&\qquad 
\pp\ll + \tilde{\psb}{}\uu - = - \hh M \psi\uu + \ ,
\\ & \\
\pp\ll + \pp\ll - \phi\uu - =  {{i\apr \TachGrad}\over 4}
\bar{M} \psb\uu + \tilde{\psb}{}\uu + \ ,
&\qquad 
\pp\ll + \pp\ll - \phb\uu - =  {{i\apr\TachGrad}\over 4}
M\psi\uu + \pst\uu + \ ,
\\ & \\
\pp\ll - \psi\uu + = \pp\ll - \psb\uu + = \pp\ll +
\pst\uu + = \pp\ll + \tilde{\psb}{}\uu + = 0 \ ,
&\qquad
\pp\ll + \pp\ll - \phi\uu + = \pp\ll + \pp\ll - \phb\uu + = 0\ .
\end{array}
\eee

The stress tensor can also be decomposed into contributions
from the lightcone and transverse sectors of the theory:
\bbb
T\equiv \tlc + T\uu{\perp}\ ,
\eee
where
\bbb
\tlc \equiv T\uu{\phi\uu\pm} + T\uu{\psi\uu\pm}\ .
\eee
Explicitly, we obtain following:
\bbb
T\uu{\phi\uu\pm} &\equiv&
 {2\over\apr} 
	\left(
	\pp\ll + \phi\uu + \pp\ll + \phb\uu - 
	+ \pp\ll + \phi\uu - \pp\ll + \phb\uu + 
	\right)
\xxx
&&
- {q\over 2} \left( \pp\sqd\ll + \phi\uu +
+ \pp\sqd\ll + \phi\uu -
+ \pp\sqd\ll + \phb\uu +
+ \pp\sqd\ll + \phb\uu - \right)  \ ,
\xxx
T\uu{\psi\uu\pm}  &\equiv&
 - {i\over 2} 
	\left(
	\psi\uu + \pp\ll + \psb\uu - 
  	+ \psi\uu - \pp\ll + \psb\uu + 
	+ \psb\uu + \pp\ll + \psi\uu - 
  	+ \psb\uu - \pp\ll + \psi\uu + 
	\right) \ ,
\xxx
T\uu\perp &\equiv &- {2\over\apr} \pp\ll + \phi\uu i \pp\ll + \phb\uu i  
	+ {i\over 2} \left(
	\psi\uu i \pp\ll + \psb\uu i 
	+  \psb\uu i \pp\ll + \psi\uu i  
	\right) \ .
\eee
The stress tensor $\tlc$ is conserved in the 
presence of the interaction, even if $T\uu{\phi\uu\pm}$ and
$T\uu{\psi\uu\pm}$ are not separately conserved.

As with the c-duality transitions starting from type 0 string theory, 
the perturbing superpotential induces worldsheet interaction terms
that become infinitely strong in the distant future.  Working by analogy from
\cite{previous3}, we perform a canonical variable redefinition
to render these interaction terms weakly coupled in the distant future:  
\bbb
\begin{array}{cc}
\psi\uu - \equiv \bar{M}\cc\tilde{\bar{c}}\ll 6 \ ,
& \qquad 
\pst\uu -\equiv - \bar{M} \cc\bar{c}\ll 6 \ ,
\\ & \\
\psb\uu - \equiv M \cc\tilde{c}\ll 6 \ ,
&  \qquad 
\tilde{\psb}\uu - \equiv - M c\ll 6 \ ,
\\ & \\
\psi\uu + \equiv
2 c\ll 6 \pr - {1\over M} \tilde{\bar{b}}\ll 6  + 2\TachGrad (\pp\ll +
\phi\uu +) c\ll 6 \ ,
&  \qquad 
\pst\uu +\equiv 2 \tilde{c}\pr\ll 6 + {1\over M} \bar{b}
\ll 6
+ 2\TachGrad (\pp\ll - \phi\uu +) \tilde{c}\ll 6  \ ,
\\ & \\
\psb\uu +\equiv
2\bar{c}\ll 6 \pr - {1\over{\bar{M}}} \tilde{b}\ll 6
 + 2 \bar{\TachGrad}
(\pp\ll + \phb\uu +) \bar{c}\ll 6 \ ,
& \qquad 
\tilde{\psb}\uu + \equiv
2\tilde{\bar{c}}\ll 6\pr 
+ {1\over{\bar{M}}} b\ll 6
+ 2\bar{\TachGrad} (\pp\ll - \phb\uu +)
\tilde{\bar{c}}\ll 6  \ ,
\\ & \\
\phi\uu - \equiv \chi\uu - - i \bar{\TachGrad} \apr \bar{M}
\bar{\tilde{c}}\ll 6 \bar{c}\ll 6  \ ,
& \qquad 
\phb\uu - = \chb\uu -  - i \TachGrad \apr M \tilde{c}\ll 6  c\ll 6  \ ,
\\ & \\
\phi\uu + = \chi\uu + \ ,
& 
\phb\uu + = \chb\uu +  \ .
\end{array}
\eee
We have traded the light-cone fermions $\psi\uu\pm,\psi\uu{\pm\dagger}$ for
a complex $bc$ ghost system labeled 
by $b_6$, $c_6$ and  $\bar b_6$, $\bar c_6$ (and their left-moving counterparts,
which remain suppressed).  For lack of better terminology, we will refer to these
objects as {\it antighosts} (though not to be confused with antighosts in the sense of 
BV quantization).  We will also count the imaginary parts $u\uu \pm$ of the
lightcone bosons as antighosts, for reasons that will become clear.

To preserve Lorentz invariance, these antighost fields have spins that are shifted
from the more familiar Fadeev-Popov ghosts associated with worldsheet reparametrization
symmetry.  For reference, we record in Tab.~\ref{TAB1} 
the weights and R-charges of the various objects we have introduced thus far,
including the shifted antighost system.  Furthermore, in accordance with 
Ref.~\cite{previous3}, we refer to the new set of variables as 
{\it infrared} (IR) variables (while the original variables are denoted as {\it ultraviolet},
or UV).  In terms of IR variables, the lightcone action takes the form
\bbb
{\cal L}\ll{\rm LC} &=& 
 {1\over{\pi\apr}}
\biggl(
(\pp\ll + \chi\uu +)(\pp\ll - \chb\uu -) 
-  (\pp\ll + \chi\uu -)(\pp\ll - \chb\uu + ) 
-  (\pp\ll + \chb\uu + )(\pp\ll - \chi\uu - )
-  (\pp\ll + \chb\uu -)(\pp\ll - \chi\uu +)
\biggr)
\xxx
&&
+ \frac{i}{\pi} \biggl(
- \bar{b}\ll 6 \pp\ll - c\ll 6 
-  b\ll 6 \pp\ll - \bar{c}\ll 6
-  \tilde{b}\ll 6 \pp\ll + \tilde{\bar{c}}\ll 6
-  \tilde{\bar{b}}\ll 6 \pp\ll + \tilde{c}\ll 6
+  {1\over{2 \bar{M}}} \tilde{b}\ll 6
b\ll 6 +  {1\over{2 M}}\tilde{\bar{b}}\ll 6 \bar{b}
\ll 6
\biggr)\ .
\eee
As intended, the IR theory becomes free in 
the $|M|\to\infty$ limit.

\begin{table}
\begin{center}
\begin{tabular}{|c|c|c|}
\hline
${\rm object}$ & $(\tilde{h}, h)$ & $(\tildercharge ,~  \rcharge)$ \\
\hline
\hline
$Q$ & $(~[0]~, ~[+\hh]~ )$ & (0,1) \\
$\tilde{Q}$  &   $(~[+\hh]~, ~[0]~)$ & (1,0) \\
$\bar{Q}$    &   $(~[0]~, ~[+\hh]~ )$ & (0,-1) \\
$\bar{\tilde{Q}}$   &   $(~[+\hh]~, ~[0]~)$ & (-1,0) \\
$W$   &   $(\hh, \hh)$   &   (-1,-1) \\
$\bar{W}$   &   $(\hh, \hh)$ & (1,1) \\
\hline
$\phi,~\phb$   &   (0,0) & (0,0) \\
$\psi$ & $(0,\hh)$   &   (0,1) \\
$ \psb$ & $(0,\hh)$ & (0,-1) \\
$\pst$ & $(\hh,0)$ & (1,0) \\
$\bar{\pst}$ & $(\hh, 0 )$ & (-1,0) \\
$\m~\exp{\TachGrad\phi\uu +}$ & 
$(+ {1\over 4} \TachGrad q\apr, +{1\over 4} \TachGrad q\apr)$
 & $(-\hh\TachGrad q\apr, -\hh\TachGrad q\apr)$ \\
\hline
$b\ll 6$  & $(0,{3\over 2})$ & (0,1)\\
$\tilde{b}\ll 6$ & $({3\over 2} , 0 )$  & (1,0) \\
$c\ll 6$ & $(0 , - {1\over 2} )$  & (0,1) \\
$\tilde{c}\ll 6$ & $( - {1\over 2} , 0 )$ & (1,0) \\
$\bar{b}\ll 6 $ & $(0,{3\over 2})$ & (0,-1)\\
$\bar{\tilde{b}}\ll 6$ & $({3\over 2} , 0 )$  & (-1,0) \\
$\bar{c}\ll 6$ & $(0 , - {1\over 2} )$  & (0,-1) \\
$\bar{\tilde{c}}\ll 6$ & $( - {1\over 2} , 0 )$ & (-1,0) \\
\hline
\end{tabular}
\end{center}
\caption{The weights and R-charges of fundamental fields.}
\label{TAB1}
\end{table}

Just as with the case described in \cite{previous3},
we will need to define a normal-ordering prescription that is appropriate for
the IR system.   In the UV, OPEs of field monomials involve subtractions
of terms that are proportional to $|M|$, and hence infinite in the late-time limit.
The natural normal-ordering prescription for fields in the IR regime differs from
that in the UV, but only by terms that are independent of $M$ or $\bar M$.  These finite 
differences amount to quantum corrections to the classical expressions for
the superconformal generators in IR variables.  In turn, these corrections
lead to a finite renormalization of the dilaton gradient.  
As in \cite{previous3}, the theory can be seen to admit no 
nontrivial Feynman graphs, and yet quantum corrections arise in moving to the IR description.

Henceforth, operators and expressions in IR variables are assumed to be
normal-ordered under the proper IR ordering scheme.  In \cite{previous3}, different 
operator orderings were made explicit by using the usual $::$ normal-ordering symbols 
in the UV regime, and introducing the bubble notation $\nno \nno$ for use with
IR variables.  Since we are keeping normal-ordering implicit, we will not need to
revert to this strategy in the present paper.  (I.e., we will rely entirely on the
results presented in \cite{previous3}, with no need to compute OPEs directly in the 
IR regime.)

Performing the classical transformations alone, the 
lightcone stress tensors in the $b_6 c_6$-antighost and bosonic sectors of the theory
appear as
\bbb
T^{\rm LC}_{b\ll 6 c\ll 6} &=& 
 - {{3 i}\over 2} \left(
	    (\pp\ll + \bar{c}\ll 6) b\ll 6  
	+   (\pp\ll + c \ll 6) \bar{b}\ll 6  
	\right)
- {i\over 2}   \left(
	  \bar{c}\ll 6 (\pp\ll + b\ll 6)     
 	+   c \ll 6 (\pp\ll + \bar{b}\ll 6)  
	\right)      \ ,
\xxx
T^{\rm LC}_{\chi} &=& 
-\frac{q}{2}\left( \del_+^2 \chi^- + \del_+^2\bar \chi^-  
+ \del_+^2\chi^+  +  \del_+^2\bar\chi^+ \right)
\xxx
&&
+ \frac{2}{\apr} \left(
	\del_+ \bar\chi^- \del_+ \chi^+
	+\del_+\chi^- \del_+\bar\chi^+ 
	\right)\ .
\eee
To capture quantum corrections arising from the transformation to IR variables,
it is useful to move to two real bosonic coordinates 
$u\ll 6$ and $y\ll 6$ (these coordinates carry a subscript, since they
will undergo further redefinitions in the following section):
\be  
\chi^\pm = \frac{1}{\sqrt{2}}(y\ll 6^\pm + i u\ll 6^\pm ) \  .
\ee
This breaks the bosonic lightcone stress tensor into two pieces:
\be
T^{\rm LC}_u & = & \frac{2}{\apr} \del_+ u\ll 6^- \del_+ u\ll 6^+ \ ,
\xxx
(T^{\rm LC}_{y})_{\rm classical} & = & \frac{2}{\apr} \del_+ y\ll 6^- \del_+ y\ll 6^+
- \frac{q}{\sqrt{2}}\left( \del_+^2 y\ll 6^- + \del_+^2 y\ll 6^+ \right) \ .
\ee
Quantum corrections to the classical transformations 
can be computed by referring directly to the calculations in \cite{previous3}
(here we essentially have two real copies of the system in \cite{previous3}).  
In the $y$ sector of the bosonic stress 
tensor, quantum effects contribute the following correction:
\be
\Delta T^{\rm LC}_{y} = \frac{2\sqrt{2}}{\apr q} \del_+^2 y\ll 6^+ \ .
\label{yquantum}
\ee
Below we will collect the classical and quantum contributions in the 
bosonic $y$ sector into the single expression
\be
\tlcbos \equiv  T^{\rm LC}_y  + \Delta T^{\rm LC}_y \ .
\label{tlcbose}
\ee
From the results in \cite{previous3}, it is straightforward to compute the renormalization of the dilaton 
gradient $\Delta V^\mu$ arising from the shift to IR variables.   We find
\bbb
\Delta V\ll + = \sqrt{2}\cc\TachGrad\ , 
\qquad \Delta V\st\ll + = \sqrt{2}\cc\bar{\TachGrad} \ ,
\xxx
\Delta V\ll - = 
\Delta V\st\ll - = 
\Delta V\ll i = \Delta V\st\ll i =  0  \ .
\eee


For the sake of presentation, we compute the transformed lightcone supercurrent
in complex bosonic $\chi$ coordinates:
\bbb
\glc_6 &=&
{1\over 2} q\sqrt{\apr}~ b\ll 6
 - {{8\OMEGA}\over{q\apr{}\uu{3/2}}} c\ll 6
  \pp\ll + \chi\uu + \pp\ll + \chb\uu -  
+ {{4 i \OMEGA}\over{q\sqrt{\apr}}}
   \bar{b}\ll 6 c\ll 6 \pp\ll + c\ll 6  
\xxx
&&
+ {{4 \OMEGA}\over{\sqrt{\apr}}}
\lrdd c\ll 6 \pp\ll + \sqd \chi \uu +
+ (\pp\ll + c\ll 6)\pp\ll + \chi\uu + - (\pp\ll + c\ll 6)
\pp\ll + \chb\uu - \rrdd + 2 \OMEGA q\sqrt{\apr}\cc \pp\ll + \sqd c\ll 6
\xxx
&&
-\OMEGA\TachGrad \sqrt{\apr} \pp\ll + \sqd c\ll 6 - 2\OMEGA
\TachGrad \sqd \sqrt{\apr} c\ll 6\cc
\pp\ll + \sqd \chi\uu + \ .
\label{supercurrentsix}
\eee
The two terms in the last line are quantum corrections.
We note that neither $\chb\uu +$ nor $\chi\uu -$ 
appears in $\glc_6$.

The R-current transforms entirely classically:
\bbb
J_6 = b\ll 6 \bar{c}\ll 6  + c\ll 6
\bar{b}\ll 6  
	- i  q  \lrdd  
	\pp\ll + \chi\uu +
	+  \pp\ll + \chi\uu - 
	-  \pp\ll + \chb\uu +
	-  \pp\ll + \chb\uu - 
	\rrdd \ .
\eee
In particular, the \it original \rm dilaton gradient
$V\ll\m$ and its complex conjugate, rather than
the renormalized dilaton gradient $V\ll\m + \Delta
V\ll\m$, enter the R-current in the IR variables.
This is necessary for consistency, since
the complex $bc$ antighosts and their conjugates
contribute to the R-current central term in exactly the 
same fashion as the complex $\psi\uu\pm$ fermions.
The dilaton contribution to the central term in $J$
must therefore remain unchanged.  We conclude that $V_\m$, rather than
$V_\m + \Delta V_\m$, is the appropriate quantity appearing in $J$.

At this stage we have established the existence of a consistent limiting
description at late target-space time.  The tachyon perturbation has added
a nonzero $F$ term \rr{sbosfterm} to the supersymmetry algebra, 
$F\uu - = {{\bar M}\over{\bar\TachGrad \sqrt{\apr}}}$,
and deep in the IR regime worldsheet supersymmetry is spontaneously broken by an infinite
amount.  We therefore
expect that the late-time limit of the theory comprises an embedding of
bosonic string theory in the solution space of the $\cn = 2$ string.
In the following section we will show that this is indeed the case:
the deep IR is in fact described by purely bosonic string theory.

\section{Embedding of bosonic string theory}
\label{cov}
Now that we have moved to infrared variables and have accounted for all possible
quantum effects resulting from the canonical variable transformation, we would like to
understand in detail the late-time limit of the theory.  In this section we will
motivate and introduce a series of further canonical transformations that will
bring the theory into a recognizable form.  As noted above, we will use numerical
subscripts on the shifted $bc$ antighosts to keep track of sequential variable transformations, 
such that the final shifted $bc$ variables will be labeled as $b_1$, $c_1$, etc. 
We will also find it useful to decompose expressions into lightcone and 
transverse sectors: $G\equiv \glc + G\uu\perp$, $J = J\uu{\rm LC} +
J\uu\perp$ and $T = \tlc + T\uu\perp$, where  $\tlc
\equiv T\uu{\rm \chi\uu\pm} + T\uu{\rm antighost}$.

We also find it convenient to explicitly 
separate the bosonic sector of the theory into
lightcone and transverse contributions:
\be
T_{\rm bose} \equiv T^{\rm LC}_{\rm bose} + T^\perp_{\rm bose} \ ,
\ee
where $T^{\rm LC}_{\rm bose}$ has been defined above in Eqn.~\rr{tlcbose}, 
and includes the lightcone contributions from the bosonic $y$ sector, plus 
the quantum correction computed in Eqn.~\rr{yquantum}.
The stress tensor $T_{\rm bose}$ denotes the bosonic sector of the theory
with central charge exactly 
\be
c_{\rm bose} = \clcbos + \centperp = 26 \ .
\ee
The condition that the total central charge of the $\cn = 2$ parent theory 
takes the critical value $c_{\rm total} = 6$ sets the value of the transverse
component to
\be
\centperp = \frac{24}{\apr\TachGrad^2} = 3(D\ll c -2) = - c^{\rm dilaton}  \ .
\label{centperpdef}
\ee
For purposes of illustration and for future reference, we will keep
$c_{\rm bose}$, $\clcbos$, and $\centperp$ 
explicit until after the final variable redefinition below.

\heading{Rescaling}
To begin, we perform a rescaling of the $bc$ antighost system to simplify 
subsequent manipulations:
\bbb
b\ll 6 &=& {{2\OMEGABAR}\over{q\sqrt{\apr}}} b\ll 5 = \TachGrad \sqrt{\apr}
\OMEGABAR b\ll 5 \ ,
\xxx
c\ll 6 &=& {{q\sqrt{\apr}\OMEGABAR}\over 2} c\ll 5
= {{1}\over {\TachGrad \sqrt{\apr}}} c\ll 5 \ ,
\xxx
\bar{b}\ll 6 &=& {{2\OMEGA}\over{q\sqrt{\apr}}} \bar{b}\ll 5 = \TachGrad \sqrt{\apr}
\OMEGA \bar{b}\ll 5 \ ,
\xxx
\bar{c}\ll 6 &=& {{q\sqrt{\apr}\OMEGA}\over 2} \bar{c}\ll 5
= {{1}\over {\TachGrad \sqrt{\apr}}} \bar{c}\ll 5 \ .
\label{rescalings}
\eee
Even though the bosonic directions remain unchanged at this stage,
we choose to relabel the subscripts to keep the presentation uniform:
\be
u\ll 6 = u\ll 5 \ , \qquad
y\ll 6 = y\ll 5 \ .
\ee

Again, we are assuming real values for $\TachGrad $ and $q$, satisfying
$\TachGrad q = {2\over{\apr}}$.

Under this rescaling, the stress tensor $T$ and R-current $J$ do not change.
In terms of the real bosonic coordinates $u\ll 5$ and $y\ll 5$, the complex 
supercurrent takes the form:
\bbb
G_5 & = & 
   \bfive 
	+\frac{2 \sqrt{2}}{\TachGrad \apr}  
	\bigl(
	i \cfive\,\dupfive'
	+\cfive\,\dypfive'
	+i    \cfive'\,\dumfive
	+i \cfive'\,\dupfive
	-  \cfive'\,\dymfive
	+ \cfive'\,\dypfive
	\bigr)
	-2 i\cfive'\,\cfive\,\bar{b}_5
\xxx
&&
	-\frac{2}{\apr}\cfive
	\Bigl(
	\frac{i \TachGrad \apr}{\sqrt{2} } \dupfive'
	+\dupfive\,\dumfive 
	+\dypfive\,\dymfive 
	+{i}\dymfive\,\dupfive 
	- i\dypfive\,\dumfive 
	+\frac{ \TachGrad\apr}{\sqrt{2} }\dypfive'
	\Bigr)
\xxx
&&
	+ \Bigl( \frac{4}{\TachGrad^2 \apr^2} \alpha ' - 1\Bigr) \cfive''  \ .
\label{Gfive}
\eee
Likewise, the R-current and stress tensor are
\be
J_5 &=& 
-\frac{2 \sqrt{2}}{\TachGrad \apr}  \left( \dumfive   +  \dupfive \right) +
 c_5 \, \bar{b}_5 - \bar{c}_5\, b_5  \ ,
\xxx
T_5 & =  &
-\frac{i}{2} \left( c_5\, \bar{b}_5'+ \bar{c}_5\, b_5' \right)
-\frac{3 i}{2} \left( c_5'\, \bar{b}_5 + \bar{c}_5'\, b_5 \right)
-\frac{\sqrt{2} }{ \TachGrad \apr } \left( \dymfive' +  \dypfive' \right)
\xxx
&&
+ \frac{2}{\apr} \left(\dupfive \,  \dumfive   
+\dypfive\, \dymfive \right)   +   \sqrt{2} \TachGrad  \dypfive'   \ ,
\xxx
	 & = & 
	T_{\rm bose}^{\rm LC}
	-\frac{i}{2} \left( c_5\, \bar{b}_5'+ \bar{c}_5\, b_5' \right)
	-\frac{3 i}{2} \left( c_5'\, \bar{b}_5 + \bar{c}_5'\, b_5 \right)
	+ \frac{2}{\apr} 
	\dupfive \,  \dumfive   
\ .
\ee
Note that $T_{\rm bose}^{\rm LC}$ now appears explicitly in the stress tensor.

\heading{Reflection symmetry}
Now we wish to perform a canonical transformation to
make the bosonic the $(y,u)$ theory reflection symmetric about
the little group of the renormalized linear dilaton.
We therefore introduce the infinitesimal generator
\bbb
{\bf g}\ll 5 \equiv \int d\s\ll 1\,   g\ll 5 (\s) \ .
\eee
The expression for $g\ll 5$ is
given by the sum of an antighost-number two
piece and antighost-number four piece:
\be
g_5 = g_5^{(2)} + g_5^{(4)}\ ,
\ee
where 
\be
g_5^{(2)} & = &
    \frac{\sqrt{2} }{ \TachGrad \apr } 
	\biggl(
		\bar{c}_5\, c_5\, \dumfive'
	+ {3 }
   		\bar{c}_5\, c_5\, \dupfive'
	+ 
		\bar{c}_5\, c_5'\, \dumfive
	+
   		\bar{c}_5\, c_5'\, \dupfive
	+{i } 
		\bar{c}_5\, c_5'\, \dymfive
\xxx
&&
	-{i }
   		\bar{c}_5\, c_5'\, \dypfive
	+ 
		\bar{c}_5'\, c_5\, \dumfive
	+
   		\bar{c}_5'\, c_5\, \dupfive
	-{i } 
		\bar{c}_5'\, c_5\, \dymfive
	+{i }
   		\bar{c}_5'\, c_5\, \dypfive
	\biggr)
\xxx
&&	-\frac{2}{\alpha '}\left(
		\bar{c}_5\, c_5\, \dymfive\, \dupfive 
	- \bar{c}_5\, c_5\, \dypfive\, \dumfive 
	+\frac{\TachGrad \apr}{\sqrt{2} }  \bar{c}_5\, c_5\, \dupfive' \right) \ ,
\xxx
g_5^{(4)} & = & \frac{\sqrt{2} }{ \TachGrad \apr }  \left(
	\bar{c}_5'\, \bar{c}_5\, c_5'\, c_5\, \dymfive-
   \bar{c}_5'\, \bar{c}_5\, c_5'\, c_5\, \dypfive \right)  \ .
\label{genfive}
\ee
%
We generate finite transformations by taking
\bbb
U\ll 5 \equiv \exp{{i\over{2\pi}}{\bf g}\ll 5} \ .
\eee
We thereby obtain the following transformed supercurrent,
now written in terms of ``4'' variables:
\bbb
G_4 = U_5 G_5 U_5^\dagger & = & 
  \bfour -\cfour\, \tlcbos  + G^\perp 
  -i \tlcbos \cfourbar\,\cfour'\,\cfour
   +2 i \cfourbar\,\cfour''\,\cfour'-\frac{7 i}{6}
   \cfourbar\,\cfour\tripleprime\,\cfour
\xxx
&&
\kern-45pt
	-2 i \cfour'\,\cfour\,\bfourbar
	-\frac{3 i}{2} \cfourbar'\,\cfour''\,\cfour
	-\frac{3 i}{2} \cfourbar''\,\cfour'\,\cfour
	-4\cfourbar'\,\cfourbar\,\cfour''\,\cfour'\,\cfour
	 -\frac{2}{\alpha '}
	\bigl(
	\cfour\,\dupfour\,\dumfour
\xxx
&&
\kern-45pt
   + i\cfourbar\,\cfour'\,\cfour\,\dupfour\,\dumfour
	\bigr)	
	+ \frac{1}{\TachGrad ^2 \alpha '}
	\bigl(
	i\cfourbar\,\cfour\tripleprime\,\cfour
	+{i} \cfourbar'\,\cfour''\,\cfour 
	 -{2 i}\cfourbar\,\cfour''\,\cfour' 
  	+ {5 i}\cfourbar''\,\cfour'\,\cfour
\xxx
&&
\kern-45pt
	+{4}\cfourbar'\,\cfourbar\,\cfour''\,\cfour'\,\cfour 
	 + 4 \cfour'' \bigr)
	+\frac{2  \sqrt{2}}{\TachGrad  \alpha '}
	\bigl(
	i \cfour'\,\dumfour 
	+ i\cfour'\,\dupfour
	+\cfourbar\,\cfour'\,\cfour\,\dumfour' 
	+ \cfourbar\,\cfour'\,\cfour\,\dupfour'
\xxx
&&
\kern-45pt
	+\cfourbar'\,\cfour'\,\cfour\,\dumfour 
	+\cfourbar'\,\cfour'\,\cfour\,\dupfour 
	\bigr)
	-  \cfour'' \ .
\eee
Throughout these intermediate stages, we will record successive transformations
of the stress tensor and the R-current in Appendix~\ref{appeqns}.

\heading{R-current rotation}
At this stage we want to formulate a transformation that puts the R-current into
a universal form in which the worldsheet fermions do not appear 
(which are contained in $J^\perp$).  
We therefore define a second canonical transformation
generated by ${\bf g}\ll 4$:
\bbb
{\bf g}\ll 4 \equiv \int d\s\ll 1\,   g\ll 4 (\s) \ ,
\qquad 
g\ll 4 \equiv - {\TachGrad \over{\sqrt{2}}} J\uu\perp u\uu + \ .
\eee
Forming the finite transformation
\bbb
U\ll 4 \equiv \exp{{i\over{2\pi}} {\bf g}\ll 4} \ ,
\eee
we obtain
\bbb
G_3 	&=& U_4 G_4 U_4^\dagger
\xxx
  	& = & \bthree  -\cthree \, \tlcbos  + G^\perp \normx +
	-\frac{\jperp \TachGrad }{\sqrt{2}} \cthree\,\dupthree
	+\frac{i \centperp \TachGrad}{6 \sqrt{2}} \cthree'\,\dupthree-\frac{\centperp \TachGrad ^2}{12}
   \cthree\,\dupthree\,\dupthree  
\xxx
&&
	- 2 i \cthree'\,\cthree\,\bthreebar
	-\frac{2}{\alpha '} \cthree\,\dupthree\,\dumthree 
	+\frac{2 i \sqrt{2}}{\TachGrad  \alpha '}
	\bigl(
	\cthree'\,\dumthree 
	+\cthree'\,\dupthree 
	\bigr)
	+i \jperp \cthree'
	- \cthree''
\xxx
&&
	+\frac{4}{\TachGrad ^2 \alpha '} \cthree''
	-i \tlcbos \cthreebar\,\cthree'\,\cthree+\jperp' \cthreebar\,\cthree'\,\cthree
	+2 i  \cthreebar\,\cthree''\,\cthree'
	-\frac{7 i}{6} \cthreebar\,\cthree\tripleprime\,\cthree
	+\jperp\cthreebar'\,\cthree'\,\cthree
\xxx
&&
	-\frac{3 i}{2} \cthreebar'\,\cthree''\,\cthree-\frac{3 i}{2}
   \cthreebar''\,\cthree'\,\cthree-\frac{i \jperp \TachGrad }{\sqrt{2}}
   \cthreebar\,\cthree'\,\cthree\,\dupthree
	+\frac{\centperp \TachGrad }{6 \sqrt{2}}
   	\bigl(
	\cthreebar\,\cthree'\,\cthree\,\dupthree'
	+\cthreebar'\,\cthree'\,\cthree\,\dupthree
	\bigr)
\xxx
&&
	-\frac{1}{12} i \centperp \TachGrad ^2 \cthreebar\,\cthree'\,\cthree\,\dupthree\,\dupthree
	-\frac{2 i}{\alpha '}\cthreebar\,\cthree'\,\cthree\,\dupthree\,\dumthree
	+\frac{i}{\TachGrad^2\apr}\bigl(
	\cthreebar\,\cthree\tripleprime\,\cthree 
	+\cthreebar'\,\cthree''\,\cthree 
	-{2}\cthreebar\,\cthree''\,\cthree' 
\xxx
&&
	+{5}\cthreebar''\,\cthree'\,\cthree
	\bigr) 
	+\frac{2 \sqrt{2}}{\TachGrad  \alpha '}
	\bigl(
	 \cthreebar\,\cthree'\,\cthree\,\dumthree' 	
	+\cthreebar\,\cthree'\,\cthree\,\dupthree' 	
	+\cthreebar'\,\cthree'\,\cthree\,\dumthree 	
	+\cthreebar'\,\cthree'\,\cthree\,\dupthree 
	\bigr)
\xxx
&&
	+4\bigl( \frac{1}{\TachGrad^2\apr} - 1 \bigr)	
	\cthreebar'\,\cthreebar\,\cthree''\,\cthree'\,\cthree \ .
\eee

To illustrate the action of this transformation on the R-current explicitly, we
record the form of $J_4$, prior to the linear redefinition defined by $g\ll 4$ above,
decomposed by antighost number:
\be
J_4^{(0)} & = &    
J^\perp + 
\cfour\,\bfourbar- \cfourbar\,\bfour
	-\frac{2 \sqrt{2}}{\TachGrad  \alpha'}
	\bigl(
	\dumfour 
	+\dupfour 
	\bigr) \ ,
\xxx
J_4^{(2)} & = & 
	\frac{1}{8}\bigl(
	-{42} 
	+{\cbos}
	-{\centperp} 
	\bigr)\cfourbar\,\cfour''
	-4 \cfourbar'\,\cfour'
	-2 \cfourbar''\,\cfour
	+\frac{1}{\TachGrad^2\apr}
	\bigl(
	{7}\cfourbar\,\cfour''
	+4 \cfourbar''\,\cfour
	+8 \cfourbar'\,\cfour' 
	\bigr) \ ,
\xxx
J_4^{(4)} & = & -\frac{i}{8\apr\TachGrad^2}
	\bigl(	24 + \apr\TachGrad^2(\cbos - \centperp -26) \bigr)
	\bigl(
	3 \cfourbar'\, \cfourbar\, \cfour''\, \cfour
	+ \cfourbar''\, \cfourbar\, \cfour'\, \cfour
	\bigr) \ .
\ee
As intended, the transformed version of the R-current 
(under $U_4$) is completely independent of $J\uu\perp$:
\be
J_3 & = & 
	\frac{\centperp \TachGrad }{6 \sqrt{2}}\dupthree 
	+ \cthree\,\bthreebar
	- \cthreebar\,\bthree
	-\frac{21}{4} \cthreebar\,\cthree''
	+\frac{\clcbos}{8} \cthreebar\,\cthree''
	-4 \cthreebar'\,\cthree'
	-2 \cthreebar''\,\cthree
	+\frac{4}{\TachGrad ^2 \alpha '} \cthreebar''\,\cthree 
\xxx
&&
	+\frac{7}{\TachGrad ^2 \alpha'} \cthreebar\,\cthree'' 
	+\frac{8}{\TachGrad ^2 \alpha '} \cthreebar'\,\cthree' 
	-\frac{2\sqrt{2}}{\TachGrad  \alpha '} \left( \dumthree  + \dupthree \right) \ .
\ee

\heading{Hermitian universal form}
At this stage we see that the lightcone component of the bosonic stress tensor
$\tlcbos $ appears explicitly in the supercurrent.  We now want to move to a
universal formulation written strictly in terms of a {\it generic} bosonic stress tensor
$T_{\rm bose}$, describing a bosonic theory of critical central charge $\cbos = 26$. 
We therefore define the generating function 
${\bf g}\ll 3 \equiv \int d\s\ll 1\,   g\ll 3 (\s)$, with
\be
g\ll 3 & = & 
	\frac{J^\perp}{2} 
	\del_+ \left(
		\cthreebar\, \cthree
		\right)
	-\frac{1}{12}\left(
		\bar G^\perp \normxb \cthreebar\, \cthree'\, \cthree
		+ {G^\perp \normx} \cthreebar'\, \cthreebar\, \cthree
		\right)
	- \cthreebar\, \cthree'\, \cthree\, \bar{b}_2
	+ \cthreebar'\, \cthreebar\, b_2\, \cthree
\xxx
&&
	-i\bar G^\perp \normxb \cthree
	-i G^\perp \normx \cthreebar
	+\frac{2 \sqrt{2}}{\TachGrad  \alpha'}
		\left(
		\cthreebar\, \cthree'\, \dupthree
		+\cthreebar'\, \cthree\, \dupthree 
		\right) \ .
\label{genthree}
\ee
Acting with the finite unitary transformation 
$U\ll 3\equiv \exp{{i\over{2\pi}}{\bf g}\ll 3}$
puts the supercurrent into the following form:
\bbb
G_2 	&=& U_3 G_3 U_3^\dagger
\xxx
	& = &  \btwo - \ctwo  \, T_{\text{bose}} 
      -\frac{i \centperp \TachGrad }{6 \sqrt{2}} \del_+ \left(
		\ctwo\,\duptwo  \right)
	+\frac{2 i\sqrt{2}}{\TachGrad  \alpha '} \ctwo\,\duptwo'
	+i \ctwobar\,\btwo\,\ctwo'+i \ctwobar\,\btwo'\,\ctwo-i
   	\ctwo'\,\ctwo\,\btwobar
\xxx
&&
\kern-10pt
	+2 i \ctwobar'\,\btwo\,\ctwo
	-\frac{2}{\alpha '}\ctwo\,\duptwo\,\dumtwo 
	+ \frac{2 i \sqrt{2}}{\TachGrad  \alpha '}
	\bigl(
	\ctwo'\,\dumtwo 
	+\ctwo'\,\duptwo
   	\bigr) 
	+\left(\frac{4}{\TachGrad ^2 \alpha '} - \frac{\centperp}{6}  - 1\right)\ctwo'' 
\xxx
&&
\kern-10pt
	+\frac{i \centperp}{12} \ctwobar\,\ctwo''\,\ctwo'
	-\frac{2 i}{\TachGrad ^2 \alpha '}
   		\ctwobar\,\ctwo''\,\ctwo'
	+\frac{i \centperp}{8} \ctwobar\,\ctwo\tripleprime\,\ctwo
	+\frac{7 i \centperp}{24} \ctwobar'\,\ctwo''\,\ctwo
	+\frac{i \centperp}{8} \ctwobar''\,\ctwo'\,\ctwo
\xxx
&&
\kern-10pt
	-\frac{i}{\TachGrad^2\apr}
		\bigl(
		3 \ctwobar\,\ctwo\tripleprime\,\ctwo
		+3 \ctwobar''\,\ctwo'\,\ctwo 
		+7 \ctwobar'\,\ctwo''\,\ctwo 
		\bigr)
	+\frac{2 \sqrt{2}}{\TachGrad  \alpha   '}
	\bigl(
	\ctwobar\,\ctwo'\,\ctwo\,\dumtwo' 
	+\ctwobar\,\ctwo''\,\ctwo\,\dumtwo 
\xxx
&&
\kern-10pt
	+2\ctwobar'\,\ctwo'\,\ctwo\,\dumtwo 
	\bigr)
	-\frac{\centperp}{24} \ctwobar'\,\ctwobar\,\ctwo''\,\ctwo'\,\ctwo
	+\frac{1}{\TachGrad ^2 \alpha '} \ctwobar'\,\ctwobar\,\ctwo''\,\ctwo'\,\ctwo   \ .
\eee
We see that $T_{\rm bose}$ appears explicitly in the second term.

If we assign the appropriate value 
$\centperp = 24/(\apr B^2)$ (which is set by the condition on the parent $\cn = 2$
theory that $c_{\rm total} = 6$; see Eqn.~\rr{centperpdef}) and define
\be
v\uu \pm \equiv \TachGrad \uu{\pm 1} u\uu\pm\ll 2 \ ,
\ee
we find that, written in the bosonic variables $v^\pm$,
the tachyon gradient $\TachGrad $ scales out of the supercurrent entirely:
\be
G\ll 2 & \equiv &
	 \btwo
	- \ctwo \, T_{\text{bose}}	
	- \ctwo''
	+ i \ctwobar\, \btwo\, \ctwo'
	+i \ctwobar\, \btwo'\, \ctwo
	-i \ctwo'\, \ctwo\, \btwobar
	+ 2 i \, \ctwobar'\, \btwo\, \ctwo
\xxx
&&
\kern-25pt
	+\frac{2 \sqrt{2}}{\alpha '} 
		\left( 
		i\ctwo'\, \vminus' 
		+ \ctwobar\, \ctwo'\, \ctwo\, \vminus'' 
		+\ctwobar\, \ctwo''\, \ctwo\, \vminus' 
		+2\, \ctwobar'\, \ctwo'\, \ctwo\, \vminus' 
		\right)
	-\frac{2 }{\alpha '} \ctwo\, \vplus' \, \vminus'   \ .
\label{lasthermitian}
\ee

\heading{Final transformation}
The physics of this system is not yet clear.  Ultimately, we
are not interested in the details of the supercurrent at all, but only in 
its role in defining physical states via the OCQ or BRST procedure.
The OCQ or BRST cohomology is covariant not only
under unitary transformations, but under all
similarity transformations as well.  At this stage we would like to
place the conjugate supercurrent $\bar G$ in the simplest form possible,
and this can be achieved via a particular similarity transformation.
We therefore define a non-Hermitian
generator ${\bf g}\ll 2 \equiv \int d\s\ll 1\,   g\ll 2 (\s)$, where
\be
g\ll 2 & = & 
	i \ctwobar'\, \ctwo'
	+\frac{i c^\perp}{6} \ctwobar'\, \ctwo'
	+\frac{c^\perp \TachGrad }{6 \sqrt{2}}\ctwobar\, \ctwo'\, \duptwo
	+ \ctwobar\, \ctwo'\, \ctwo\, \bar{b}_2
	+ \ctwobar'\, \ctwobar\, b_2\, \ctwo
	+\frac{13}{8}\ctwobar'\, \ctwobar\, \ctwo''\, \ctwo
\xxx
&&
	-\frac{c_{\rm bose} }{16} \ctwobar'\, \ctwobar\, \ctwo''\, \ctwo
	+\frac{11 c^\perp}{48}\ctwobar'\, \ctwobar\, \ctwo''\, \ctwo
	+\frac{i c^\perp \TachGrad }{12 \sqrt{2}} \ctwobar'\, \ctwobar\, \ctwo'\, \ctwo\, \duptwo
	-i T_{\text{bose}} \ctwobar\, \ctwo 
	-\frac{2 i}{\alpha '} \ctwobar\, \ctwo\, \duptwo\, \dumtwo 
\xxx
&&
	+\frac{1}{\TachGrad\apr}\biggl(
		{2 \sqrt{2} } \ctwobar'\, \ctwo\, \dumtwo 
		-\frac{4 i}{\TachGrad} \ctwobar'\, \ctwo'
		-\frac{11}{2 \TachGrad } \ctwobar'\, \ctwobar\, \ctwo''\, \ctwo 
		-{2 \sqrt{2}} \ctwobar\, \ctwo'\, \duptwo 
\xxx
&&
		-{i \sqrt{2}} \ctwobar'\, \ctwobar\, \ctwo'\, \ctwo\, \duptwo 
		-{3 i \sqrt{2}} \ctwobar'\, \ctwobar\, \ctwo'\, \ctwo\, \dumtwo 
		\biggr) \ .
\label{gentwo}
\ee
The finite similarity transformation of interest takes the form
\bbb
S_2\equiv \exp{{i\over{2\pi}} {\bf g}\ll 2} \ ,
\een{similar}
So that conjugating the supercurrent with $S_2$ yields
\bbb
G_1 & = & S_2 G_2 S_2^{-1}
\xxx
	& = & b_1  -2 \cone\, T_{\rm bose} 
   -\frac{i \centperp \TachGrad }{6 \sqrt{2}} \cone\,\dupone'+\frac{2 i
   \sqrt{2}}{\TachGrad  \alpha '} \cone\,\dupone'-\frac{i \centperp \TachGrad }{3
   \sqrt{2}} \cone'\,\dupone+2 i \conebar\,\bone\,\cone'+2 i \conebar\,\bone'\,\cone
\xxx
&&
   -2 i \cone'\,\cone\,\bonebar+4 i \conebar'\,\bone\,\cone
	-\frac{4}{\apr}\cone\,\dupone\,\dumone 
	+\frac{2 i \sqrt{2}}{\TachGrad  \alpha '}
	\bigl(
	\cone\,\dumone' 
	+2 \cone'\,\dumone 
	+2 \cone'\,\dupone 
	\bigr)
\xxx
&&
	-2 \cone''
	-\frac{\centperp}{3} \cone''
	+\frac{8}{\TachGrad ^2 \alpha '}\cone''
\xxx
&&
	 - \frac{i}{16\TachGrad^2\apr}
	\bigl( 24 + \apr \TachGrad^2 (\cbos - \centperp - 26) \bigr)
	\biggl(
	\conebar\,\cone''\,\cone'
	-\conebar\,\cone\tripleprime\,\cone
	- {4} \conebar'\,\cone''\,\cone
	-6 \conebar''\,\cone'\,\cone \biggr)
\xxx
&&
	+ \frac{7}{24 \TachGrad^2\apr}
	\bigl( 24 + \apr \TachGrad^2 (\cbos - \centperp - 26) \bigr)
	\, \conebar'\,\conebar\,\cone''\,\cone'\,\cone  \ .
\eee
Since $S_2$ is generated by a non-Hermitian function, $\bar G_1$ is no
longer given simply by the conjugate of $G_1$.   
Specifically, we obtain
\bbb
\bar G_1 & = & \bar b_1
	-\frac{i}{48\TachGrad^2 \apr}
	\bigl( 24 + \apr \TachGrad^2 (\cbos - \centperp - 26) \bigr)
	\biggl(
	6 \conebar'\,\conebar\,\cone''
	-3  \conebar''\,\conebar'\,\cone
	- \conebar\tripleprime\,\conebar\,\cone
	\biggr)
\xxx
&&
	+\frac{1}{12\TachGrad^2 \apr}
	\bigl( 24 + \apr \TachGrad^2 (\cbos - \centperp - 26) \bigr) \,
	\conebar''\,\conebar'\,\conebar\,\cone'\,\cone \ .
\eee

The final expressions simplify significantly
when we replace $\cbos$ and $\centperp$ with values 
appropriate to the system at hand (see Eqn.~\rr{centperpdef}),
and move to the rescaled bosonic $v^\pm$ variables (with
$u\uu \pm \equiv \TachGrad \uu{\mp 1} v\uu\pm$).
Once again, we obtain expressions that are independent of the tachyon gradient $\TachGrad$:
\be
G_1 & = & b_1  -2 \cone \, T_{\text{bose}} 
	+2 i \left(
	\bar{c}_1\,b_1\,c_1'
	+ \bar{c}_1\,b_1'\,c_1
	- c_1'\,c_1\,\bar{b}_1
	+2  \bar{c}_1'\,b_1\,c_1 
	\right)
\xxx
&&
	-\frac{4}{\alpha '} c_1\,\dvpone\,\dvmone 
	+\frac{2 i \sqrt{2}}{ \alpha '} 
	\left(
	c_1\,\dvmone' 
	+2 c_1'\,\dvmone \right)
	-2 c_1'' \ ,
\label{fin1}
\xxx
\bar G_1 &=& \bar b_1 \ .
\label{fin2}
\ee
Likewise, the final R-current and stress tensor are as follows:
\be
J_1 & = & -\frac{2\sqrt{2}}{ \apr}\dvmone + \cone\, \bonebar - \conebar\, \bone \ ,
\xxx
T_1 & = & T_{\rm bose} 
	+ \frac{2}{\apr} \dvpone\, \dvmone
	- \biggl(
	\frac{i}{2} \cone\, \bonebar'
	+ \frac{3i}{2} \cone'\, \bonebar
	+ {\rm h.c.}
	\biggr) \ .
\label{fin3}
\ee


In summary, we began with the $\cn =2$ string in flat space with timelike linear dilaton. 
Following condensation of the closed-sting tachyon profile specified by the
superpotential in Eqn.~\rr{c_duality_superpotential}, the theory 
asymptotes at late times to a constant CFT with no
$X\uu +$ dependence in the Lagrangian, stress tensor, R-current or supercurrents.
The theory admits a simple late-time limit that is static (modulo a time-dependent
dilaton).   We would like to understand in detail what this limit actually describes.
In the next section we show that the theory deep in the IR regime 
is precisely equivalent to a certain background of bosonic string theory, 
with $D - 2  = 2 D\ll c - 2$
noncompact dimensions, a linear dilaton,
and an $SO(2 D\ll c - 4)\ll L \times SO(2 D\ll c - 4)\ll R$ current algebra.
(In this paper, $D$ and $D\ll c$ will always refer to numbers of dimensions of the
\it initial \rm string background.)
The corresponding bosonic string is precisely described by the generic stress tensor 
$T\ll{\rm bose}$.
This system stands as an analogue of the bosonic string embedding into 
$\cn = 1$ superstring theory introduced by
Berkovits and Vafa in \cite{bv}.

\section{BRST quantization }
\label{BRST}
\def\qbrst{Q}
In this Section we will establish the
equivalence of the late-time limit of our worldsheet theory
to the bosonic string in a particular background.
To do this, we quantize the string in the BRST formalism.
We then take the standard form of the BRST current
and perform a similarity transformation
that reduces the BRST charge to two anticommuting pieces
involving disjoint sets of worldsheet fields.  One of
the two pieces will have a trivially computable cohomology
consisting of a single state, the vacuum $\kket{0}$.  
The second piece is precisely
equal to the BRST charge of the bosonic string in a particular
background.  The cohomology of the full product theory thus
corresponds one-to-one with that of the bosonic string.\footnote{This statement is
strictly true modulo a subtlety concerning the zero mode of the $v\uu -$ antighost, which
we will discuss below.}

In total, the BRST current of our original theory takes the form
\be
j_{\rm BRST} = c \, {\bf T}
 	+ \frac{1}{2} \calt \, {\bf J}
 	+ \frac{i}{\sqrt{2}} \bar\gamma \, {\bf G}
 	-\frac{i}{\sqrt{2}} \gamma \,  \bar{\bf G} \ ,
\ee
with the quantities ${\bf T},~{\bf J},~{\bf G}$ and $\bar{\bf G}$
defined above in Eqn.~\rr{moddefs}.
We want to show that this form can be brought into a final universal form
by the action of a similarity transformation.
One way to find such a transformation is to create an exhaustive
list of possible operator-valued terms that could appear in a candidate
generating function.  These consist of field monomials that have zero
total R-charge, zero ghost number and total weight one.
Given the field content of the theory, there are a large number of 
possible terms (several hundred) that satisfy these conditions.  Upon constructing a 
candidate generating function $g_0$ from these terms, we can solve for 
free coefficients by demanding that the full transformation yield the 
desired final incarnation of the BRST current, up to a total derivative:
\bbb
S_0 \, j_{\rm BRST} \, S_0^{-1} = j\uu{\rm new} \ll{\rm BRST} + j\uu{\rm deriv} \ll{\rm BRST} \ ,
\een{brstcomm}
where $j\uu{\rm new} \ll{\rm BRST}$ is described below.
Here, $S_0$ comprises a similarity transformation, as opposed to a unitary 
transformation.  In other words, it is generated by a non-Hermitian generator, 
according to
\be
S_0 = \exp{ i g_0 } \ ,
\ee
with $g_0 \neq g_0\dag$.  
In the end, the final transformation rules are still somewhat complicated.
We present these rules, including the complete form of the generating 
function $g_0$, in Appendix~\ref{brst}.

Under the full transformation in Eqn.~\rr{brstcomm}, the BRST current 
becomes
\be
j\uu{\rm new}\ll{\rm BRST}
 	 &\equiv&
 	j\uu{\rm triv}\ll{\rm BRST} + j\uu{\rm bose}\ll {\rm BRST}
\xxx
&&
\kern-40pt
 	 =
 	-\frac{i}{\sqrt{2}\omega} \gamma \, \bar b_1
 	+ \frac{i \omega}{\sqrt{2}} \bar\gamma \, b_1
 	-\frac{\sqrt{2}}{\apr} \calt \, \vminus '
 	+ i b \, c' \, c
 	+ \frac{3}{2} c''
 	+ c\, T_{\rm bose} \ .
\label{BRSTfinal}
\ee
Since the overall transformation is rather complicated, we have included a 
Mathematica notebook in the arXiv source package of this paper that can be used to 
verify the result.\footnote{This notebook can also be found online at the URL
\href{http://sns.ias.edu/~swanson/BRST.nb}{http://sns.ias.edu/$\sim$swanson/BRST.nb}.}

The total derivative term $j\uu{\rm deriv}\ll{\rm BRST}$ does not contribute to the
BRST charge, and we can ignore it for the purpose of computing the physical
state spectrum.
The first term $j\uu{\rm triv}\ll{\rm BRST}$ is a quadratic
piece involving the variables $c\ll 1,~b\ll 1,~v\uu\pm$, and the
Fadeev-Popov ghosts for the R-symmetry and local supersymmetry:
\be
j\uu{\rm triv}\ll{\rm BRST} = -\frac{i}{\sqrt{2}\omega} \gamma \, \bar b_1
 	+ \frac{i \omega}{\sqrt{2}} \bar\gamma \, b_1
 	-\frac{\sqrt{2}}{\apr} \calt \, \vminus ' \ .
\ee
The only cohomology in this sector is the vacuum $\kket{ 0 }$, corresponding
to the operator 
\be
\kket{0} \leftrightharpoons c_1\, \bar c_1\, \delta(\g ) \delta(\bar\g) \ .
\ee
For oscillators with nonvanishing mode number,
this statement follows from a straightforward application of the BRST quartet principle.
In particular, one should consider the three field quartets
\bbb
\lrdd b\ll 1, \bar{\g}, \bar{c}\ll 1 , \b \rrdd \ ,
\qquad
\lrdd \bar{b}\ll 1, \g,c\ll 1, \bar{\b} \rrdd \ ,
\qquad
\lrdd v\uu{-'}, \calt,v\uu +,\balt \rrdd \ .
\eee

One subtlety in this approach is that the
field $v\uu -$ never appears undifferentiated in
the current $j\uu{\rm triv}\ll{\rm BRST}$.  This implies
that, while $v\uu {-'}$ is the $\qbrst$-image of
a local operator (namely $\balt$), the $\qbrst$-closed operator
$v\uu -$ is not.  This leads to a strange situation in which
$v\uu -(\s\uu \pm)$ is an element of the BRST cohomology, but
$v\uu{-'}$ is BRST-exact.

Since $v\uu -(\s)$ is independent of $\s$ in cohomology,
operators constructed from undifferentiated $v\uu -$ fields
behave analogously to picture-changing operators.  That is,
$v\uu -$ momentum labels an infinite number of disjoint copies of
the physical state cohomology, between which the exponentials
${\bf X}\ll {{\bf P}} \equiv \exp{i {\bf P} v\uu -}$
interpolate.  The operators ${\bf X}\ll{{\bf P}}$
functioning as picture-changing operators for ``$v\uu -$-picture.''
All PCOs for $v\uu -$-picture are invertible and position-independent, so it
is manifest that all $v\uu -$-pictures are completely equivalent.
It seems likely that a careful treatment
of picture- and instanton-number-changing operators in the
${\cal N} = 2$ string framework will shed light on the correct 
treatment of $v\uu -$-pictures at the level of interacting strings.



The third piece, $j\uu{\rm bose}\ll{\rm BRST}$,
is a BRST operator in the standard form, describing a bosonic
string with a $c=26$ conformal stress tensor:
\bbb
j\uu{\rm bose}\ll{\rm BRST} = c\, T_{\rm bose}  
	+ i b \, c' \, c
 	+ \frac{3}{2} c''\ .
\eee
The stress tensor $T_{\rm bose}$ describes
$2 D\ll c - 4$ flat transverse real coordinates $\phi\uu a,~\phb\uu a$,
as well as a pair of light-cone coordinates $y\uu\pm$, also
with flat metric $\eta\ll{+-} = \eta\ll{-+} = -1,~\eta\ll{++} = \eta
\ll{--} = 0$.  There is a varying dilaton whose
contribution to the central charge is given by
\bbb
c\uu{\rm dilaton} = -\frac{3}{2}( D - 20 ) \ ,
\eee
where $D = 2 D_c$ is the number of real noncompact dimensions of the
initial configuration.

The physical state cohomology ${\cal V}$ in the bosonic sector is characterized,
as usual, by the $bc$ ghost vacuum times a matter primary of weight one.
Given a matter primary of weight one, the corresponding vertex operator 
will be of the form $c\, c_1\, \bar c_1\, \delta(\g )\, \delta(\bar\g)\, {\cal V}$
(along with the appropriate left-moving operators).

Finally, there is a stress tensor present for the $\psi\uu a,~\psi\uu{a\dag}$ 
degrees of freedom.  These variables no longer have any significance as
superpartners of the $\phi$ variables.  
In particular, currents
generated by $\psi\uu a,~\psi\uu{a\dagger}$
do not give rise to gauge bosons in the spectrum of the initial $\cn = 2$
theory.  In the late-time theory, however,
these variables are primary operators of weight one, and therefore
enter the theory on the same footing as any other such operator, giving
rise to gauge bosons propagating in spacetime.  The same is
true for the currents generated by $\tilde\psi\uu a,~\tilde\psi\uu {a\dagger}$.
Together, these variables generate an $SO(D-4)\ll L \times SO(D-4)\ll R$ current algebra, with one
real fermion for each real dimension transverse to the light cone.

The complex structure of the $\phi$ and $\psi$ degrees of freedom
has decoupled completely, along with the original supersymmetry.  By 
decomposing
$\phi\uu a \equiv {1\over{\sqrt{2}}} \lrdd y\uu {2a} + i y\uu{2a+1} \rrdd$
and $\psi\uu a\equiv {1\over{\sqrt{2}}} \lrdd \l\uu {2 a} + i \l\uu{2 a + 1}
\rrdd $ for $a = 1,\cdots , D\ll c - 2 $, it is clear that
the theory has an $SO(D-4)$ symmetry rotating the $\l\uu A$
(with the index $A$ labeling real directions), as well as
an unrelated $SO(D-3,1)$ spatial symmetry rotating the spatial
directions $(y\uu 0, y\uu 1,y\uu A)$, if one ignores the dilaton.  The dilaton
gradient breaks this spatial symmetry down to $SO(D-3)$ if the
final dilaton gradient is timelike, $SO(D-4)$ if the final dilaton 
gradient
is lightlike, and $SO(D-4,1)$ if the final dilaton gradient is
spacelike.  The total central charge of the bosonic stress tensor 
$T\ll{\rm bose}$
is $D-2$ from the scalars $(y\uu 0, y\uu 1, y\uu A)$, $\hh (D - 4)$ from the
fermions $\l\uu A$ and $ -3( D/2 - 10 ) $ from the dilaton, for a total
of $c = 26$.  Of course, this is the correct central charge for a
consistent bosonic string theory.

\section{Summary and conclusions}
The $\cn = 2$ string exhibits a number of interesting properties in 
supercritical dimensions.  In fact, because the presence of a linear dilaton background
renders one of the two timelike directions unphysical, it is somewhat natural to
study this string theory in dimensions above the critical value $D=4$.  In line with 
previous studies of supercritical strings, we have found that closed-string tachyon
condensation leads to a number of interesting dynamical transitions among
different string theories.  In certain examples, these transitions 
connect the $\cn = 2$ string in various
spacetime dimensions via dynamical dimensional reduction.  We have also shown that
closed-string tachyon condensation can drive a dynamical transition directly
to bosonic string theory.  

Although the existence of consistent string theories 
in supercritical dimensions was established long ago \cite{Polyakov:1981rd,Myers:1987fv,ali},
it is apparent that such theories harbor a number of novel and surprising features. 
It seems likely that further exploration of supercritical
string backgrounds will yield additional connections among
theories that were previously thought to be completely distinct.

\section*{Acknowledgments}
S.H.~is the D.~E.~Shaw \& Co.,~L.~P.~Member
at the Institute for Advanced Study.
S.H.~is also supported by U.S.~Department of Energy grant DE-FG02-90ER40542. 
I.S.~is the Marvin L.~Goldberger Member
at the Institute for Advanced Study, and is supported additionally
by U.S.~National Science Foundation grant PHY-0503584. 
We thank Nathan Berkovits for useful discussions regarding BRST 
quantization and the quartet principle.
We thank the 2007 Simons Workshop in Mathematics and Physics for hospitality
during the completion of portions of this work.  We also thank the
Natural Sciences computing staff at the Institute for Advanced Study for technical
assistance.

\renewcommand{\thefigure}{A-\arabic{equation}} 
\setcounter{equation}{0}
\numberwithin{equation}{section}
\appendix{$\cn = 2$ supersymmetry algebra}
\label{SUSY}
In this appendix we record the off-shell supersymmetry transformations of the 
$D\ll c$ chiral multiplets of our theory.  To begin, the $\cn = 2$ superalgebra
provides the following:
\be\{Q, \qqb\} = - 2 P\ll +  \ ,
\qquad
\{\qqt, \qtb\} = - 2 P\ll -  \ ,
\ee
along with the vanishing quantities
\bbb
\qqt\sqd = \qtb\sqd = Q\sqd = \qqb\sqd =
\{Q, \qqt\} = \{\qqb, \qbt\} = \{Q, \qbt\} = 
\{\qqb, \qqt\} = 0 \ .
\eee
We recall that $P\ll \pm = \hh\lrdd - P\ll 0 \pm P\ll 1 \rrdd$,
so both $P\ll\pm$ are negative-definite in a unitary
theory.  The transformation laws for a chiral multiplet $\phi,\psi$ are as
follows: 
\begin{eqnarray}
\fourtrans{\phi}{i \sqrt{\apr} \cc\psi}{i \sqrt{\apr} \cc\pst} 0 0 \ ,
\xxx
\fourtrans{\psi} 0 F {{2\over{\sqrt{\apr}}}\cc\pp\ll + \phi} 0  \ ,
\xxx
\fourtrans{\pst} {- F} 0 0 { {2\over{\sqrt{\apr}}}\cc\pp\ll - 
\phi} \ ,
\xxx
\fourtrans{F} 0 0 {- 2 i \cc \pp\ll + \pst} {+ 2 i\cc
\pp\ll - \psi} \ .
\eee
The transformations of the conjugate multiplet are
\bbb
\fourtrans{\phb} 0 0 {i \sqrt{\apr} \cc\psb}{i \sqrt{\apr} \cc
\bar{\pst}} \ ,
\xxx
\fourtrans{\psb}  {{2\over{\sqrt{\apr}}}\cc\pp\ll + \phb} 0
 0 {\bar{F}}  \ , 
\xxx
\fourtrans{\bar{\pst}} 0 { {2\over{\sqrt{\apr}}}\cc\pp\ll - 
\phb}  {- \bar{F}} 0 \ ,
\xxx
\fourtrans{\bar{F}} {- 2 i \cc \pp\ll + \bar{\pst}} {+ 2 i\cc
\pp\ll - \psb} 0 0 \ .
\een{chirtrans}

\appendix{OPEs of basic objects}
\label{appope}
In this appendix we collect the OPEs of both fundamental and
composite operators in our theory.
The OPE of the fundamental $\phi$ fields reads:
\be
\phi^\m (\sigma) \bar \phi^\n (\tau) \sim -\frac{\apr}{2}
	\log 
	\left| (\sigma^+ - \tau^+) (\sigma^- - \tau^-) \right| \eta^{\m\n} \ ,
\ee
where, as usual, $\sim$ indicates equivalence up to nonsingular terms.
Similarly, the $\psi$ fields admit the OPE
\be
\psi(\s)^\m \psb(\t)^\n \sim \psb(\s)^\m \psi(\t)^\n \sim \frac{i}{\s^+ - \t^+ }\eta^{\m\n}\ .
\ee
Furthermore, the OPEs of the various composite operators in the theory (namely,
the complex supercurrent, the R-current and the stress tensor) are
\bbb
J(\s) J(\t) &\sim & - {c\over{3 (\s\uu + - \t\uu +)\sqd}} \ ,
\xxx
T(\s)
 J(\t) &\sim &  {1\over{(\s\uu + - \t\uu + )\sqd}} J(\t) 
+ {1\over{\s\uu + - \t\uu +} } \pp\ll + J(\t)  \ ,
\xxx
T (\s) \cc G(\t) &\sim & \cc  {3\over{2 (\s\uu + - \t\uu +)\sqd}} 
\cc G(\t) + {1\over {\s\uu + - \t\uu +}}
\pp\ll + G (\t) \ ,
\xxx
T (\s) \cc \ggb(\t) &\sim & \cc {3\over{2 (\s\uu + - \t\uu +)
\sqd}} \cc \ggb(\t) + {1\over {\s\uu + - \t\uu +}}
\pp\ll + \ggb (\t) \ ,
\xxx
J(\s) G(\t) &\sim &  {i\over {\s\uu + - \t\uu + }} G(\t) \ ,
\xxx
J(\s) \ggb (\t) &\sim &  - {i\over {\s\uu + - \t\uu +}} \ggb(\t) \ ,
\xxx
G(\s) \ggb (\t) &\sim & - {{2 i c}\over{3\cc(\s\uu + - \t\uu +)\uu 3}}
- {2\over{(\s\uu + - \t\uu +)\sqd}} J(\t)
- {{2 i }\over{\s\uu + - \t\uu +}}
T(\t) 
- {1\over{\s\uu + - \t\uu +}} \pp\ll + J(\t) \ ,
\xxx
T(\s) T(\t) &\sim &  {c\over{2\cc(\s\uu + - \t\uu +)\uu 4}}
+ {2\over{(\s\uu + - \t\uu +)\sqd}} T(\t)
+ {1\over{\s\uu + - \t\uu +}} \pp\ll + T (\t) \ .
\een{n2ope}

\appendix{Details of the bosonic string embedding}
\label{appeqns}
In this appendix we record in detail each step of the multi-part variable transformation that
brings us to the final incarnation of the IR theory in 
Eqns.~(\ref{fin1}-\ref{fin3}) above.
We start just after the rescaling transformation that introduces the set of variables
labeled by the subscript 5.  The supercurrent at this stage 
is presented above in Eqn.~\rr{Gfive} with
bosonic coordinates expressed in $\chi$ variables.  It can be reached from the supercurrent
$G_6$ in Eqn.~\rr{supercurrentsix} by performing the rescalings in Eqn.~\rr{rescalings}.
For completeness, we present $G_5$ in real $u$ and $y$ bosonic variables
(as opposed to the complex $\chi$ system):
\bbb
G_5 & = & 
   \bfive 
	+\frac{2 \sqrt{2}}{\TachGrad \apr}  
	\bigl(
	i \cfive\,\dupfive'
	+\cfive\,\dypfive'
	+i    \cfive'\,\dumfive
	+i \cfive'\,\dupfive
	-  \cfive'\,\dymfive
	+ \cfive'\,\dypfive
	\bigr)
	-2 i\cfive'\,\cfive\,\bar{b}_5
\xxx
&&
	-\frac{2}{\apr}\cfive
	\bigl(
	\frac{i \TachGrad \apr}{\sqrt{2} } \dupfive'
	+\dupfive\,\dumfive 
	+\dypfive\,\dymfive 
	+{i}\dymfive\,\dupfive 
	- i\dypfive\,\dumfive 
	+\frac{ \TachGrad\apr}{\sqrt{2} }\dypfive'
	\bigr)
\xxx
&&
 	  - \cfive''
	+ \frac{4}{\TachGrad^2 \apr^2} \alpha '\, \cfive''  \ .
\eee
For the sake of collecting all of the relevant quantities in one place, we will also 
record the R-current and stress tensor after each step.  
The R-current at this stage takes the form
\be
J_5 & = & J^\perp 
	+\cfive\, \bfivebar
	-1 \cfivebar\, \bfive-\frac{21}{4} \cfivebar\, \cfive''
	+\frac{c_{\rm bose}}{8} \cfivebar\, \cfive''
	-\frac{c^\perp}{8} \cfivebar\, \cfive''
	-4\cfivebar'\, \cfive'
	-2 \cfivebar''\, \cfive
	+\frac{39 i}{4} \cfivebar'\, \cfivebar\, \cfive''\, \cfive
\xxx
&&
	-\frac{3 i c_{\rm bose}}{8} \cfivebar'\, \cfivebar\, \cfive''\, \cfive	
	+\frac{3 i c^\perp}{8} \cfivebar'\, \cfivebar\, \cfive''\, \cfive
	+\frac{13 i}{4} \cfivebar''\, \cfivebar\, \cfive'\, \cfive	
	-\frac{i c_{\rm bose}}{8}\cfivebar''\, \cfivebar\, \cfive'\, \cfive
	+\frac{i c^\perp}{8} \cfivebar''\, \cfivebar\, \cfive'\, \cfive
\xxx
&&
	+\frac{1}{\TachGrad^2\apr}\biggl(
		{4}\cfivebar''\, \cfive 
		-{3 i}\cfivebar''\, \cfivebar\, \cfive'\, \cfive 
		-{9 i}\cfivebar'\, \cfivebar\, \cfive''\, \cfive 
		+{7}\cfivebar\, \cfive'' 
		+{8}\cfivebar'\, \cfive' 
		\biggr)
\xxx
&&
		-\frac{2  \sqrt{2}}{\TachGrad \apr} \left( \dumfive + \dupfive \right)    \ ,
\ee
and the stress tensor is given by
\be
T_5 & = & 
	-\frac{i}{2} \cfive\, \bfivebar'
	-\frac{i}{2} \cfivebar\, \bfive'
	+\frac{11 i}{24} \cfivebar\, \cfive\tripleprime
	+\frac{i c_{\rm bose}}{48} \cfivebar\, \cfive\tripleprime
	-\frac{ic^\perp}{48} \cfivebar\, \cfive\tripleprime
	-\frac{3 i}{2} \cfive'\, \bfivebar
	-\frac{3 i}{2} \cfivebar'\, \bfive
	-\frac{5 i}{8} \cfivebar'\, \cfive''	
\xxx
&&
	+\frac{i c_{\rm bose}}{16} \cfivebar'\, \cfive''
	-\frac{i c^\perp}{16} \cfivebar'\, \cfive''
	-i \cfivebar''\, \cfive'
	-i \cfivebar\tripleprime\, \cfive
	+\frac{3}{8}\cfivebar'\, \cfivebar\, \cfive''\, \cfive'
	+\frac{c_{\rm bose}}{16} \cfivebar'\, \cfivebar\, \cfive''\, \cfive'
	-\frac{c^\perp}{16}\cfivebar'\, \cfivebar\, \cfive''\, \cfive'
\xxx
&&
	+\frac{35}{24} \cfivebar'\, \cfivebar\, \cfive\tripleprime\, \cfive
	+\frac{c_{\rm bose}}{48}\cfivebar'\, \cfivebar\, \cfive\tripleprime\, \cfive
	-\frac{c^\perp}{48} \cfivebar'\, \cfivebar\, \cfive\tripleprime\, \cfive
	+\frac{3}{4}\cfivebar''\, \cfivebar\, \cfive''\, \cfive
	+\frac{c_{\rm bose}}{8} \cfivebar''\, \cfivebar\, \cfive''\, \cfive
\xxx
&&
	-\frac{c^\perp}{8}\cfivebar''\, \cfivebar\, \cfive''\, \cfive
	+\frac{3}{8} \cfivebar''\, \cfivebar'\, \cfive'\, \cfive
	+\frac{c_{\rm bose}}{16}\cfivebar''\, \cfivebar'\, \cfive'\, \cfive
	-\frac{c^\perp}{16} \cfivebar''\, \cfivebar'\, \cfive'\, \cfive
	+\frac{35}{24} \cfivebar\tripleprime\, \cfivebar\, \cfive'\, \cfive
\xxx
&&
	+\frac{c_{\rm bose}}{48} \cfivebar\tripleprime\, \cfivebar\, \cfive'\, \cfive
	-\frac{c^\perp}{48} \cfivebar\tripleprime\, \cfivebar\, \cfive'\, \cfive
	+T^{\rm LC}_{\rm bose} 
	+T^\perp
	+\dupfive\, \dumfive \frac{2}{\alpha'}	
\xxx
&&
	-\frac{3}{2 \TachGrad ^2 \alpha '}
		\biggl(
		\cfivebar'\, \cfivebar\, \cfive\tripleprime\, \cfive 
		+\cfivebar\tripleprime\, \cfivebar\, \cfive'\, \cfive
		\biggr) 
	-\frac{1}{\TachGrad ^2 \alpha '}
		\biggl(
		\cfivebar''\, \cfivebar\, \cfive''\, \cfive 
		+\frac{1}{2 }\cfivebar'\, \cfivebar\, \cfive''\, \cfive' 
		+\frac{1}{2 }\cfivebar''\, \cfivebar'\, \cfive'\, \cfive 
\xxx
&&
		-\frac{i}{2 } \cfivebar\, \cfive\tripleprime 
		-\frac{3 i}{2 } \cfivebar'\, \cfive'' 
		\biggr)
	-\frac{2 \sqrt{2}}{\TachGrad  \alpha '}
		\biggl(
	\cfivebar\, \cfive'\, \dumfive' 
	+\cfivebar\, \cfive'\, \dupfive'
	+ \cfivebar'\, \cfive\, \dumfive' 
	+ \cfivebar'\, \cfive\, \dupfive' 
\xxx
&&
	+\cfivebar'\, \cfive'\, \dumfive 
	+\cfivebar'\, \cfive'\, \dupfive 
		\biggr)
	-\frac{\sqrt{2}}{\TachGrad  \alpha   '}
		\biggl(
	\cfivebar\, \cfive\, \dumfive'' 
	+\cfivebar\, \cfive\, \dupfive'' 
	+\cfivebar\, \cfive''\, \dumfive 
	+\cfivebar\, \cfive''\, \dupfive 
\xxx
&&
	+\cfivebar''\, \cfive\, \dumfive 
	+\cfivebar''\, \cfive\, \dupfive
		\biggr)  \ .
\ee


The next transformation renders the bosonic $y,~u$ theory 
reflection symmetric about the little group of the renormalized 
linear dilaton. This is achieved by
acting with the unitary transformation $U_5$:
\bbb
U\ll 5 \equiv \exp{{i\over{2\pi}}{\bf g}\ll 5} \ ,
\eee
where
\bbb
{\bf g}\ll 5 \equiv \int d\s\ll 1  g\ll 5 (\s) \ .
\eee
The explicit form of $g_5$ is given above in Eqn.~\rr{genfive}.
This transformation moves us to variables marked with the subscript 4.
The transformed supercurrent takes the form
\bbb
G_4 = U_5 G_5 U_5^\dagger & = & 
  \bfour + G^\perp 
  -i \tlcbos \cfourbar\,\cfour'\,\cfour
   +2 i \cfourbar\,\cfour''\,\cfour'-\frac{7 i}{6}
   \cfourbar\,\cfour\tripleprime\,\cfour
	-2 i \cfour'\,\cfour\,\bfourbar
\xxx
&&
\kern-45pt
	-\frac{3 i}{2}
   \cfourbar'\,\cfour''\,\cfour
	-\frac{3 i}{2} \cfourbar''\,\cfour'\,\cfour
	-4\cfourbar'\,\cfourbar\,\cfour''\,\cfour'\,\cfour
	-\tlcbos \cfour
	 -\frac{2}{\alpha '}
	\bigl(
	\cfour\,\dupfour\,\dumfour
\xxx
&&
\kern-45pt
   + i\cfourbar\,\cfour'\,\cfour\,\dupfour\,\dumfour
	\bigr)	
	+ \frac{1}{\TachGrad ^2 \alpha '}
	\bigl(
	i\cfourbar\,\cfour\tripleprime\,\cfour
	+{i} \cfourbar'\,\cfour''\,\cfour 
	 -{2 i}\cfourbar\,\cfour''\,\cfour' 
  	+ {5 i}\cfourbar''\,\cfour'\,\cfour
\xxx
&&
\kern-45pt
	+{4}\cfourbar'\,\cfourbar\,\cfour''\,\cfour'\,\cfour 
	 + 4 \cfour'' \bigr)
	+\frac{2  \sqrt{2}}{\TachGrad  \alpha '}
	\bigl(
	i \cfour'\,\dumfour 
	+ i\cfour'\,\dupfour
	+\cfourbar\,\cfour'\,\cfour\,\dumfour' 
	+ \cfourbar\,\cfour'\,\cfour\,\dupfour'
\xxx
&&
\kern-45pt
	+\cfourbar'\,\cfour'\,\cfour\,\dumfour 
	+\cfourbar'\,\cfour'\,\cfour\,\dupfour 
	\bigr)
	-  \cfour'' \ .
\eee
We now record the R-current, decomposed according to 
antighost number (denoted by the superscript index on the left-hand side): 
\bbb
J_4^{(0)} & = &    
J^\perp + 
\cfour\,\bfourbar- \cfourbar\,\bfour
	-\frac{2 \sqrt{2}}{\TachGrad  \alpha'}
	\bigl(
	\dumfour 
	+\dupfour 
	\bigr) \ ,
\xxx
J_4^{(2)} & = & 
	\frac{1}{8}\bigl(
	-{42} 
	+{\cbos}
	-{\centperp} 
	\bigr)\cfourbar\,\cfour''
	-4 \cfourbar'\,\cfour'
	-2 \cfourbar''\,\cfour
	+\frac{1}{\TachGrad^2\apr}
	\bigl(
	{7}\cfourbar\,\cfour''
	+4 \cfourbar''\,\cfour
	+8 \cfourbar'\,\cfour' 
	\bigr) \ ,
\xxx
J_4^{(4)} & = & -\frac{i}{8\apr\TachGrad^2}
	\bigl(	24 + \apr\TachGrad^2(\cbos - \centperp -26) \bigr)
	\bigl(
	3 \cfourbar'\, \cfourbar\, \cfour''\, \cfour
	+ \cfourbar''\, \cfourbar\, \cfour'\, \cfour
	\bigr) \ .
\eee
Similarly, the stress tensor is given by
\bbb
T_4^{(0)} & = &  
	\tlcbos + \tperp 
   -\frac{i}{2} \bigl(
	\cfour\,\bfourbar'
	+ \cfourbar\,\bfour'
	\bigr)
	-\frac{3 i}{2}
	\bigl(
   	\cfour'\,\bfourbar
	+ \cfourbar'\,\bfour
	\bigr)
	+ \frac{2}{\alpha'}\dupfour\,\dumfour \ ,
\eee
\bbb
T_4^{(2)} & = & \frac{11 i}{24} \cfourbar\,\cfour\tripleprime
	+\frac{i \clcbos}{48}\cfourbar\,\cfour\tripleprime
	+\frac{i}{2 \TachGrad ^2 \alpha '} \cfourbar\,\cfour\tripleprime
	-\frac{5 i}{8} \cfourbar'\,\cfour''
	+\frac{i \clcbos}{16} \cfourbar'\,\cfour''
	-i \cfourbar''\,\cfour'
\xxx
&&
	-i \cfourbar\tripleprime\,\cfour
	+\frac{3 i}{2 \TachGrad ^2\alpha '} \cfourbar'\,\cfour'' 
	    -\frac{2 \sqrt{2}}{\TachGrad  \alpha'} 
	\bigl(
	\cfourbar\,\cfour'\,\dumfour' 
	+\cfourbar\,\cfour'\,\dupfour' 
	+\cfourbar'\,\cfour\,\dumfour' 
	+\cfourbar'\,\cfour\,\dupfour' 
	+\cfourbar'\,\cfour'\,\dumfour
\xxx
&&
	+\cfourbar'\,\cfour'\,\dupfour 
	\bigr)
	    -\frac{\sqrt{2}}{\TachGrad  \alpha '}
	\bigl(
	 \cfourbar\,\cfour\,\dumfour'' 
	+\cfourbar\,\cfour\,\dupfour'' 
	+\cfourbar\,\cfour''\,\dumfour 
	+\cfourbar\,\cfour''\,\dupfour 
\xxx
&&
	+\cfourbar''\,\cfour\,\dumfour 
	+\cfourbar''\,\cfour\,\dupfour 
	\bigr) \ ,
\eee
\bbb
T_4^{(4)} & = &
\frac{3}{8} \cfourbar'\,\cfourbar\,\cfour''\,\cfour'
	+\frac{\cbos}{16}\cfourbar'\,\cfourbar\,\cfour''\,\cfour'
	-\frac{\centperp}{16}\cfourbar'\,\cfourbar\,\cfour''\,\cfour'
	-\frac{1}{2 \TachGrad ^2 \alpha '}\cfourbar'\,\cfourbar\,\cfour''\,\cfour'
	+\frac{35}{24}\cfourbar'\,\cfourbar\,\cfour\tripleprime\,\cfour
\xxx
&&
	+\frac{\cbos}{48}\cfourbar'\,\cfourbar\,\cfour\tripleprime\,\cfour
	-\frac{\centperp}{48}\cfourbar'\,\cfourbar\,\cfour\tripleprime\,\cfour
	+\frac{3}{4}\cfourbar''\,\cfourbar\,\cfour''\,\cfour
	+\frac{\cbos}{8}\cfourbar''\,\cfourbar\,\cfour''\,\cfour
	-\frac{\centperp}{8}\cfourbar''\,\cfourbar\,\cfour''\,\cfour
\xxx
&&
	+\frac{3}{8}\cfourbar''\,\cfourbar'\,\cfour'\,\cfour
	+\frac{\cbos}{16}\cfourbar''\,\cfourbar'\,\cfour'\,\cfour
	-\frac{\centperp}{16}\cfourbar''\,\cfourbar'\,\cfour'\,\cfour
	+\frac{35}{24}\cfourbar\tripleprime\,\cfourbar\,\cfour'\,\cfour
	+\frac{\cbos}{48}\cfourbar\tripleprime\,\cfourbar\,\cfour'\,\cfour
\xxx
&&
	-\frac{\centperp}{48}\cfourbar\tripleprime\,\cfourbar\,\cfour'\,\cfour
	-\frac{1}{\TachGrad^2\apr}
		\bigl(
		\frac{3}{2}\cfourbar'\,\cfourbar\,\cfour\tripleprime\,\cfour 
		+\frac{3}{2}\cfourbar\tripleprime\,\cfourbar\,\cfour'\,\cfour
\xxx
&&
		+\cfourbar''\,\cfourbar\,\cfour''\,\cfour  
		+\frac{1}{2}\cfourbar''\,\cfourbar'\,\cfour'\,\cfour  
		\bigr) \ .
\eee

The next step is to define a unitary transformation that rotates the
R-current into a particular universal form, in which $J^\perp$ is absent entirely.
This is achieved by defining the generating function
\bbb
g\ll 4 \equiv - {\TachGrad \over{\sqrt{2}}} J\uu\perp u\uu + \ .
\eee
Under the corresponding finite unitary transformation, we obtain
the transformed supercurrent in ``3'' variables:
\bbb
G_3 & = & b + G^\perp \normx +
	-\frac{\jperp \TachGrad }{\sqrt{2}} \cthree\,\dupthree
	+\frac{i \centperp \TachGrad}{6 \sqrt{2}} \cthree'\,\dupthree
	-\frac{\centperp \TachGrad ^2}{12}\cthree\,\dupthree\,\dupthree  
	- 2 i \cthree'\,\cthree\,\bthreebar
\xxx
&&
	-\tlcbos \cthree
	-\frac{2}{\alpha '} \cthree\,\dupthree\,\dumthree 
	+\frac{2 i \sqrt{2}}{\TachGrad  \alpha '}
	\bigl(
	\cthree'\,\dumthree 
	+\cthree'\,\dupthree 
	\bigr)
	+i \jperp \cthree'
	- \cthree''
\xxx
&&
	+\frac{4}{\TachGrad ^2 \alpha '} \cthree''
	-i \tlcbos \cthreebar\,\cthree'\,\cthree	
	+\jperp' \cthreebar\,\cthree'\,\cthree
	+2 i  \cthreebar\,\cthree''\,\cthree'
	-\frac{7 i}{6} \cthreebar\,\cthree\tripleprime\,\cthree
	+\jperp\cthreebar'\,\cthree'\,\cthree
\xxx
&&
	-\frac{3 i}{2} \cthreebar'\,\cthree''\,\cthree
	-\frac{3 i}{2}\cthreebar''\,\cthree'\,\cthree
	-\frac{i \jperp \TachGrad }{\sqrt{2}}\cthreebar\,\cthree'\,\cthree\,\dupthree
	+\frac{\centperp \TachGrad }{6 \sqrt{2}}
   		\bigl(
		\cthreebar\,\cthree'\,\cthree\,\dupthree'
		+\cthreebar'\,\cthree'\,\cthree\,\dupthree
		\bigr)
\xxx
&&
	-\frac{1}{12} i \centperp \TachGrad ^2 \cthreebar\,\cthree'\,\cthree\,\dupthree\,\dupthree
	-\frac{2 i}{\alpha '}\cthreebar\,\cthree'\,\cthree\,\dupthree\,\dumthree
	+\frac{i}{\TachGrad^2\apr}
		\bigl(
		\cthreebar\,\cthree\tripleprime\,\cthree 
		+\cthreebar'\,\cthree''\,\cthree 
		-{2}\cthreebar\,\cthree''\,\cthree' 
\xxx
&&
		+{5}\cthreebar''\,\cthree'\,\cthree
		\bigr) 
	+\frac{2 \sqrt{2}}{\TachGrad  \alpha '}
		\bigl(
	 	\cthreebar\,\cthree'\,\cthree\,\dumthree' 	
		+\cthreebar\,\cthree'\,\cthree\,\dupthree' 	
		+\cthreebar'\,\cthree'\,\cthree\,\dumthree 	
		+\cthreebar'\,\cthree'\,\cthree\,\dupthree 
		\bigr)
\xxx
&&
	+4\bigl( \frac{1}{\TachGrad^2\apr} - 1 \bigr)	
	\cthreebar'\,\cthreebar\,\cthree''\,\cthree'\,\cthree \ .
\eee
The corresponding R-current is
\bbb
J_3 & = & 
	\frac{\centperp \TachGrad }{6 \sqrt{2}} \dupthree 
	+ \cthree\,\bthreebar
	-\cthreebar\,\bthree
	-\frac{21}{4} \cthreebar\,\cthree''
	+\frac{\clcbos}{8}\cthreebar\,\cthree''
	-4 \cthreebar'\,\cthree'
	-2 \cthreebar''\,\cthree
\xxx
&&
	+\frac{1}{\TachGrad ^2 \alpha '}
		\left(
	{4} \cthreebar''\,\cthree
	+{7} \cthreebar\,\cthree'' 
	+{8} \cthreebar'\,\cthree' 
		\right)
	-\frac{2\sqrt{2}}{\TachGrad  \alpha '}
		\left( \dumthree +\dupthree \right) \ .
\eee
Decomposing by antighost number, the stress tensor appears as
\bbb
T_3 ^{(0)} & = & 
	-\frac{i}{2} 
		\left(
		\cthree\,\bthreebar' +  \cthreebar\,\bthree'
		\right)
	-\frac{3 i}{2} 
		\left(\cthree'\,\bthreebar
		+ \cthreebar'\,\bthree
		\right)
	+\frac{2}{\alpha '} \dupthree\,\dumthree \ ,
\eee
\bbb
T_3 ^{(2)} & = & 
	-\frac{1}{2}\del^2_+ \left( \jperp\, \cthreebar\, \cthree \right)
	+\frac{11 i}{24} \cthreebar\,\cthree\tripleprime
	+\frac{i\cbos}{48} \cthreebar\,\cthree\tripleprime
	-\frac{i \centperp}{48}\cthreebar\,\cthree\tripleprime
	-\frac{5 i}{8}\cthreebar'\,\cthree''
	+\frac{i \cbos}{16} \cthreebar'\,\cthree''
\xxx
&&
	-\frac{i\centperp}{16} \cthreebar'\,\cthree''
	-i\cthreebar''\,\cthree'
	-i \cthreebar\tripleprime\,\cthree
	-\frac{\centperp \TachGrad }{12\sqrt{2}} 
		\bigl(
	\cthreebar\,\cthree\,\dupthree''
	+2 \cthreebar\,\cthree'\,\dupthree'
\xxx
&&
	+\cthreebar\,\cthree''\,\dupthree
	+2\cthreebar'\,\cthree\,\dupthree'
	+2\cthreebar'\,\cthree'\,\dupthree
	+\cthreebar''\,\cthree\,\dupthree
		\bigr)
	+\frac{i}{2 \TachGrad ^2 \alpha'} 
		\bigl(
	\cthreebar\,\cthree\tripleprime 
	+{3 } \cthreebar'\,\cthree'' 
		\bigr)
\xxx
&&
	-\frac{2 \sqrt{2}}{\TachGrad  \alpha '}
		\bigl(
   	\cthreebar\,\cthree'\,\dumthree'
   	+\cthreebar\,\cthree'\,\dupthree'
   	+\cthreebar'\,\cthree\,\dumthree'
   	+\cthreebar'\,\cthree\,\dupthree'
   	+\cthreebar'\,\cthree'\,\dumthree
   	+\cthreebar'\,\cthree'\,\dupthree
		\bigr)
\xxx
&&
	-\frac{\sqrt{2}}{\TachGrad\apr}
		\bigl(
  	\cthreebar\,\cthree\,\dumthree''
  	+\cthreebar\,\cthree\,\dupthree''
  	+\cthreebar\,\cthree''\,\dumthree
\xxx
&&
\kern+40pt
  	+\cthreebar\,\cthree''\,\dupthree
  	+\cthreebar''\,\cthree\,\dumthree
  	+\cthreebar''\,\cthree\,\dupthree
		\bigr)   \ ,
\eee
\bbb
T_3 ^{(4)} & = & \frac{3}{8} \cthreebar'\,\cthreebar\,\cthree''\,\cthree'
	+\frac{\cbos}{16}
   \cthreebar'\,\cthreebar\,\cthree''\,\cthree'
	-\frac{\centperp}{16}
   \cthreebar'\,\cthreebar\,\cthree''\,\cthree'
	-\frac{1}{2 \TachGrad ^2 \alpha '}
   \cthreebar'\,\cthreebar\,\cthree''\,\cthree'
	+\frac{35}{24}
   \cthreebar'\,\cthreebar\,\cthree\tripleprime\,\cthree
\xxx
&&
	+\frac{\cbos}{48}
   \cthreebar'\,\cthreebar\,\cthree\tripleprime\,\cthree
	-\frac{\centperp}{48}
   \cthreebar'\,\cthreebar\,\cthree\tripleprime\,\cthree
	+\frac{3}{4}
   \cthreebar''\,\cthreebar\,\cthree''\,\cthree
	+\frac{\cbos}{8}
   \cthreebar''\,\cthreebar\,\cthree''\,\cthree
	-\frac{\centperp}{8}
   \cthreebar''\,\cthreebar\,\cthree''\,\cthree
\xxx
&&
	+\frac{3}{8}
   \cthreebar''\,\cthreebar'\,\cthree'\,\cthree
	+\frac{\cbos}{16}
   \cthreebar''\,\cthreebar'\,\cthree'\,\cthree
	-\frac{\centperp}{16}
   \cthreebar''\,\cthreebar'\,\cthree'\,\cthree
	+\frac{35}{24}
   \cthreebar\tripleprime\,\cthreebar\,\cthree'\,\cthree
	+\frac{\cbos}{48}
   \cthreebar\tripleprime\,\cthreebar\,\cthree'\,\cthree
\xxx
&&
	-\frac{\centperp}{48}\cthreebar\tripleprime\,\cthreebar\,\cthree'\,\cthree
	-\frac{3}{2 \TachGrad ^2 \alpha '}
		\bigl(
	\cthreebar'\,\cthreebar\,\cthree\tripleprime\,\cthree 
	+\cthreebar\tripleprime\,\cthreebar\,\cthree'\,\cthree 
		\bigr)
\xxx
&&
	-\frac{1}{\TachGrad ^2 \alpha '}
		\bigl(
	\cthreebar''\,\cthreebar\,\cthree''\,\cthree 
	+\frac{1}{2}\cthreebar''\,\cthreebar'\,\cthree'\,\cthree 
		\bigr) \ .
\eee

At this stage we need to bring the supercurrent itself into
a universal form in which only $ T_{\text{bose}} $ appears explicitly.
To this end, we employ the generating 
function defined above in Eqn.~\rr{genthree}.  The corresponding 
finite transformation yields the supercurrent  
\bbb
G_2 	& = &  b 
      -\frac{i \centperp \TachGrad }{6 \sqrt{2}} \del_+ \left(
		\ctwo\,\duptwo  \right)
	+\frac{2 i\sqrt{2}}{\TachGrad  \alpha '} \ctwo\,\duptwo'
	+i \ctwobar\,\btwo\,\ctwo'+i \ctwobar\,\btwo'\,\ctwo-i
   	\ctwo'\,\ctwo\,\btwobar
\xxx
&&
\kern-10pt
	+2 i \ctwobar'\,\btwo\,\ctwo
	-T_{\text{bose}} \ctwo 
	-\frac{2}{\alpha '}\ctwo\,\duptwo\,\dumtwo 
	+ \frac{2 i \sqrt{2}}{\TachGrad  \alpha '}
	\bigl(
	\ctwo'\,\dumtwo 
	+\ctwo'\,\duptwo
   	\bigr) 
	+\left(\frac{4}{\TachGrad ^2 \alpha '} - \frac{\centperp}{6}  - 1\right)\ctwo'' 
\xxx
&&
\kern-10pt
	+\frac{i \centperp}{12} \ctwobar\,\ctwo''\,\ctwo'
	-\frac{2 i}{\TachGrad ^2 \alpha '}
   		\ctwobar\,\ctwo''\,\ctwo'
	+\frac{i \centperp}{8} \ctwobar\,\ctwo\tripleprime\,\ctwo
	+\frac{7 i \centperp}{24} \ctwobar'\,\ctwo''\,\ctwo
	+\frac{i \centperp}{8} \ctwobar''\,\ctwo'\,\ctwo
\xxx
&&
\kern-10pt
	-\frac{i}{\TachGrad^2\apr}
		\bigl(
		3 \ctwobar\,\ctwo\tripleprime\,\ctwo
		+3 \ctwobar''\,\ctwo'\,\ctwo 
		+7 \ctwobar'\,\ctwo''\,\ctwo 
		\bigr)
	+\frac{2 \sqrt{2}}{\TachGrad  \alpha   '}
	\bigl(
	\ctwobar\,\ctwo'\,\ctwo\,\dumtwo' 
	+\ctwobar\,\ctwo''\,\ctwo\,\dumtwo 
\xxx
&&
\kern-10pt
	+2\ctwobar'\,\ctwo'\,\ctwo\,\dumtwo 
	\bigr)
	-\frac{\centperp}{24} \ctwobar'\,\ctwobar\,\ctwo''\,\ctwo'\,\ctwo
	+\frac{1}{\TachGrad ^2 \alpha '} \ctwobar'\,\ctwobar\,\ctwo''\,\ctwo'\,\ctwo   \ .
\eee
The R-current takes the form
\bbb
J_2 & = & 
	\frac{\centperp \TachGrad }{6 \sqrt{2}}\duptwo 
	+ \ctwo\,\btwobar
	-\ctwobar\,\btwo
	-\frac{13}{4} \ctwobar\,\ctwo''
	+\frac{\cbos}{8}\ctwobar\,\ctwo''
	-\frac{\centperp}{8} \ctwobar\,\ctwo''
	+\frac{3}{\TachGrad ^2 \alpha '} \ctwobar\,\ctwo'' 
\xxx
&&
	-\frac{2 \sqrt{2}}{\TachGrad  \alpha '} \left(\dumtwo  +\duptwo  \right) \ .
\eee
Broken up by antighost number, the stress tensor takes the form
\bbb
T_2^{(0)} & = &  
	T^{\rm bose} 
	-\frac{i}{2} \left( \ctwo\,\btwobar'+ \ctwobar\,\btwo' \right)
	-\frac{3 i}{2} \left( \ctwo'\,\btwobar +  \ctwobar'\,\btwo\right)
	+\duptwo\,\dumtwo \frac{2}{\alpha '} \ ,
\eee
\bbb
T_2^{(2)} & = &
	-\frac{13 i}{24} \ctwobar\,\ctwo\tripleprime
	+\frac{i \cbos}{48} \ctwobar\,\ctwo\tripleprime
	-\frac{i \centperp}{48} \ctwobar\,\ctwo\tripleprime
	+\frac{i}{2\TachGrad ^2 \alpha '} \ctwobar\,\ctwo\tripleprime
	-\frac{13 i}{8} \ctwobar'\,\ctwo''
	+\frac{i\cbos}{16} \ctwobar'\,\ctwo''
\xxx
&&
	-\frac{i \centperp}{16}\ctwobar'\,\ctwo''        
	-\frac{\centperp \TachGrad }{12 \sqrt{2}}
   		\bigl(
   	\ctwobar\,\ctwo\,\duptwo''
   	+2 \ctwobar\,\ctwo'\,\duptwo'
   	+ \ctwobar\,\ctwo''\,\duptwo
   	+ 2 \ctwobar'\,\ctwo\,\duptwo'
   	+ 2 \ctwobar'\,\ctwo'\,\duptwo
\xxx
&&
   	+ \ctwobar''\,\ctwo\,\duptwo
		\bigr)
	+\frac{3 i}{2 \TachGrad ^2 \alpha'} \ctwobar'\,\ctwo'' 
	- \frac{ \sqrt{2}}{\TachGrad \apr}
		\biggl(
	2\ctwobar\,\ctwo'\,\dumtwo' 
	+2\ctwobar'\,\ctwo\,\dumtwo' 
	+2\ctwobar'\,\ctwo'\,\dumtwo 
	+\ctwobar\,\ctwo\,\dumtwo'' 
\xxx
&&
	+\ctwobar\,\ctwo''\,\dumtwo 
	+\ctwobar''\,\ctwo\,\dumtwo 
	-\ctwobar\,\ctwo\,\duptwo'' 
	-\ctwobar\,\ctwo''\,\duptwo
	-\ctwobar''\,\ctwo\,\duptwo       
\xxx
&&
	-2\ctwobar\,\ctwo'\,\duptwo'        
	-2\ctwobar'\,\ctwo\,\duptwo'       
	-2\ctwobar'\,\ctwo'\,\duptwo 
	\biggr)  \ ,
\eee
\bbb
T_2^{(4)} & = &  
   -\frac{1}{12 \TachGrad^2\apr} 
	\bigl(\TachGrad^2\apr \centperp - 24\bigr)
	\biggl(
	 \ctwobar'\,\ctwobar\,\ctwo''\,\ctwo'
	+\ctwobar'\,\ctwobar\,\ctwo\tripleprime\,\ctwo
\xxx
&&
\kern+60pt
	+2 \ctwobar''\,\ctwobar\,\ctwo''\,\ctwo
	+\ctwobar''\,\ctwobar'\,\ctwo'\,\ctwo 
	+\ctwobar\tripleprime\,\ctwobar\,\ctwo'\,\ctwo
	\biggr)\ .
\eee

The final step is to render the supercurrent in a complex
universal form, in which the conjugate current $\bar G_1$ is given simply by
\be
\bar G_1 = \bar b_1 \ .
\ee
This final transformation is not unitary.  Instead, it comprises a similarity transformation
$S_2$, generated by the function in Eqn.~\rr{gentwo} above.  We obtain
\bbb
G_1 & = & b_1 + 
   -\frac{i \centperp \TachGrad }{6 \sqrt{2}} \cone\,\dupone'+\frac{2 i
   \sqrt{2}}{\TachGrad  \alpha '} \cone\,\dupone'-\frac{i \centperp \TachGrad }{3
   \sqrt{2}} \cone'\,\dupone+2 i \conebar\,\bone\,\cone'+2 i \conebar\,\bone'\,\cone
\xxx
&&
   -2 i \cone'\,\cone\,\bonebar+4 i \conebar'\,\bone\,\cone
	-2 T_{\rm bose} \cone 
	-\frac{4}{\apr}\cone\,\dupone\,\dumone 
	+\frac{2 i \sqrt{2}}{\TachGrad  \alpha '}
	\bigl(
	\cone\,\dumone' 
	+2 \cone'\,\dumone 
	+2 \cone'\,\dupone 
	\bigr)
\xxx
&&
	-2 \cone''
	-\frac{\centperp}{3} \cone''
	+\frac{8}{\TachGrad ^2 \alpha '}\cone''
\xxx
&&
	 - \frac{i}{16\TachGrad^2\apr}
	\bigl( 24 + \apr \TachGrad^2 (\cbos - \centperp - 26) \bigr)
	\biggl(
	\conebar\,\cone''\,\cone'
	-\conebar\,\cone\tripleprime\,\cone
	- {4} \conebar'\,\cone''\,\cone
	-6 \conebar''\,\cone'\,\cone \biggr)
\xxx
&&
	+ \frac{7}{24 \TachGrad^2\apr}
	\bigl( 24 + \apr \TachGrad^2 (\cbos - \centperp - 26) \bigr)
	\,
	\conebar'\,\conebar\,\cone''\,\cone'\,\cone  \ .
\eee
The conjugated supercurrent now appears as
\bbb
\bar G_1 & = & \bar b_1
	-\frac{i}{48\TachGrad^2 \apr}
	\bigl( 24 + \apr \TachGrad^2 (\cbos - \centperp - 26) \bigr)
	\biggl(
	6 \conebar'\,\conebar\,\cone''
	-3  \conebar''\,\conebar'\,\cone
	- \conebar\tripleprime\,\conebar\,\cone
	\biggr)
\xxx
&&
	+\frac{1}{12\TachGrad^2 \apr}
	\bigl( 24 + \apr \TachGrad^2 (\cbos - \centperp - 26) \bigr)
	\, \conebar''\,\conebar'\,\conebar\,\cone'\,\cone \ .
\eee
Decomposed by antighost number, the R-current takes the form
\bbb
J_1^{(0)} & = & 
	\dupone \frac{\centperp \TachGrad }{6 \sqrt{2}}
	+ \cone\,\bonebar
	- \conebar\,\bone
	- \frac{2 \sqrt{2}}{\TachGrad  \alpha'} 
		\biggl( 
		\dumone + \dupone \biggr) \ ,
\xxx
J_1^{(2)} & = & 
	\frac{1}{8\TachGrad^2 \apr}
	\bigl( 24 + \apr \TachGrad^2 (\cbos - \centperp - 26) \bigr) 
	\, \conebar\,\cone''  \ ,
\xxx
J_1^{(4)} & = & 
	-\frac{i}{4\TachGrad^2 \apr}
	\bigl( 24 + \apr \TachGrad^2 (\cbos - \centperp - 26) \bigr)
	\bigl(
	\conebar'\,\conebar\,\cone''\,\cone
	+ \conebar''\,\conebar\,\cone'\,\cone
	\bigr) \ .
\eee
Similarly, the stress tensor becomes
\bbb
T_1^{(0)} & = & -\frac{i}{2} \cone\,\bonebar'-\frac{i}{2} \conebar\,\bone'-\frac{3 i}{2}
   \cone'\,\bonebar-\frac{3 i}{2} \conebar'\,\bone+\dupone\,\dumone \frac{2}{\alpha
   '} \ ,
\eee
\bbb
T_1^{(2)} & = & 
	-\frac{i}{48 \TachGrad^2 \apr}
	\bigl( 24 + \apr \TachGrad^2 (\cbos - \centperp - 26) \bigr)
\xxx
&&
\kern+40pt
	\times \bigl(
	3 \conebar\,\cone\tripleprime
	+ 9 \conebar'\,\cone''
	+ 12 \conebar''\,\cone' 
	+ 4 \conebar'''\,\cone
	\bigr) \ ,
\eee
\bbb
T_1^{(4)} & = & 
	-\frac{1}{48 \TachGrad^2 \apr}
	\bigl( 24 + \apr \TachGrad^2 (\cbos - \centperp - 26) \bigr)
	\biggl(
	9 \conebar'\,\conebar\,\cone''\,\cone'
	+ 7 \conebar'\,\conebar\,\cone'''\,\cone
\xxx
&&
\kern+50pt
	+ 12 \conebar''\,\conebar\,\cone''\,\cone
	+ 3 \conebar''\,\conebar'\,\cone'\,\cone
	+ 5 \conebar'''\,\conebar\,\cone'\,\cone
	\biggr) \ .
\eee

As noted above, the final expressions simplify greatly
upon assigning the values
\be
\cbos = 26 \ , \qquad \centperp = \frac{24}{\TachGrad^2 \apr} \ .
\ee
We obtain the final supercurrent in the form
\be
G_1 & = & b_1 +
2 i \left(
	\bar{c}_1\,b_1\,c_1'
	+ \bar{c}_1\,b_1'\,c_1
	- c_1'\,c_1\,\bar{b}_1
	+2  \bar{c}_1'\,b_1\,c_1 
	\right)
    -2 T_{\text{bose}}  b_1+c_1 
\xxx
&&
	-\frac{4}{\alpha '} c_1\,\dupone\,\dumone 
	+\frac{2 i \sqrt{2}}{\TachGrad  \alpha '} 
	\left(
	c_1\,\dumone' 
	+2 c_1'\,\dumone \right)
	-2 c_1''   \ ,
\xxx
\bar G_1 &=& \bar b_1 \ .
\ee
The R-current and stress tensor take the final forms
\be
J & = & -\frac{2\sqrt{2}}{\TachGrad \apr}\dumone + \cone\, \bonebar - \conebar\, \bone \ ,
\xxx
T & = & T_{\rm bose} + \frac{2}{\apr} \dupone\, \dumone
	- \biggl(
	\frac{i}{2} \cone\, \bonebar'
	+ \frac{3i}{2} \cone' \bonebar'
	+ {\rm h.c.}
	\biggr) \ .
\ee


\appendix{Explicit similarity transformation of the BRST current}
\label{brst}
Here we present the explicit canonical variable redefinition that 
renders the BRST current in the universal form presented in Eqn.~\rr{BRSTfinal}.
As described above, because of the presence of a number of currents that
commute with the BRST current, there is a certain amount of freedom to
choose a generating function that yields Eqn.~\rr{BRSTfinal}.
The following choice has the property that it admits particularly simple
transformation rules for the reparametrization $c$ ghosts, and the 
shifted-spin $c_1$ antighosts:
\be
g_0   & = & 
2 i \bar{c}_1\,c_1''
-\sqrt{2} \OMEGA   b\,\bar{\gamma }\,c_1
+\frac{3}{\sqrt{2} \OMEGA } c\,\beta \,\bar{c}_1'
-\frac{3 \OMEGA }{\sqrt{2}} c\,\bar{\beta }\,c_1' 
-i \sqrt{2}  c\,\balt\, {\vplus }'
\nn\\&&
+\frac{1}{\sqrt{2} \OMEGA } c\,\beta '\,\bar{c}_1
-\frac{1 }{\sqrt{2}}   c\,\bar{\beta }'\,c_1
+{2 i \sqrt{2}} \balt\,\gamma \,\bar{c}_1'
-i \sqrt{2} \OMEGA   \balt\,\bar{\gamma }\,c_1'
+ 3   i \sqrt{2} \OMEGA   \balt\,\bar{\gamma }'\,c_1
\nn\\&&
-\frac{i}{\sqrt{2} \OMEGA } \calt\,\beta \,\bar{c}_1
-\frac{i \OMEGA }{\sqrt{2}}\calt\,\bar{\beta }\,c_1
-\frac{4 \sqrt{2}}{\apr } \bar{c}_1\,c_1\, {\vminus }''
+\frac{1}{3 \sqrt{2}} \bar{c}_1\,c_1\, {\vplus }''
-\frac{4 \sqrt{2}}{\apr }\bar{c}_1\,c_1'\, {\vminus }'
\nn\\&&
-\frac{1}{3 \sqrt{2}} \bar{c}_1\,c_1'\, {\vplus }'
-1 b\,c\,\bar{c}_1'\,c_1
+\frac{i}{2} b\,\calt\,\bar{c}_1\,c_1
+\frac{1}{2}  b\,c'\,\bar{c}_1\,c_1
-\frac{i}{2 \OMEGA } c\,\beta \,\bar{c}_1\, {\vplus }'
\nn\\&&
-\frac{i \OMEGA }{2} c\,\bar{\beta }\,c_1\, {\vplus }'
+i c\,\balt\,\bar{c}_1'\,c_1'
-i c\,\balt\,\bar{c}_1''\,c_1
-\frac{i}{2} c\,\balt'\,\bar{c}_1\,c_1'
-\frac{i}{2}c\,\balt'\,\bar{c}_1'\,c_1
\nn\\&&
-\frac{i}{2} c\,\balt''\,\bar{c}_1\,c_1
+{i} \beta \,\gamma \,\bar{c}_1'\,\bar{c}_1
+(3 i) \beta   \,\bar{\gamma }\,\bar{c}_1'\,c_1
-2 i \beta \,\bar{\gamma }'\,\bar{c}_1\,c_1
-i \bar{\beta }\,\gamma \,\bar{c}_1'\,c_1
\nn\\&&
+\OMEGA  \balt\,\bar{\gamma }\,c_1\, {\vplus }'
-\frac{1}{2} \balt\,\calt\,\bar{c}_1\,c_1'
+\frac{1}{2} \balt\,\calt\,\bar{c}_1'\,c_1
-\frac{1}{2} \balt\,\calt'\,\bar{c}_1\,c_1
+2 \bar{c}_1'\,\bar{c}_1\,b_1\,c_1
\nn\\&&
+\frac{5}{6} \bar{c}_1'\,\bar{c}_1\,c_1''\,c_1
+\frac{i \sqrt{2}}{3}  b\,c\,\bar{c}_1\,c_1\, {\vplus }'
-\frac{i \sqrt{2}}{3 \OMEGA } b\,\gamma \,\bar{c}_1'\,\bar{c}_1\,c_1
+\frac{i \sqrt{2}}{3 \OMEGA } c\,\beta \,\bar{c}_1'\,\bar{c}_1\,c_1'
+\frac{i \OMEGA }{3 \sqrt{2}} c\,\bar{\beta }\,\bar{c}_1\,c_1''\,c_1
\nn\\&&
+\frac{i \OMEGA }{3 \sqrt{2}} c\,\bar{\beta  }\,\bar{c}_1'\,c_1'\,c_1
-\frac{1}{3 \sqrt{2}} c\,\balt\,\bar{c}_1\,c_1\, {\vplus }''
+\frac{1}{3 \sqrt{2}}   c\,\balt\,\bar{c}_1\,c_1'\, {\vplus }'
-\frac{5}{3 \sqrt{2}} c\,\balt\,\bar{c}_1'\,c_1\, {\vplus }'
\nn\\&&
+\frac{i \OMEGA }{3 \sqrt{2}} c\,\bar{\beta   }'\,\bar{c}_1\,c_1'\,c_1
-\frac{1}{3 \sqrt{2}} c\,\balt'\,\bar{c}_1\,c_1\, {\vplus }'
+\frac{1}{3 \sqrt{2}} \beta \,\bar{\gamma   }\,\bar{c}_1\,c_1\, {\vplus }'
-\frac{5 \sqrt{2}}{3 \OMEGA } \balt\,\gamma \,\bar{c}_1'\,\bar{c}_1\,c_1'
\nn\\&&
-{\sqrt{2}} \balt\,\gamma   \,\bar{c}_1''\,\bar{c}_1\,c_1
+\frac{4 \sqrt{2} \OMEGA }{3} \balt\,\bar{\gamma }\,\bar{c}_1'\,c_1'\,c_1
-\frac{i}{3 \sqrt{2}}   \balt\,\calt\,\bar{c}_1\,c_1\, {\vplus }'
+{\sqrt{2}} \balt\,\gamma '\,\bar{c}_1'\,\bar{c}_1\,c_1
\nn\\&&
-\frac{4 \sqrt{2} \OMEGA }{3}   \balt\,\bar{\gamma }'\,\bar{c}_1\,c_1'\,c_1
-\frac{\sqrt{2}}{3 \OMEGA } \calt\,\beta \,\bar{c}_1'\,\bar{c}_1\,c_1
+\frac{1 }{3 \sqrt{2}}   \calt\,\bar{\beta }\,\bar{c}_1\,c_1'\,c_1
+\frac{4 i \sqrt{2}}{\apr } \bar{c}_1'\,\bar{c}_1\,c_1'\,c_1\, {\vminus }'
\nn\\&&
+\frac{i}{5 \sqrt{2}}   \bar{c}_1'\,\bar{c}_1\,c_1'\,c_1\, {\vplus }'
-\frac{1}{6 \OMEGA } c\,\beta \,\bar{c}_1'\,\bar{c}_1\,c_1\, {\vplus }'
+\frac{1 }{3} c\,\bar{\beta   }\,\bar{c}_1\,c_1'\,c_1\, {\vplus }'
+\frac{i}{6} c\,\balt\,\bar{c}_1\,c_1\, {\vplus }'\, {\vplus }'
\nn\\&&
-\frac{1}{3}   c\,\balt\,\bar{c}_1'\,\bar{c}_1\,c_1''\,c_1
-\frac{1}{3} c\,\balt\,\bar{c}_1''\,\bar{c}_1\,c_1'\,c_1
-\frac{1}{3}   c\,\balt'\,\bar{c}_1'\,\bar{c}_1\,c_1'\,c_1
-\frac{2}{3} \beta \,\bar{\gamma }\,\bar{c}_1'\,\bar{c}_1\,c_1'\,c_1
\nn\\&&
+\frac{2}{3} \bar{\beta }\,\gamma   \,\bar{c}_1'\,\bar{c}_1\,c_1'\,c_1
+\frac{i}{3 \OMEGA } \balt\,\gamma \,\bar{c}_1'\,\bar{c}_1\,c_1\, {\vplus }'
-\frac{i}{3}   \balt\,\calt\,\bar{c}_1'\,\bar{c}_1\,c_1'\,c_1
+\frac{2 i \sqrt{2}}{15} c\,\balt\,\bar{c}_1'\,\bar{c}_1\,c_1'\,c_1\, {\vplus }'
\ .
\nn\\
&&
\ee

When this generating function is 
promoted to the finite similarity transformation $S$, we obtain the following
explicit transformation rules for each of the fields of interest:
\be
b & \to & 
b-\frac{3 i}{\sqrt{2} \OMEGA } \beta \,\bar{c}_1'
+\frac{3 i \OMEGA }{\sqrt{2}} \bar{\beta }\,c_1'
-\sqrt{2}   \balt\, {\vplus}'-\frac{i}{\sqrt{2} \OMEGA } \beta '\,\bar{c}_1
+\frac{i \OMEGA }{\sqrt{2}} \bar{\beta }'\,c_1
-2 \balt\,\bar{c}_1\,c_1''-4 \balt\,\bar{c}_1'\,c_1'
\zzz
-2 \balt\,\bar{c}_1''\,c_1-4 \balt'\,\bar{c}_1\,c_1'
-4   \balt'\,\bar{c}_1'\,c_1-2 \balt''\,\bar{c}_1\,c_1 
\ ;
\ee
\be
c & \to & 
c+ i \sqrt{2} \OMEGA \bar{\gamma }\,c_1
+(2 i) c\,\bar{c}_1'\,c_1
+\calt\,\bar{c}_1\,c_1
-i c'\,\bar{c}_1\,c_1
+\sqrt{2}  c\,\bar{c}_1\,c_1\, {\vplus}'  \ ;
\ee
\be
b_1 & \to & 
b_1
-{i \sqrt{2}} c\,\beta '
+{2 \sqrt{2}} \balt\,\gamma '
+\frac{1}{\sqrt{2} \OMEGA } \calt\,\beta 
-\frac{3   i}{\sqrt{2} \OMEGA } c'\,\beta 
+{2 \sqrt{2}} \balt'\,\gamma 
+ c\,\beta \, {\vplus}'
+4 c\,\balt\,c_1''
\zzz
+8   c\,\balt'\,c_1'
+4 c\,\balt''\,c_1
-{2} \beta \,\gamma \,\bar{c}_1'
-2 \beta \,\bar{\gamma }\,c_1'
-{2} \beta   \,\gamma '\,\bar{c}_1
-2 \beta \,\bar{\gamma }'\,c_1
-2 i \balt\,\calt\,c_1'
\zzz
+2 i \bar{c}_1\,b_1\,c_1'
+2 i \bar{c}_1\,b_1'\,c_1
+2 i   \bar{c}_1\,c_1''\,c_1'
+\frac{4 i}{3} \bar{c}_1\,c_1\tripleprime\,c_1
+6 c'\,\balt\,c_1'
+6 c'\,\balt'\,c_1
-{2} \beta '\,\gamma   \,\bar{c}_1
\zzz
-2 \beta '\,\bar{\gamma }\,c_1
-2 i \balt'\,\calt\,c_1
+4 i \bar{c}_1'\,b_1\,c_1
+4 i \bar{c}_1'\,c_1''\,c_1
-{2   \sqrt{2}} c\,\beta \,\bar{c}_1\,c_1''
-{4 \sqrt{2}} c\,\beta \,\bar{c}_1'\,c_1'
\zzz
-{2 \sqrt{2}} c\,\beta   \,\bar{c}_1''\,c_1
+2 i \sqrt{2} c\,\balt\,c_1'\, {\vplus}'
-{4 \sqrt{2}} c\,\beta '\,\bar{c}_1\,c_1'
-{6   \sqrt{2}} c\,\beta '\,\bar{c}_1'\,c_1
+2 i \sqrt{2} c\,\balt'\,c_1\, {\vplus}'
\zzz
-{2 \sqrt{2}} c\,\beta   ''\,\bar{c}_1\,c_1
+{4 i \sqrt{2}} \balt\,\gamma \,\bar{c}_1\,c_1''
+{4 i \sqrt{2}} \balt\,\gamma   \,\bar{c}_1'\,c_1'
-4 i \sqrt{2} \OMEGA  \balt\,\bar{\gamma }\,c_1''\,c_1
-{8 i \sqrt{2}} \balt\,\gamma   '\,\bar{c}_1'\,c_1
\zzz
-4 i \sqrt{2} \OMEGA  \balt\,\bar{\gamma }'\,c_1'\,c_1
-{4 i \sqrt{2}} \balt\,\gamma   ''\,\bar{c}_1\,c_1
-{i \sqrt{2}} \calt\,\beta \,\bar{c}_1\,c_1'
-{2 i \sqrt{2}} \calt\,\beta   \,\bar{c}_1'\,c_1
-{i \sqrt{2}} \calt\,\beta '\,\bar{c}_1\,c_1
\zzz
-{3 \sqrt{2}} c'\,\beta \,\bar{c}_1\,c_1'
-{6   \sqrt{2}} c'\,\beta \,\bar{c}_1'\,c_1
-{3 \sqrt{2}} c'\,\beta '\,\bar{c}_1\,c_1
+{4 i \sqrt{2}}   \balt'\,\gamma \,\bar{c}_1\,c_1'
-{4 i \sqrt{2}} \balt'\,\gamma \,\bar{c}_1'\,c_1
\zzz
-4 i \sqrt{2} \OMEGA    \balt'\,\bar{\gamma }\,c_1'\,c_1
-{4 i \sqrt{2}} \balt'\,\gamma '\,\bar{c}_1\,c_1
-{2 i} c\,\beta   \,\bar{c}_1\,c_1'\, {\vplus}'
-{4 i} c\,\beta \,\bar{c}_1'\,c_1\, {\vplus}'
+8 i c\,\balt\,\bar{c}_1\,c_1''\,c_1'
\zzz
+24 i c\,\balt\,\bar{c}_1'\,c_1''\,c_1
+8 i c\,\balt\,\bar{c}_1''\,c_1'\,c_1
-{2 i}   c\,\beta '\,\bar{c}_1\,c_1\, {\vplus}'
+16 i c\,\balt'\,\bar{c}_1\,c_1''\,c_1
+32 i c\,\balt'\,\bar{c}_1'\,c_1'\,c_1
\zzz
+8 i   c\,\balt''\,\bar{c}_1\,c_1'\,c_1
-4 i \beta \,\bar{\gamma }\,\bar{c}_1'\,c_1'\,c_1
-{4 i} \beta \,\gamma   '\,\bar{c}_1'\,\bar{c}_1\,c_1
+4 \balt\,\calt\,\bar{c}_1\,c_1''\,c_1
+8 \balt\,\calt\,\bar{c}_1'\,c_1'\,c_1
\zzz
+12 i   c'\,\balt\,\bar{c}_1\,c_1''\,c_1
+24 i c'\,\balt\,\bar{c}_1'\,c_1'\,c_1
+12 i c'\,\balt'\,\bar{c}_1\,c_1'\,c_1
-{4 i}   \beta '\,\gamma \,\bar{c}_1'\,\bar{c}_1\,c_1
+4 \balt'\,\calt\,\bar{c}_1\,c_1'\,c_1
\zzz
-4 \bar{c}_1'\,\bar{c}_1\,b_1\,c_1'\,c_1
-4   \bar{c}_1'\,\bar{c}_1\,c_1''\,c_1'\,c_1
+{4 i \sqrt{2}} c\,\beta \,\bar{c}_1'\,\bar{c}_1\,c_1''\,c_1
-4 \sqrt{2}   c\,\balt\,\bar{c}_1\,c_1''\,c_1\, {\vplus}'
\zzz
-8 \sqrt{2} c\,\balt\,\bar{c}_1'\,c_1'\,c_1\, {\vplus}'
+{8 i \sqrt{2}}   c\,\beta '\,\bar{c}_1'\,\bar{c}_1\,c_1'\,c_1
-4 \sqrt{2} c\,\balt'\,\bar{c}_1\,c_1'\,c_1\, {\vplus}'
+{8 \sqrt{2}}   \balt\,\gamma \,\bar{c}_1'\,\bar{c}_1\,c_1''\,c_1
\zzz
-{2 \sqrt{2}} \calt\,\beta \,\bar{c}_1'\,\bar{c}_1\,c_1'\,c_1
+{6 i   \sqrt{2}} c'\,\beta \,\bar{c}_1'\,\bar{c}_1\,c_1'\,c_1
+{8 \sqrt{2}} \balt'\,\gamma   \,\bar{c}_1'\,\bar{c}_1\,c_1'\,c_1
-{4} c\,\beta \,\bar{c}_1'\,\bar{c}_1\,c_1'\,c_1\, {\vplus}'
\zzz
-16   c\,\balt\,\bar{c}_1'\,\bar{c}_1\,c_1''\,c_1'\,c_1
-\frac{16 \sqrt{2}}{\alpha   '}\bar{c}_1'\,c_1'\,c_1\,\vminus' 
-\frac{8 \sqrt{2}}{\alpha '}\bar{c}_1\,c_1'\,c_1\,\vminus'' 
-\frac{8   \sqrt{2}}{\alpha '}\bar{c}_1\,c_1''\,c_1\,\vminus' 
\zzz
+c_1\,\vminus'' \frac{4 i \sqrt{2}}{\alpha '}
+c_1'\,\vminus' \frac{4 i \sqrt{2}}{\alpha '}
+8 i   c\,\balt\,\bar{c}_1\,c_1\tripleprime\,c_1
+c_1'' \ ;
\ee
\be
\bar{b}_1 & \to & 
\bar{b}_1
-i \sqrt{2} \OMEGA  b\,\bar{\gamma }
+i \sqrt{2} \OMEGA  c\,\bar{\beta }'
-4 \sqrt{2} \OMEGA    \balt\,\bar{\gamma }'
+\frac{1 }{\sqrt{2}} \calt\,\bar{\beta }
+i \sqrt{2} \bar{c}_1\,\vplus''
+\frac{3 i \OMEGA   }{\sqrt{2}} c'\,\bar{\beta }
-\sqrt{2} \OMEGA  \balt'\,\bar{\gamma }
\zzz
-2 i b\,c\,\bar{c}_1'
- b\,\calt\,\bar{c}_1
+i   b\,c'\,\bar{c}_1
+\OMEGA  c\,\bar{\beta }\,\vplus'
+8 c\,\balt\,\bar{c}_1''
+6 c\,\balt'\,\bar{c}_1'
-4 \beta \,\bar{\gamma }\,\bar{c}_1'
+4   \beta \,\bar{\gamma }'\,\bar{c}_1
\zzz
+2 \bar{\beta }\,\gamma \,\bar{c}_1'
+2 i \OMEGA  \balt\,\bar{\gamma }\,\vplus'
+4 i   \balt\,\calt\,\bar{c}_1'
+2 i \balt\,\calt'\,\bar{c}_1
+8 c'\,\balt\,\bar{c}_1'
-1 c'\,\balt'\,\bar{c}_1
+2  \bar{\gamma }'\,\bar{\beta }\,c_1
+i \balt'\,\calt\,\bar{c}_1
\zzz
+2 i \bar{c}_1'\,\bar{c}_1\,b_1
-2 i   \bar{c}_1'\,\bar{c}_1\,c_1''
-2 c''\,\balt\,\bar{c}_1
+2 i \bar{c}_1''\,\bar{c}_1\,c_1'
-4 i \bar{c}_1''\,\bar{c}_1'\,c_1
+\frac{2 i}{3}   \bar{c}_1^{3}\,\bar{c}_1\,c_1
-\sqrt{2} b\,c\,\bar{c}_1\,\vplus'
\zzz
+2 \sqrt{2} \OMEGA  b\,\bar{\gamma }\,\bar{c}_1'\,c_1
+{4   \sqrt{2}} c\,\beta \,\bar{c}_1''\,\bar{c}_1
+2 \sqrt{2} \OMEGA  c\,\bar{\beta }\,\bar{c}_1'\,c_1'
+2 \sqrt{2} \OMEGA    c\,\bar{\beta }\,\bar{c}_1''\,c_1
-2 i \sqrt{2} c\,\balt\,\bar{c}_1\,\vplus''
\zzz
-6 i \sqrt{2}   c\,\balt\,\bar{c}_1'\,\vplus'
+{2 \sqrt{2}} c\,\beta '\,\bar{c}_1'\,\bar{c}_1
-i \sqrt{2}   c\,\balt'\,\bar{c}_1\,\vplus'
+8 i \sqrt{2} \OMEGA  \balt\,\bar{\gamma }\,\bar{c}_1'\,c_1'
+\sqrt{2}   \balt\,\calt\,\bar{c}_1\,\vplus'
\zzz
+{4 i \sqrt{2}} \balt\,\gamma '\,\bar{c}_1'\,\bar{c}_1
-8 i \sqrt{2} \OMEGA    \balt\,\bar{\gamma }'\,\bar{c}_1\,c_1'
-8 i \sqrt{2} \OMEGA  \balt\,\bar{\gamma }'\,\bar{c}_1'\,c_1
-{3 i   \sqrt{2}} \calt\,\beta \,\bar{c}_1'\,\bar{c}_1
+{3 \sqrt{2}} c'\,\beta \,\bar{c}_1'\,\bar{c}_1
\zzz
-i \sqrt{2}   c'\,\balt\,\bar{c}_1\,\vplus'
+{4 i \sqrt{2}} \balt'\,\gamma \,\bar{c}_1'\,\bar{c}_1
+6 i \sqrt{2} \OMEGA    \balt'\,\bar{\gamma }\,\bar{c}_1'\,c_1
-8 i \sqrt{2} \OMEGA  \balt'\,\bar{\gamma }'\,\bar{c}_1\,c_1
\zzz
+2 \sqrt{2}   \bar{c}_1'\,\bar{c}_1\,c_1\,\vplus''
+2 i b\,\calt\,\bar{c}_1'\,\bar{c}_1\,c_1
+2 b\,c'\,\bar{c}_1'\,\bar{c}_1\,c_1
-{6 i} c\,\beta   \,\bar{c}_1'\,\bar{c}_1\,\vplus'
-2 c\,\balt\,\bar{c}_1\,\vplus'\,\vplus'
\zzz
+8 i c\,\balt\,\bar{c}_1'\,\bar{c}_1\,c_1''
+16 i   c\,\balt\,\bar{c}_1''\,\bar{c}_1\,c_1'
+16 i c\,\balt\,\bar{c}_1''\,\bar{c}_1'\,c_1
+16 i c\,\balt'\,\bar{c}_1'\,\bar{c}_1\,c_1'
+16 i   c\,\balt'\,\bar{c}_1''\,\bar{c}_1\,c_1
\zzz
+8 i c\,\balt''\,\bar{c}_1'\,\bar{c}_1\,c_1
+4 i \beta \,\bar{\gamma }\,\bar{c}_1'\,\bar{c}_1\,c_1'
-4   i \beta \,\bar{\gamma }'\,\bar{c}_1'\,\bar{c}_1\,c_1
-4 \OMEGA  \balt\,\bar{\gamma }\,\bar{c}_1'\,c_1\,\vplus'
-12   \balt\,\calt\,\bar{c}_1'\,\bar{c}_1\,c_1'
\zzz
+4 \balt\,\calt'\,\bar{c}_1'\,\bar{c}_1\,c_1
+12 i   c'\,\balt\,\bar{c}_1'\,\bar{c}_1\,c_1'
+14 i c'\,\balt'\,\bar{c}_1'\,\bar{c}_1\,c_1
+4 i \beta '\,\bar{\gamma   }\,\bar{c}_1'\,\bar{c}_1\,c_1
-10 \balt'\,\calt\,\bar{c}_1'\,\bar{c}_1\,c_1
\zzz
+4 i c''\,\balt\,\bar{c}_1'\,\bar{c}_1\,c_1
+4   \bar{c}_1''\,\bar{c}_1'\,\bar{c}_1\,c_1'\,c_1
+2 i \sqrt{2} b\,c\,\bar{c}_1'\,\bar{c}_1\,c_1\,\vplus'
-{4 i \sqrt{2}}   c\,\beta \,\bar{c}_1''\,\bar{c}_1'\,\bar{c}_1\,c_1
\zzz
-4 \sqrt{2} c\,\balt\,\bar{c}_1'\,\bar{c}_1\,c_1\,\vplus''
+12   \sqrt{2} c\,\balt\,\bar{c}_1'\,\bar{c}_1\,c_1'\,\vplus'
+10 \sqrt{2}   c\,\balt'\,\bar{c}_1'\,\bar{c}_1\,c_1\,\vplus'
\zzz
-8 \sqrt{2} \OMEGA  \balt\,\bar{\gamma   }\,\bar{c}_1'\,\bar{c}_1\,c_1''\,c_1
-2 i \sqrt{2} \balt\,\calt\,\bar{c}_1'\,\bar{c}_1\,c_1\,\vplus'
+8 \sqrt{2} \OMEGA    \balt\,\bar{\gamma }'\,\bar{c}_1'\,\bar{c}_1\,c_1'\,c_1
\zzz
-2 \sqrt{2}   c'\,\balt\,\bar{c}_1'\,\bar{c}_1\,c_1\,\vplus'
-8 \sqrt{2} \OMEGA  \balt'\,\bar{\gamma }\,\bar{c}_1'\,\bar{c}_1\,c_1'\,c_1
+4   i c\,\balt\,\bar{c}_1'\,\bar{c}_1\,c_1\,\vplus'\,\vplus'
-16   c\,\balt\,\bar{c}_1''\,\bar{c}_1'\,\bar{c}_1\,c_1'\,c_1
\zzz
+\bar{c}_1'\,\bar{c}_1\,c_1\,\vminus'' 
-\frac{8 \sqrt{2}}{\alpha   '}
+\bar{c}_1'\,\bar{c}_1\,c_1'\,\vminus' 
-\frac{8 \sqrt{2}}{\alpha '}
+\bar{c}_1'\,\vminus' \frac{4 i \sqrt{2}}{\alpha   '}-3 \bar{c}_1'' \ ;
\ee
\be
\bar c_1 \to  \bar{c}_1
-2 i \bar{c}_1'\,\bar{c}_1\,c_1  \ ;
\ee
\be
\gamma & \to & 
\gamma 
-\sqrt{2} \OMEGA c\,c_1'
-\frac{i \OMEGA }{\sqrt{2}} \calt\,c_1
+\frac{1 }{\sqrt{2}} c'\,c_1
-i \OMEGA  c\,c_1\,\vplus'
-2 i \gamma \,\bar{c}_1'\,c_1
+ 2 i  \bar{\gamma }\,c_1'\,c_1
\xxx
&&
+\sqrt{2} \OMEGA \calt\,\bar{c}_1\,c_1'\,c_1+\left ( 
-i \sqrt{2} \OMEGA \right ) c'\,\bar{c}_1\,c_1'\,c_1+2 \OMEGA  c\,\bar{c}_1\,c_1'\,c_1\,\vplus' \ ;
\ee
\be
\bar \gamma  & \to & \bar\gamma +
	\biggl(
	{\sqrt{2}} c\, \bar{c}_1'
	-\frac{i}{\sqrt{2}  } \calt\, \bar{c}_1
	-\frac{1}{\sqrt{2}  } c'\, \bar{c}_1
	- {i} c\, \bar{c}_1\, \vplus'
	-{\sqrt{2}} \calt\, \bar{c}_1'\, \bar{c}_1\, c_1
\xxx
&&
	+ {i \sqrt{2}} c'\, \bar{c}_1'\, \bar{c}_1\, c_1
	- {2} c\, \bar{c}_1'\, \bar{c}_1\, c_1\, \vplus'
	\biggr)
	+ 2 i  \bar{\gamma }\, \bar{c}_1'\, c_1 \ ;
\ee
\be
\beta & \to & 
	\beta 
	+ \sqrt{2} \OMEGA b\, c_1
	+ 4 i \sqrt{2} \OMEGA   \balt\, c_1'
	+ 3 i \sqrt{2} \OMEGA  \balt'\, c_1
	-4 i \beta \, \bar{c}_1\, c_1'
	-8 i \beta \, \bar{c}_1'\, c_1	
\xxx
&&
	+ 2 i  \bar{\beta }\, c_1'\, c_1
	-2 \OMEGA  \balt\, c_1\, \vplus'
	-4 i \beta '\, \bar{c}_1\, c_1
	-8 \sqrt{2} \OMEGA \balt\, \bar{c}_1\, c_1''\, c_1
	-16 \sqrt{2}  \OMEGA  \balt\, \bar{c}_1'\, c_1'\, c_1
\xxx
&&
	-8 \sqrt{2} \OMEGA \balt'\, \bar{c}_1\, c_1'\, c_1
	-8 \beta \, \bar{c}_1'\, \bar{c}_1\, c_1'\, c_1 \ ;
\ee
\be
\bar\beta & \to & 
	\bar{\beta }
	-{2 i \sqrt{2}} \balt\, \bar{c}_1'
	-{2 i} \beta \, \bar{c}_1'\, \bar{c}_1
	+{4 \sqrt{2}}   \balt\, \bar{c}_1'\, \bar{c}_1\, c_1'
	+{4 \sqrt{2}} \balt'\, \bar{c}_1'\, \bar{c}_1\, c_1 \ ;
\ee
\be
\vplus & \to & 
\vplus + 2 \sqrt{2}  \bar{c}_1'\, c_1 \ ;
\ee
\be
\vminus & \to &
\vminus 
	-\frac{i \alpha '}{\sqrt{2}} c\, \balt  \ ; 
\ee
\be
\balt & \to & 
	\balt
	-\frac{1}{\sqrt{2} \OMEGA } \beta \, \bar{c}_1
	-\frac{1 }{\sqrt{2}} \bar{\beta }\, c_1
	+ 2 i  \left( \balt\, \bar{c}_1\, c_1'
	+  \balt\, \bar{c}_1'\, c_1
	+   \balt'\, \bar{c}_1\, c_1 \right) \ ; 
\ee
\be
\calt & \to & 
	\calt
	+  \sqrt{2}   c\, \vplus'
	+{2 \sqrt{2}} \gamma \, \bar{c}_1'
	-\sqrt{2} \OMEGA   \bar{\gamma }\, c_1'
	+ 3 \sqrt{2} \OMEGA   \bar{\gamma }'\, c_1
	+ 2  c\, \bar{c}_1'\, c_1'
	+ 6  c\, \bar{c}_1''\, c_1
	+i \calt\, \bar{c}_1\, c_1'
\xxx
&&
	-i \calt\, \bar{c}_1'\, c_1
	+ c'\, \bar{c}_1\, c_1'
	+ c'\, \bar{c}_1'\, c_1
	-i \calt'\, \bar{c}_1\, c_1
	- c''\, \bar{c}_1\, c_1
	-i \sqrt{2} c\, \bar{c}_1\, c_1\, \vplus''
	+ i \sqrt{2}  c\, \bar{c}_1\, c_1'\, \vplus'
\xxx
&&
	-i \sqrt{2}  c\, \bar{c}_1'\, c_1\, \vplus'
	+ 4  i \sqrt{2} \OMEGA   \bar{\gamma }\, \bar{c}_1'\, c_1'\, c_1
	-i \sqrt{2}  c'\, \bar{c}_1\, c_1\, \vplus'
	-4   \calt\, \bar{c}_1'\, \bar{c}_1\, c_1'\, c_1
\xxx
&&
	+ 4 i c'\, \bar{c}_1'\, \bar{c}_1\, c_1'\, c_1
	-4 \sqrt{2}  c\, \bar{c}_1'\, \bar{c}_1\, c_1'\, c_1\, \vplus'  \ .
\ee

\appendix{Commuting currents}
\label{currents}
It turns out that there are a number of linearly independent currents that
commute with the BRST current.  Here, we present these currents explicitly
after our final variable redefinition, in
which the BRST current appears in the universal form in Eqn.~\rr{BRSTfinal}.  
The existence of these currents 
implies that the procedure for finding the generating function $g_0$
leaves a number of free coefficients undetermined.  In obtaining the
final version of $g_0$ used to generate the 
transformation in Eqn.~\rr{brstcomm}, we can use this freedom to choose particularly
simple transformations rules for a subset of the total field content.
The currents appear as follows:
\be
\jcurrent\ll 1 & = & i \bar \beta \, \gamma + c_1 \, \bar b_1
\ ; 
\XXX
\jcurrent\ll 2 & = & \bar{c}_1\,b_1+ i \beta \,\bar{\gamma }
\ ; 
\XXX
\jcurrent\ll 3 & = & 
	i c\,\balt\,\bar{\beta }\,\gamma 
	+ c\,\balt\,c_1\,\bar{b}_1
	-i \sqrt{2} \OMEGA  c'\,c\,\balt\,\bar{\beta }\,c_1
	-\frac{2 \OMEGA }{\alpha '} c\,\bar{\beta }\,c_1\,\vminus' 
\ ; 
\XXX
\jcurrent\ll 4 & = & 	
	{i} c\,\beta \,\bar{\beta }\,\gamma \,\bar{c}_1
	+i c\,\beta \,\bar{\gamma }\,\bar{\beta }\,c_1
	+ c\,\beta\,\bar{c}_1\,c_1\,\bar{b}_1
\xxx
&&
	+ c\,\bar{\beta }\,\bar{c}_1\,b_1\,c_1
	-{i \sqrt{2}} c'\,c\,\beta \,\bar{\beta }\,\bar{c}_1\,c_1
\ ; 
\XXX
\jcurrent\ll 5 & = &  \balt\,\bar{\gamma }\,c_1'
	+ \balt\,\gamma '\,\bar{c}_1
	-\frac{2 \OMEGA }{\alpha '} \bar{c}_1\,c_1'\,\vminus' 
\ ; 
\XXX
\jcurrent\ll 6 & = &  \balt\,\gamma \,\bar{c}_1'
	+ \balt\,\bar{\gamma }'\,c_1
	+\frac{2 i}{\OMEGA  \alpha'} \bar{c}_1\,c_1\,\vminus'' 
	+\frac{2 i}{\OMEGA  \alpha '} \bar{c}_1\,c_1'\,\vminus' 
\ ; 
\XXX
\jcurrent\ll 7 & = & \balt\,\gamma \,\bar{c}_1'- \balt\,\bar{\gamma }\,c_1'
	+ \balt'\,\gamma \,\bar{c}_1+\bar{c}_1\,c_1'\,\vminus' \frac{2 i \OMEGA}{\alpha '}
\ ; 
\XXX
\jcurrent\ll 8 & = &  - \balt\,\gamma \,\bar{c}_1'
	+ \balt\,\bar{\gamma }\,c_1'
	+ \balt'\,\bar{\gamma }\,c_1
	-\frac{2 i}{\OMEGA  \alpha '} \bar{c}_1\,c_1\,\vminus'' 
	-\frac{2 i}{\OMEGA  \alpha '} \bar{c}_1\,c_1'\,\vminus' 
\ ; 
\XXX
\jcurrent\ll 9 & = &  \balt\,\gamma \,\bar{c}_1\,\vminus'
	+ \balt\,\bar{\gamma }\,c_1\,\vminus'
	-\frac{2 i}{\OMEGA  \alpha '}  \bar{c}_1\,c_1\,\vminus'\,\vminus'  
\ ; 
\XXX
\jcurrent\ll {10} & = &  \balt\,\gamma \,\bar{c}_1\,\vplus'
	+ \balt\,\bar{\gamma }\,c_1\,\vplus'
	+ \balt\,\calt'\,\bar{c}_1\,c_1
	-\frac{2 i}{\OMEGA  \alpha '}\bar{c}_1\,c_1\,\vplus'\,\vminus' 
\ ; 
\XXX
\jcurrent\ll {11} & = & -{i} \balt\,\gamma \,\bar{c}_1\,c_1\,\bar{b}_1
	+ \balt\,\bar{\beta }\,\gamma \,\gamma \,\bar{c}_1
	+\balt\,\bar{\gamma }\,\bar{\beta }\,\gamma \,c_1
	-\frac{2 i}{\OMEGA  \alpha '}\bar{\beta }\,\gamma \,\bar{c}_1\,c_1\,\vminus' 
\ ; 
\XXX
\jcurrent\ll {12} & = & {i} \balt\,\beta \,\bar{\gamma }\,\gamma \,\bar{c}_1
	+i \balt\,\beta \,\bar{\gamma }\,\bar{\gamma }\,c_1
	+ \balt\,\bar{\gamma}\,\bar{c}_1\,b_1\,c_1
	+ \frac{2 \beta \,\bar{\gamma }\,\bar{c}_1\,c_1\,\vminus'}{\OMEGA  \alpha '}
\ ; 
\XXX
\jcurrent\ll {13} & = & \frac{1 }{\sqrt{2}} c\,\balt\,\beta \,\bar{\gamma }
	-\frac{i \OMEGA }{\sqrt{2}} c\,\balt\,\bar{c}_1\,b_1
	+ c'\,c\,\balt\,\beta \,\bar{c}_1
	-\frac{i \sqrt{2}}{\alpha '}c\,\beta \,\bar{c}_1\,\vminus' 
\ ; 
\XXX
\jcurrent\ll {14} & = & - \bar{c}_1\,c_1''+  \bar{c}_1''\,c_1
\ ; 
\XXX
\jcurrent\ll{ 15} & = & - \bar{\beta }\,\gamma \,\bar{c}_1\,c_1'
	+ \bar{\beta }\,\gamma '\,\bar{c}_1\,c_1
	+ \bar{\gamma }\,\bar{\beta}\,c_1'\,c_1
	+{i} \bar{c}_1\,c_1'\,c_1\,\bar{b}_1
\ ; 
\XXX
\jcurrent\ll {16} & = & \bar{c}_1'\,\bar{c}_1\,b_1\,c_1 
	+ {i \beta \,\gamma \,\bar{c}_1'\,\bar{c}_1} 
	+ i \beta \,\bar{\gamma }\,\bar{c}_1'\,c_1 
   -i \beta \,\bar{\gamma }'\,\bar{c}_1\,c_1  
\ ; 
\XXX
\jcurrent\ll {17} & = &  c\,\balt'\,\balt\,\gamma \,\bar{c}_1
	+ c\,\balt'\,\balt\,\bar{\gamma }\,c_1
	-{\sqrt{2}}c'\,c\,\balt'\,\balt\,\bar{c}_1\,c_1
\xxx
&&
	-\frac{2 i}{\OMEGA  \alpha'} c\,\balt'\,\bar{c}_1\,c_1\,\vminus' 
	+\frac{2 i}{\OMEGA  \alpha '} c\,\balt\,\bar{c}_1\,c_1\,\vminus'' 
\ ; 
\XXX
\jcurrent\ll {18} & = & - c\,\balt\,\bar{\beta }\,\gamma \,\bar{c}_1\,c_1'
	+ c\,\balt\,\bar{\beta }\,\gamma '\,\bar{c}_1\,c_1
	+c\,\balt\,\bar{\gamma }\,\bar{\beta }\,c_1'\,c_1
	+{i} c\,\balt\,\bar{c}_1\,c_1'\,c_1\,\bar{b}_1
\xxx
&&
	+{\sqrt{2}} c'\,c\,\balt\,\bar{\beta }\,\bar{c}_1\,c_1'\,c_1
	-\frac{2 i}{\OMEGA  \alpha '} c\,\bar{\beta }\,\bar{c}_1\,c_1'\,c_1\,\vminus' 
\ ; 
\XXX
\jcurrent\ll {19} & = & {i} c\,\balt\,\beta \,\gamma \,\bar{c}_1'\,\bar{c}_1
	+i c\,\balt\,\beta \,\bar{\gamma }\,\bar{c}_1'\,c_1
	-i c\,\balt\,\beta\,\bar{\gamma }'\,\bar{c}_1\,c_1
	+ c\,\balt\,\bar{c}_1'\,\bar{c}_1\,b_1\,c_1
\xxx
&&
	-{i \sqrt{2}} c'\,c\,\balt\,\beta \,\bar{c}_1'\,\bar{c}_1\,c_1
	-\frac{2}{\OMEGA  \alpha '}c\,\beta \,\bar{c}_1'\,\bar{c}_1\,c_1\,\vminus' 
\ ; 
\XXX
\jcurrent\ll {20} & = &  \balt\,\gamma \,\bar{c}_1'\,\bar{c}_1\,c_1'
	- \balt\,\bar{\gamma }\,\bar{c}_1'\,c_1'\,c_1
	- \balt\,\gamma '\,\bar{c}_1'\,\bar{c}_1\,c_1
\xxx
&&
	+ \balt\,\bar{\gamma }'\,\bar{c}_1\,c_1'\,c_1
	-\frac{2 i}{\OMEGA  \alpha '} \bar{c}_1'\,\bar{c}_1\,c_1'\,c_1\,\vminus'  
\ ; 
\XXX
\jcurrent\ll {21} & = & - \balt\,\gamma \,\bar{c}_1'\,\bar{c}_1\,c_1'
	+ \balt\,\bar{\gamma }\,\bar{c}_1\,c_1''\,c_1
	+2 \balt\,\bar{\gamma  }\,\bar{c}_1'\,c_1'\,c_1
	+ \balt\,\gamma '\,\bar{c}_1'\,\bar{c}_1\,c_1
\xxx
&&
	+ \balt'\,\bar{\gamma }\,\bar{c}_1\,c_1'\,c_1
	+\frac{2 i}{\OMEGA  \alpha '} \bar{c}_1'\,\bar{c}_1\,c_1'\,c_1\,\vminus' 
\ . 
\ee


\bibliographystyle{utcaps}
\bibliography{dimchange}

\end{document}